\documentclass[aps,prd,reprint]{revtex4-1}

\usepackage{graphicx}
\usepackage{enumerate}
\usepackage{hyperref}
\usepackage{textcomp}
\usepackage{amsmath}
\usepackage{amssymb}
\usepackage{pgf}
\usepackage{bm}
\usepackage{epsfig}
\usepackage{figsize}

\begin{document}

\title{Consistent combination of truncated-unity functional renormalization group and mean-field theory}

\author{Song-Jin O}
\email{sj.o@ryongnamsan.edu.kp}
\affiliation{Institute of Theoretical Physics, Faculty of Physics, Kim Il Sung University, Ryongnam-Dong, Taesong District, Pyongyang, Democratic People's Republic of Korea}
\date{\today}

\begin{abstract}
We propose a novel scheme for combining efficiently the truncated-unity functional renormalization group (TUFRG) and the mean-field theory. It follows the method of Wang, Eberlein, and Metzner that uses only the two-particle channel-irreducible part of the vertex as an input for the mean-field treatment. In the TUFRG, the neglect of fluctuation effects from other channels in the symmetry-broken regime is represented by applying the random phase approximation (RPA) in each individual channel, below the divergence scale. Then the Bethe-Salpeter equation for the four-point vertex is translated into the RPA matrix equations for the bosonic propagators that relates the singular and irreducible singular modes of the propagators. The universal symmetries for the irreducible singular modes are obtained from the antisymmetry of Grassmann variables. The mean-field equation based on these modes is derived by the saddle-point approximation in the framework of the path-integral formalism. By using our scheme, the power of the TUFRG, as a diagrammatically unbiased tool for identifying the many-body instabilities, could be elevated to a quantitatively reasonable level, and its application would be extended to a quantitatively reasonable analysis of the coexisting orders. As an illustration, we employ this scheme to study the coexistence phase of the chiral superconductivity and the chiral spin-density wave, predicted near van Hove filling of the honeycomb lattice.
\end{abstract}


\maketitle

\section{Introduction}

Intermediately correlated electron systems exhibit a plethora of fascinating properties and their theoretical analysis continue to be one of the important tasks in condensed matter physics. The systems can be described within both itinerant and strong coupling approaches. Among the itinerant approaches, the functional renormalization group (FRG) has proven to be an effective and reliable tool to study Fermi-surface instabilities in a diagrammatically unbiased way \cite{ref01, ref02, ref03}. It treats the pairing, spin, and charge channels on equal footing, and thus enables to investigate the fluctuation-driven instabilities and competing or coexisting orders \cite{ref04}. The FRG method has been successfully applied to capture the $d$-wave pairing instability in the two-dimensional (2D) repulsive Hubbard model \cite{ref05, ref06, ref07, ref08} and the extended $s$-wave superconductivity (SC) in Fe-based superconductors \cite{ref09, ref10, ref11}.

In these FRG studies the evolution of the four-point vertex was approximated by a one-loop truncation, where the six-point and the higher-order vertices were completely neglected. Furthermore, in many FRG calculations one applies additional approximations of discarding the frequency dependence of and the self-energy feedback to the four-point vertex. They are justified in the weak coupling regime by the fact that the frequency dependence appears, in its power counting, to be irrelevant \cite{ref03, ref12}, and the higher-order vertices and the self-energy correction can only make contributions of third order in the bare interaction \cite{ref01, ref02}. We will also use these approximations in the present work.

In the FRG method, Fermi-surface instabilities are signaled by divergences of the four-point vertex at a divergence energy scale $\Omega_D$. The FRG flow of the four-point vertex should be stopped at this scale because, when the vertex becomes very large, the one-loop approximation is no longer valid. To complete the calculation and perform a quantitative analysis of the resulting ordered phases, one has to continue the flow below the divergence scale $\Omega_D$, allowing the emergence of the symmetry-broken phase. This task can be accomplished by using either of two approaches, within a purely fermionic or a partially bosonized formalism.

In the first approach \cite{ref13}, one inserts an infinitesimal symmetry-breaking component into the initial action and tracks the FRG flow of purely fermionic vertex functions. It was applied to obtain the exact solutions of the mean-field (MF) models for superconductivity \cite{ref13, ref14}, and to describe a formation of the $s$-wave superfluid phase in the 2D attracting Hubbard model \cite{ref15, ref16}. That scheme was also used to analyze a $d$-wave superconducting ($d$-wave SC) state in the 2D repulsive Hubbard model \cite{ref08} and an antiferromagnetic phase in a two-pocket model for the iron pnictides \cite{ref17}. In this approach, since the effective action is no longer charge/spin-conserving, the FRG flow equations would involve the abnormal vertices and become very complicated. Particularly, in complex systems with several potential instabilities, introducing various kinds of seeds for those into the flow is a formidable task.

In the second approach \cite{ref18}, the symmetry-broken phases in interacting fermion systems are treated by introducing bosonic order-parameter fields via the Hubbard-Stratonovich transformation and solving partially bosonized FRG flow equations. It was applied to describe the antiferromagnetic phase \cite{ref18} and the $d$-wave SC induced by antiferromagnetic fluctuation \cite{ref19}, both in the 2D repulsive Hubbard model. By using that partial bosonization approach, the competition of the antiferromagnetic and $d$-wave SC orders in the Hubbard model was analyzed \cite{ref20, ref21}, and the exact solution of the Tomonaga-Luttinger model was reproduced \cite{ref22}. It has also been employed to analyze the superfluid ground state of the 2D attracting Hubbard model, taking into account the effect of the order-parameter fluctuations \cite{ref23, ref24}. When distinct instabilities are competing, one should decouple the bare interaction in various channels, involving several bosonic fields, which leads to an ambiguity in splitting the interaction \footnote{Since this ambiguity is closely related to the Fierz identity $\delta _{\alpha \mu } \delta _{\beta \nu }  = \frac{1}{2}\left[ {\delta _{\alpha \nu } \delta _{\beta \mu }  + \vec \sigma _{\alpha \nu }  \cdot \vec \sigma _{\beta \mu } } \right]$, it is dubbed the \emph{Fierz ambiguity}. For a detailed discussion of it, see Refs. \cite{ref18, ref67}.} and may introduce a certain bias. Furthermore, the fluctuation effects associated with other channels appear to become more complicated than in the purely fermionic approach.

Although the above-mentioned schemes are logically reasonable and make it possible to continue the flow into the symmetry-broken phase, their implementation for multiband systems with competing orders are numerically demanding. In this case, one may neglect low-energy fluctuations and combine the FRG flow at high scales with a mean-field treatment of low-energy modes. In this renormalized mean-field theory \cite{ref26}, the flow of the four-point vertex is stopped near the divergence scale, i.e., before entering the symmetry-broken regime, and the remaining low-energy modes are treated in mean-field approximation, with a reduced effective interaction extracted from resulting vertex. With this scheme, the competitions of the antiferromagnetic and superconducting instabilities have been considered quantitatively in the Hubbard model for the cuprates \cite{ref26} and in an eight-band model for the iron arsenides \cite{ref27}. It has also been applied to describe the $s+id$ pairing state in a five-band model for the iron pnictides \cite{ref28} and a mixed state of the spin-singlet and triplet SC in an attracting Rashba model on the triangular lattice \cite{ref29}.

This approach makes sense in the FRG flow with the sharp momentum cutoff regulator, due to clear separation of the high- and low-energy modes. However, in the case of the frequency-dependent regulation scheme like the $\Omega$ scheme \cite{ref30} and the sharp frequency cutoff, or the temperature cutoff scheme \cite{ref31}, the renormalized MF approach is not applicable, as the high- and low-energy modes cannot entirely be separated. In this regard, it is, strictly speaking, not suitable even for the FRG study with the smooth momentum cutoff regulation as performed in Ref. [\onlinecite{ref29}]. In these cases, one may try to simply use the mean-field approach, plugging the effective interaction at the divergence scale into the MF equation, as done in Ref. [\onlinecite{ref32}], but it would give rise to double counting of the contributions from high-energy modes, leading to overestimation of the order parameters.

To extend the renormalized MF approach to the generic cases, Wang {\it et al.} \cite{ref33} developed a sophisticated FRG + MF procedure that is applicable to the FRG study with a general regulator. Unlike the renormalized MF theory, in this approach, only the part of resulting vertex, which is two-particle irreducible in a given channel, is inserted as the effective interaction into the MF equation. In the following we will call this part the channel-irreducible one. Moreover, the mean-field calculation here involves all the fermion degrees of freedom, which differs from the previous theory where only the low-energy modes are considered. It was applied to analyze the coexistence of $d$-wave SC and antiferromagnetism \cite{ref33} or incommensurate spin-density wave (incommensurate SDW) \cite{ref34} in the ground state of the 2D Hubbard model.

On the other hand, the FRG achieved substantial methodological developments, broadening its range of applications. A recent version of it, the truncated-unity FRG (TUFRG) approach \cite{ref35} features a high resolution of the transfer momenta and a concise matrix structure of its flow equation. The TUFRG method is based on the exchange parametrization FRG \cite{ref30} and the singular-mode FRG \cite{ref36} methods and is known to ensure a fast and highly resolved computation. It was successfully applied to analyze the electronic instabilities in several 2D single-band \cite{ref35, ref37, ref38} and multiband systems \cite{ref39, ref40, ref41, ref42, ref43}. It has also been employed to study three-dimensional systems \cite{ref44, ref45} and spin-orbit \cite{ref46, ref47, ref48} and electron-phonon \cite{ref49} coupled systems. Recently, the TUFRG has been extended to treat more complicated systems \cite{ref50, ref51, ref52}.

In this paper we propose a scheme for combining consistently the TUFRG with the MF theory for spontaneous symmetry breaking. To this end, we adopt the key idea of the efficient FRG + MF \cite{ref33}, in which only the channel-irreducible part of the four-point vertex resulting from the FRG flow is taken as an input interaction for the MF equation. In the matrix formalism of the TUFRG, the Bethe-Salpeter equation relating the channel-irreducible part and the vertex becomes simple RPA matrix equations for the bosonic propagators, by which the irreducible singular modes are extracted. We provide an explicit derivation of the MF equation built on the irreducible singular modes, by resorting to the saddle-point approximation of the path-integral formalism. As a first application to competing orders, we use our novel TUFRG + MF approach to study the coexistence phase of the chiral SC and the chiral SDW on a heavily doped honeycomb lattice.

This paper is organized as follows. In Sec.~\ref{sec2} we briefly outline the TUFRG approach. In Sec.~\ref{sec3}, we give a detailed description of our TUFRG + MF scheme and derive explicitly the MF equation using the saddle-point approximation of statistical field theory. In Sec.~\ref{sec4}, we analyze quantitatively the coexistence of chiral SC and SDW on the doped honeycomb lattice, by means of the TUFRG + MF approach. Finally, in Sec.~\ref{sec5} we draw our conclusions.

\begin{widetext}

\section{Overview of TUFRG} \label{sec2}

We start our description of the FRG with the field-theoretical expression for the grand canonical partition function of interacting electron systems \cite{ref53}:
\begin{equation}\label{eq001}
\begin{split}
\Xi  = \int {D\bar \psi {\kern 1pt} D\psi } \exp \left\{ { - \frac{1}{\hbar }\int_0^{\beta \hbar } {d\tau } \left[ {\sum\limits_l {\hbar \bar \psi _l (\tau  + 0)\frac{{d\psi _l (\tau )}}{{d\tau }} + H[\bar \psi ,\psi ]} } \right]} \right\}.
\end{split}	
\end{equation}
Here $\psi ,\bar \psi $ are fermionic Grassmann fields, $\beta$ is the inverse temperature, and $l \equiv (o,\sigma ,{\bf{k}})$ is a multi-index comprising an orbital (sublattice) index $o$, spin polarity $\sigma$, and wave vector $\bf{k}$, while the function $H[\bar \psi ,\psi ]$  is obtained by replacing $\hat c^ \dagger   \to \bar \psi ,\hat c \to \psi$ in the Hamiltonian $\hat H = H[\hat c^ \dagger  ,\hat c]$. Spin-SU(2)-invariant lattice systems of interacting electrons, having $N$ unit cells, are described by the following Hamiltonian:
\begin{equation}\label{eq002}
\begin{split}
&\hat H = \hat H_0  + \hat H_{{\mathop{\rm int}} } ,\hat H_0  = \sum\limits_{o,o'} {\sum\limits_{{\bf{k}},\sigma } {\hat c_{{\bf{k}}o\sigma }^ \dagger  (H_{oo'}^0 ({\bf{k}}) - \mu \delta _{oo'} )\hat c_{{\bf{k}}o'\sigma } } }, \\
&\hat H_{{\mathop{\rm int}} }  = \frac{1}{{2N}}\sum\limits_{o_1 ,o_2 ,o_3 ,o_4 } {\sum\limits_{{\bf{k}}_1 ,{\bf{k}}_2 ,{\bf{k}}_3 ,{\bf{k}}_4 } {\sum\limits_{\sigma ,\sigma '} {\delta _{{\bf{k}}_1  + {\bf{k}}_2 ,{\bf{k}}_3  + {\bf{k}}_4 } } } }
 V_{o_1 o_2 ,o_3 o_4 } ({\bf{k}}_1 ,{\bf{k}}_2 ;{\bf{k}}_3 ,{\bf{k}}_4 )\hat c_{{\bf{k}}_1 o_1 \sigma }^ \dagger  \hat c_{{\bf{k}}_2 o_2 \sigma '}^ \dagger  \hat c_{{\bf{k}}_4 o_4 \sigma '} \hat c_{{\bf{k}}_3 o_3 \sigma }.
\end{split}	
\end{equation}
In this case, one can perform a Fourier transformation of the Grassmann variables $\psi _l (\tau ) = \frac{1}{{\sqrt {\beta \hbar } }}\sum\limits_\omega  {\psi _l (\omega )e^{ - i\omega \tau } } $ to derive the following expression for the partition function in frequency space \cite{ref54}:
\begin{equation}\label{eq003}
\begin{split}
\Xi  =& \int {D\bar \psi {\kern 1pt} D\psi } \exp \left\{ { - S[\psi ,\bar \psi ]} \right\} = \int {D\bar \psi {\kern 1pt} D\psi } \exp \left\{ { - S_0 [\psi ,\bar \psi ] - S_{{\mathop{\rm int}} } [\psi ,\bar \psi ]} \right\},\\
S_0 [\psi ,\bar \psi ] \equiv &  - (\bar \psi ,[G^0 ]^{ - 1} \psi ) = \sum\limits_{o,o'} {\sum\limits_{k,\sigma } {\bar \psi _\sigma  (k,o)} } \left[ { - i\omega \delta _{oo'}  + \frac{1}{\hbar }(H_{oo'}^0 ({\bf{k}}) - \mu \delta _{oo'} )} \right]\psi _\sigma  (k,o'),\\
S_{{\mathop{\rm int}} } [\psi ,\bar \psi ] \equiv & \frac{1}{{2N\beta \hbar ^2 }}\sum\limits_{o_1 ,o_2 ,o_3 ,o_4 } {\sum\limits_{k_1 ,k_2 ,k_3 ,k_4 } {\sum\limits_{\sigma ,\sigma '} {V_{o_1 o_2 ,o_3 o_4 } ({\bf{k}}_1 ,{\bf{k}}_2 ;{\bf{k}}_3 ,{\bf{k}}_4 )} } } \\
& \times \delta _{k_1  + k_2 ,k_3  + k_4 } \bar \psi _\sigma  (k_1 ,o_1 )\bar \psi _{\sigma '} (k_2 ,o_2 )\psi _{\sigma '} (k_4 ,o_4 )\psi _\sigma  (k_3 ,o_3 ).
\end{split}	
\end{equation} 
 For convenience, here, we combine the momentum $\bf{k}$ and the frequency $\omega$ into a $(d+1)$-dimensional variable $k = ({\bf{k}},\omega )$. The noninteracting part ($S_0$) of the action includes the bare propagator, $G^0 ({\bf{k}},\omega ) = \left[ {i\omega  - \frac{1}{\hbar }(H^0 ({\bf{k}}) - \mu )} \right]^{ - 1}$.

The generating functional for connected Greens functions, $W[\eta ,\bar \eta ]$, is obtained from the action in Eq. (\ref{eq003}), by adding external sources $\eta ,\bar \eta$ to it and taking the logarithm of the functional integral
\begin{equation}\label{eq004}
\begin{split}
W[\eta ,\bar \eta ] =  - \ln \int {D\bar \psi {\kern 1pt} D\psi } e^{ - S[\psi ,\bar \psi ] + (\bar \eta ,\psi ) + (\bar \psi ,\eta )} .
\end{split}	
\end{equation}
Then, we obtain the generating functional of the one-particle irreducible (1PI) vertices, $\Gamma [\psi ,\bar \psi ]$, by a Legendre transformation of $W[\eta ,\bar \eta ]$:
\begin{equation}\label{eq005}
\begin{split}
\Gamma [\psi ,\bar \psi ] = W[\eta ,\bar \eta ] + (\bar \eta ,\psi ) + (\bar \psi ,\eta ),\hspace{2pc}
\psi  =  - \frac{{\partial W[\eta ,\bar \eta ]}}{{\partial \bar \eta }},\hspace{0.5pc}
\bar \psi  = \frac{{\partial W[\eta ,\bar \eta ]}}{{\partial \eta }}.
\end{split}	
\end{equation}

To set up the renormalization group flow, we introduce an artificial scale dependence to the bare propagator, i.e., $G^0 ({\bf{k}},\omega ) \to G^{0,\Omega } ({\bf{k}},\omega )$. This regularization procedure eliminates the infrared modes below the energy scale $\Omega$, and it can be implemented in different ways. In this paper we employ the $\Omega$ scheme \cite{ref30} in which the bare propagator is modified by the scale $\Omega$ as
\begin{equation}\label{eq006}
\begin{split}
G^0 ({\bf{k}},\omega ) \to G^{0,\Omega } ({\bf{k}},\omega ) = \frac{{\hbar ^2 \omega ^2 }}{{\hbar ^2 \omega ^2  + \Omega ^2 }}G^0 ({\bf{k}},\omega ).
\end{split}	
\end{equation}
The generating functional of the 1PI vertices, or the effective action $\Gamma [\psi ,\bar \psi ]$, is then defined with $G^{0,\Omega }$ and becomes scale dependent as well, $\Gamma  \to \Gamma ^\Omega$. Taking the derivative of $\Gamma ^\Omega$ with respect to $\Omega$ yields the functional flow equation. The initial condition of this flow equation is given as $\Gamma ^{\Omega  \to  + \infty }  \equiv \Gamma ^{(0)}  = S_{{\mathop{\rm int}} }$. The equation is then Taylor-expanded to provide an infinite hierarchy of flow equations for the 1PI vertices.

Within the above-mentioned approximations, namely, the one-loop truncation of the effective action and neglecting the self-energy feedback and the frequency dependence for the four-point vertex, we focus only on the evolution of the four-point part $\Gamma ^{\Omega ,(4)}$ of the action. For spin-SU(2)-invariant systems it can be expressed in terms of the effective interaction $V^\Omega$ as follows:
\begin{equation}\label{eq007}
\begin{split}
\Gamma ^{\Omega ,(4)} [\psi ,\bar \psi ] =& \frac{1}{{2N\beta \hbar ^2 }}\sum\limits_{o_1 ,o_2 ,o_3 ,o_4 } {\sum\limits_{k_1 ,k_2 ,k_3 ,k_4 } {\sum\limits_{\sigma ,\sigma '} {V_{o_1 o_2 ,o_3 o_4 }^\Omega  ({\bf{k}}_1 ,{\bf{k}}_2 ;{\bf{k}}_3 ,{\bf{k}}_4 )} } }\\
& \times \delta _{k_1  + k_2 ,k_3  + k_4 } \bar \psi _\sigma  (k_1 ,o_1 )\bar \psi _{\sigma '} (k_2 ,o_2 )\psi _{\sigma '} (k_4 ,o_4 )\psi _\sigma  (k_3 ,o_3 ).
\end{split}	
\end{equation}
The flow equation of $V^\Omega$ can be derived from the equation of the four-point vertex, and it is composed of three contributions \cite{ref07, ref55}:
\begin{equation}\label{eq008}
\begin{split}
\frac{d}{{d\Omega }}V^\Omega   = J^{{\rm{pp}}} (\Omega ) + J^{{\rm{ph,cr}}} (\Omega ) + J^{{\rm{ph,d}}} (\Omega ).
\end{split}	
\end{equation}
The concrete expressions of $J^{{\rm{pp}}} (\Omega )$, $J^{{\rm{ph,cr}}} (\Omega )$, and $J^{{\rm{ph,d}}} (\Omega )$ are as follows \cite{ref42}:
\begin{equation}\label{eq009}
\begin{split}
J_{o'_1 o'_2 ,o_1 o_2 }^{{\rm{pp}}(\Omega )} ({\bf{k'}}_1 ,{\bf{k'}}_2 ;{\bf{k}}_1 ,{\bf{k}}_2 ) =& - \sum\limits_{\mu ,\mu '} {\sum\limits_{\nu ,\nu '} {\int {dp} } } {\kern 1pt} {\kern 1pt} \frac{d}{{d\Omega }}\left[ {G_{\mu \nu }^{0,\Omega } ({\bf{p}} + {\bf{k'}}_1  + {\bf{k'}}_2 ,\omega )G_{\mu '\nu '}^{0,\Omega } ( - {\bf{p}}, - \omega )} \right]\\
& \times V_{o'_1 o'_2 ,\mu \mu '}^\Omega  ({\bf{k'}}_1 ,{\bf{k'}}_2 ;{\bf{p}} + {\bf{k'}}_1  + {\bf{k'}}_2 , - {\bf{p}})V_{\nu \nu ',o_1 o_2 }^\Omega  ({\bf{p}} + {\bf{k'}}_1  + {\bf{k'}}_2 , - {\bf{p}};{\bf{k}}_1 ,{\bf{k}}_2 ),
\end{split}	
\end{equation}
\begin{equation}\label{eq010}
\begin{split}
J_{o'_1 o'_2 ,o_1 o_2 }^{{\rm{ph,cr}}(\Omega )} ({\bf{k'}}_1 ,{\bf{k'}}_2 ;{\bf{k}}_1 ,{\bf{k}}_2 ) =& 
 - \sum\limits_{\mu ,\mu '} {\sum\limits_{\nu ,\nu '} {\int {dp} } } {\kern 1pt} {\kern 1pt} \frac{d}{{d\Omega }}\left[ {G_{\mu \nu }^{0,\Omega } ({\bf{p}} + {\bf{k'}}_1  - {\bf{k}}_2 ,\omega )G_{\nu '\mu '}^{0,\Omega } ({\bf{p}},\omega )} \right]\\
& \times V_{o'_1 \mu ',\mu o_2 }^\Omega  ({\bf{k'}}_1 ,{\bf{p}};{\bf{p}} + {\bf{k'}}_1  - {\bf{k}}_2 ,{\bf{k}}_2 )V_{\nu o'_2 ,o_1 \nu '}^\Omega  ({\bf{p}} + {\bf{k'}}_1  - {\bf{k}}_2 ,{\bf{k'}}_2 ;{\bf{k}}_1 ,{\bf{p}}),
\end{split}	
\end{equation}
\begin{equation}\label{eq011}
\begin{split}
J_{o'_1 o'_2 ,o_1 o_2 }^{{\rm{ph,d}}(\Omega )} ({\bf{k'}}_1 ,{\bf{k'}}_2 ;{\bf{k}}_1 ,{\bf{k}}_2 ) =&  - \sum\limits_{\mu ,\mu '} {\sum\limits_{\nu ,\nu '} {\int {dp} } } {\kern 1pt} {\kern 1pt} \frac{d}{{d\Omega }}\left[ {G_{\mu \nu }^{0,\Omega } ({\bf{p}} + {\bf{k'}}_1  - {\bf{k}}_1 ,\omega )G_{\nu '\mu '}^{0,\Omega } ({\bf{p}},\omega )} \right]\\
& \times [V_{o'_1 \mu ',\mu o_1 }^\Omega  ({\bf{k'}}_1 ,{\bf{p}};{\bf{p}} + {\bf{k'}}_1  - {\bf{k}}_1 ,{\bf{k}}_1 )V_{\nu o'_2 ,\nu 'o_2 }^\Omega  ({\bf{p}} + {\bf{k'}}_1  - {\bf{k}}_1 ,{\bf{k'}}_2 ;{\bf{p}},{\bf{k}}_2 )\\
& + V_{o'_1 \mu ',o_1 \mu }^\Omega  ({\bf{k'}}_1 ,{\bf{p}};{\bf{k}}_1 ,{\bf{p}} + {\bf{k'}}_1  - {\bf{k}}_1 )V_{\nu o'_2 ,o_2 \nu '}^\Omega  ({\bf{p}} + {\bf{k'}}_1  - {\bf{k}}_1 ,{\bf{k'}}_2 ;{\bf{k}}_2 ,{\bf{p}})\\
& - 2V_{o'_1 \mu ',o_1 \mu }^\Omega  ({\bf{k'}}_1 ,{\bf{p}};{\bf{k}}_1 ,{\bf{p}} + {\bf{k'}}_1  - {\bf{k}}_1 )V_{\nu o'_2 ,\nu 'o_2 }^\Omega  ({\bf{p}} + {\bf{k'}}_1  - {\bf{k}}_1 ,{\bf{k'}}_2 ;{\bf{p}},{\bf{k}}_2 )],
\end{split}	
\end{equation}
with a shorthand notation $\int {dp}  = \int {\frac{{d{\bf{p}}}}{{S_{{\rm{BZ}}} }}} \frac{1}{{\beta \hbar ^2 }}\sum\nolimits_\omega$ ($S_{{\rm{BZ}}}$ is the Brillouin zone area) and the implicit momentum conservation ${\bf{k'}}_1  + {\bf{k'}}_2  = {\bf{k}}_1  + {\bf{k}}_2$. The effective interaction is calculated by integrating Eq. (\ref{eq008}) with respect to the scale $\Omega$:
\begin{equation}\label{eq012}
\begin{split}
V^\Omega   =& V^{(0)}  + \Phi ^{{\rm{pp}}} (\Omega ) + \Phi ^{{\rm{ph,cr}}} (\Omega ) + \Phi ^{{\rm{ph,d}}} (\Omega ),\\
\Phi ^{{\rm{pp}}} (\Omega ) =& \int_{\Omega _0 }^\Omega  {d\Omega 'J^{{\rm{pp}}} (\Omega ')} ,\hspace{1pc} 
\Phi ^{{\rm{ph,cr}}} (\Omega ) = \int_{\Omega _0 }^\Omega  {d\Omega 'J^{{\rm{ph,cr}}} (\Omega ')} ,\hspace{1pc} 
\Phi ^{{\rm{ph,d}}} (\Omega ) = \int_{\Omega _0 }^\Omega  {d\Omega 'J^{{\rm{ph,d}}} (\Omega ')}.
\end{split}	
\end{equation}
Here $\Omega _0$ is the initial value of $\Omega$ (in our case of using the $\Omega$ scheme, $\Omega _0  =  + \infty $), $V^{(0)}  \equiv V^{\Omega _0 }  = V$ is the initial bare interaction, and $\Phi ^{{\rm{pp}}} (\Omega ),\Phi ^{{\rm{ph,cr}}} (\Omega )$, and $\Phi ^{{\rm{ph,d}}} (\Omega )$ are the single-channel coupling functions, respectively, in the particle-particle, crossed particle-hole, and direct particle-hole channels.

In the exchange parametrization FRG \cite{ref30}, three bosonic propagators are defined by projecting these coupling functions onto three channels:
\begin{equation}\label{eq013}
\begin{split}
P^\Omega   = {\rm{\hat P}}[\Phi ^{{\rm{pp}}} (\Omega )],\hspace{1pc} C^\Omega   = {\rm{\hat C}}[\Phi ^{{\rm{ph,cr}}} (\Omega )],\hspace{1pc} D^\Omega   = {\rm{\hat D}}[\Phi ^{{\rm{ph,d}}} (\Omega )],
\end{split}	
\end{equation}
or, more explicitly,
\begin{equation}\label{eq014}
\begin{split}
P_{o'_1 o'_2 m,o_1 o_2 n}^\Omega  ({\bf{q}}) &= \frac{1}{{S_{{\rm{BZ}}}^2 }}\int {d{\bf{p}}} \int {d{\bf{p'}}} f_m ({\bf{p}})f_n^* ({\bf{p'}})\Phi _{o'_1 o'_2 ,o_1 o_2 }^{{\rm{pp}}(\Omega )} ({\bf{p}} + {\bf{q}}, - {\bf{p}};{\bf{p'}} + {\bf{q}}, - {\bf{p'}}),\\
C_{o'_1 o_2 m,o_1 o'_2 n}^\Omega  ({\bf{q}}) &= \frac{1}{{S_{{\rm{BZ}}}^2 }}\int {d{\bf{p}}} \int {d{\bf{p'}}} f_m ({\bf{p}})f_n^* ({\bf{p'}})\Phi _{o'_1 o'_2 ,o_1 o_2 }^{{\rm{ph,cr}}(\Omega )} ({\bf{p}} + {\bf{q}},{\bf{p'}};{\bf{p'}} + {\bf{q}},{\bf{p}}),\\
D_{o'_1 o_1 m,o_2 o'_2 n}^\Omega  ({\bf{q}}) &= \frac{1}{{S_{{\rm{BZ}}}^2 }}\int {d{\bf{p}}} \int {d{\bf{p'}}} f_m ({\bf{p}})f_n^* ({\bf{p'}})\Phi _{o'_1 o'_2 ,o_1 o_2 }^{{\rm{ph,d}}(\Omega )} ({\bf{p}} + {\bf{q}},{\bf{p'}};{\bf{p}},{\bf{p'}} + {\bf{q}}).
\end{split}	
\end{equation}
Here $f_m ({\bf{p}}) = e^{i{\bf{R}}_m  \cdot {\bf{p}}} $ is the plane-wave basis with the Bravais lattice vector ${\bf{R}}_m$. The inverse transformation of Eq. (\ref{eq014}) is, e.g.,
\begin{equation}\nonumber
\Phi _{o'_1 o'_2 ,o_1 o_2 }^{{\rm{pp(}}\Omega {\rm{)}}} ({\bf{p}} + {\bf{q}}, - {\bf{p}};{\bf{k}} + {\bf{q}}, - {\bf{k}}) = \sum\limits_{m,n{\textrm{(infinite sum)}}} {P_{o'_1 o'_2 m,o_1 o_2 n}^\Omega  } ({\bf{q}})f_m^* ({\bf{p}})f_n ({\bf{k}}).
\end{equation}
In real computation one should introduce the truncation in the sum over the basis indices, which leads to somewhat approximate expressions of the single-channel coupling functions, such as, e.g.,
\begin{equation}\nonumber
\Phi _{o'_1 o'_2 ,o_1 o_2 }^{{\rm{pp(}}\Omega {\rm{)}}} ({\bf{p}} + {\bf{q}}, - {\bf{p}};{\bf{k}} + {\bf{q}}, - {\bf{k}}) \approx \sum\limits_{m,n{\textrm{(truncate sum)}}} {P_{o'_1 o'_2 m,o_1 o_2 n}^\Omega  ({\bf{q}})} f_m^* ({\bf{p}})f_n ({\bf{k}}).
\end{equation}
However, numerical experiences in the FRG studies demonstrate that the single-channel coupling functions are reproduced very well by the sum over ${\bf{R}}_m ,{\bf{R}}_n $ within the region determined by a suitable value of the cut-off radius $R_{{\rm{cut}}}$. Therefore, we simply use an equal sign, with implicit notation $\sum\nolimits_{m,n}  \equiv  \sum\nolimits_{m\left( {\left| {{\bf{R}}_m } \right| \le R_{{\rm{cut}}} } \right)} {\sum\nolimits_{n\left( {\left| {{\bf{R}}_n } \right| \le R_{{\rm{cut}}} } \right)} }$, in the inverse projections of Eq. (\ref{eq014}):
\begin{equation}\label{eq015}
\begin{split}
\Phi _{o'_1 o'_2 ,o_1 o_2 }^{{\rm{pp(}}\Omega {\rm{)}}} ({\bf{p}} + {\bf{q}}, - {\bf{p}};{\bf{k}} + {\bf{q}}, - {\bf{k}}) &= \sum\limits_{m,n} {P_{o'_1 o'_2 m,o_1 o_2 n}^\Omega  } ({\bf{q}})f_m^* ({\bf{p}})f_n ({\bf{k}}),\\
\Phi _{o'_1 o'_2 ,o_1 o_2 }^{{\rm{ph,cr(}}\Omega {\rm{)}}} ({\bf{p}} + {\bf{q}},{\bf{k}};{\bf{k}} + {\bf{q}},{\bf{p}}) &= \sum\limits_{m,n} {C_{o'_1 o_2 m,o_1 o'_2 n}^\Omega  } ({\bf{q}})f_m^* ({\bf{p}})f_n ({\bf{k}}),\\
\Phi _{o'_1 o'_2 ,o_1 o_2 }^{{\rm{ph,d}}(\Omega )} ({\bf{p}} + {\bf{q}},{\bf{k}};{\bf{p}},{\bf{k}} + {\bf{q}}) &= \sum\limits_{m,n} {D_{o'_1 o_1 m,o_2 o'_2 n}^\Omega  } ({\bf{q}})f_m^* ({\bf{p}})f_n ({\bf{k}}),
\end{split}	
\end{equation}
or, concisely,
\begin{equation}\label{eq016}
\begin{split}
\Phi ^{{\rm{pp}}} (\Omega ) = {\rm{\hat P}}^{ - 1} [P^\Omega  ],\hspace{1pc} 
\Phi ^{{\rm{ph,cr}}} (\Omega ) = {\rm{\hat C}}^{ - 1} [C^\Omega  ],\hspace{1pc} 
\Phi ^{{\rm{ph,d}}} (\Omega ) = {\rm{\hat D}}^{ - 1} [D^\Omega  ].
\end{split}	
\end{equation}
Due to the fast convergence of the expansions in Eq. (\ref{eq015}) and a feature of the bosonic propagators depending only on one momentum, not on three momenta, the parametrization of the effective interaction via the propagators seems to be efficient, requiring considerably reduced memory.

In the singular-mode FRG \cite{ref36, ref56}, the same expressions are used in the projections of the effective interaction onto three channels:
\begin{equation}\label{eq017}
\begin{split}
V^{\rm{P}} (\Omega ) = {\rm{\hat P}}[V^\Omega  ],\hspace{1pc} 
V^{\rm{C}} (\Omega ) = {\rm{\hat C}}[V^\Omega  ],\hspace{1pc} 
V^{\rm{D}} (\Omega ) = {\rm{\hat D}}[V^\Omega  ],
\end{split}	
\end{equation}
or, in more detail,
\begin{equation}\label{eq018}
\begin{split}
V_{o'_1 o'_2 m,o_1 o_2 n}^{{\rm{P}}(\Omega )} ({\bf{q}}) &= \frac{1}{{S_{{\rm{BZ}}}^2 }}\int {d{\bf{p}}} \int {d{\bf{p'}}} f_m ({\bf{p}})f_n^* ({\bf{p'}})V_{o'_1 o'_2 ,o_1 o_2 }^\Omega  ({\bf{p}} + {\bf{q}}, - {\bf{p}};{\bf{p'}} + {\bf{q}}, - {\bf{p'}}),\\
V_{o'_1 o_2 m,o_1 o'_2 n}^{{\rm{C}}(\Omega )} ({\bf{q}}) &= \frac{1}{{S_{{\rm{BZ}}}^2 }}\int {d{\bf{p}}} \int {d{\bf{p'}}} f_m ({\bf{p}})f_n^* ({\bf{p'}})V_{o'_1 o'_2 ,o_1 o_2 }^\Omega  ({\bf{p}} + {\bf{q}},{\bf{p'}};{\bf{p'}} + {\bf{q}},{\bf{p}}),\\
V_{o'_1 o_1 m,o_2 o'_2 n}^{{\rm{D}}(\Omega )} ({\bf{q}}) &= \frac{1}{{S_{{\rm{BZ}}}^2 }}\int {d{\bf{p}}} \int {d{\bf{p'}}} f_m ({\bf{p}})f_n^* ({\bf{p'}})V_{o'_1 o'_2 ,o_1 o_2 }^\Omega  ({\bf{p}} + {\bf{q}},{\bf{p'}};{\bf{p}},{\bf{p'}} + {\bf{q}}).
\end{split}	
\end{equation}
The inverse transformation of the above equation reads as follows:
\begin{equation}\label{eq019}
\begin{split}
V_{o'_1 o'_2 ,o_1 o_2 }^\Omega  ({\bf{p}} + {\bf{q}}, - {\bf{p}};{\bf{k}} + {\bf{q}}, - {\bf{k}}) = \sum\limits_{m,n} {V_{o'_1 o'_2 m,o_1 o_2 n}^{{\rm{P}}(\Omega )} } ({\bf{q}})f_m^* ({\bf{p}})f_n ({\bf{k}}),\\
V_{o'_1 o'_2 ,o_1 o_2 }^\Omega  ({\bf{p}} + {\bf{q}},{\bf{k}};{\bf{k}} + {\bf{q}},{\bf{p}}) = \sum\limits_{m,n} {V_{o'_1 o_2 m,o_1 o'_2 n}^{{\rm{C}}(\Omega )} } ({\bf{q}})f_m^* ({\bf{p}})f_n ({\bf{k}}),\\
V_{o'_1 o'_2 ,o_1 o_2 }^\Omega  ({\bf{p}} + {\bf{q}},{\bf{k}};{\bf{p}},{\bf{k}} + {\bf{q}}) = \sum\limits_{m,n} {V_{o'_1 o_1 m,o_2 o'_2 n}^{{\rm{D}}(\Omega )} } ({\bf{q}})f_m^* ({\bf{p}})f_n ({\bf{k}}),
\end{split}	
\end{equation}
or, briefly,
\begin{equation}\label{eq020}
\begin{split}
V^\Omega   = {\rm{\hat P}}^{ - 1} [V^{\rm{P}} (\Omega )] = {\rm{\hat C}}^{ - 1} [V^{\rm{C}} (\Omega )] = {\rm{\hat D}}^{ - 1} [V^{\rm{D}} (\Omega )].
\end{split}	
\end{equation}
Using the relation (\ref{eq012}) between the effective interaction and the single-channel coupling functions, we can represent the projection matrices of the effective interaction in terms of the bosonic propagators:
\begin{equation}\label{eq021}
\begin{split}
V^{\rm{P}} (\Omega ) &= V^{{\rm{P}},(0)}  + P^\Omega   + V^{{\rm{P}} \leftarrow {\rm{C}}} (\Omega ) + V^{{\rm{P}} \leftarrow {\rm{D}}} (\Omega ),\\
V^{\rm{C}} (\Omega ) &= V^{{\rm{C}},(0)}  + V^{{\rm{C}} \leftarrow {\rm{P}}} (\Omega ) + C^\Omega   + V^{{\rm{C}} \leftarrow {\rm{D}}} (\Omega ),\\
V^{\rm{D}} (\Omega ) &= V^{{\rm{D}},(0)}  + V^{{\rm{D}} \leftarrow {\rm{P}}} (\Omega ) + V^{{\rm{D}} \leftarrow {\rm{C}}} (\Omega ) + D^\Omega  ,
\end{split}	
\end{equation}
with
\begin{equation}\label{eq022}
\begin{split}
V^{{\rm{P}},(0)} &\equiv {\rm{\hat P}}[V^{(0)} ],\hspace{1pc} V^{{\rm{C}},(0)}  \equiv {\rm{\hat C}}[V^{(0)} ],\hspace{1pc} V^{{\rm{D}},(0)}  \equiv {\rm{\hat D}}[V^{(0)} ],\\
V^{{\rm{P}} \leftarrow {\rm{C}}} (\Omega ) &\equiv {\rm{\hat P}}[\Phi ^{{\rm{ph,cr}}} (\Omega )] = {\rm{\hat P}}\{ {\rm{\hat C}}^{ - 1} [C^\Omega  ]\} ,\hspace{1pc} 
V^{{\rm{P}} \leftarrow {\rm{D}}} (\Omega ) \equiv {\rm{\hat P}}[\Phi ^{{\rm{ph,d}}} (\Omega )] = {\rm{\hat P}}\{ {\rm{\hat D}}^{ - 1} [D^\Omega  ]\} ,\\
V^{{\rm{C}} \leftarrow {\rm{P}}} (\Omega ) &\equiv {\rm{\hat C}}[\Phi ^{{\rm{pp}}} (\Omega )] = {\rm{\hat C}}\{ {\rm{\hat P}}^{ - 1} [P^\Omega  ]\} ,\hspace{1pc} 
V^{{\rm{C}} \leftarrow {\rm{D}}} (\Omega ) \equiv {\rm{\hat C}}[\Phi ^{{\rm{ph,d}}} (\Omega )] = {\rm{\hat C}}\{ {\rm{\hat D}}^{ - 1} [D^\Omega  ]\} ,\\
V^{{\rm{D}} \leftarrow {\rm{P}}} (\Omega ) &\equiv {\rm{\hat D}}[\Phi ^{{\rm{pp}}} (\Omega )] = {\rm{\hat D}}\{ {\rm{\hat P}}^{ - 1} [P^\Omega  ]\} ,\hspace{1pc} 
V^{{\rm{D}} \leftarrow {\rm{C}}} (\Omega ) \equiv {\rm{\hat D}}[\Phi ^{{\rm{ph,cr}}} (\Omega )] = {\rm{\hat D}}\{ {\rm{\hat C}}^{ - 1} [C^\Omega  ]\} .
\end{split}	
\end{equation}
The detailed expressions for the crossed contributions [$V^{{\rm{P}} \leftarrow {\rm{C}}} (\Omega ),V^{{\rm{P}} \leftarrow {\rm{D}}} (\Omega )$, etc.] can be found in Ref. [\onlinecite{ref42}].

Based on these preliminaries, the TUFRG flow equations for the bosonic propagators can be set up. Taking the derivative of $P^\Omega  ,C^\Omega$, and $D^\Omega$ with respect to $\Omega$, we get the following equation:
\begin{equation}\label{eq023}
\begin{split}
\frac{d}{{d\Omega }}P^\Omega   = \frac{d}{{d\Omega }}{\rm{\hat P}}[\Phi ^{{\rm{pp}}} (\Omega )] = {\rm{\hat P}}\left[ {\frac{d}{{d\Omega }}\Phi ^{{\rm{pp}}} (\Omega )} \right] = {\rm{\hat P[}}J^{{\rm{pp}}} (\Omega ){\rm{]}},\hspace{1pc} 
\frac{d}{{d\Omega }}C^\Omega   = {\rm{\hat C}}[J^{{\rm{ph,cr}}} (\Omega )],\hspace{1pc} 
\frac{d}{{d\Omega }}D^\Omega   = {\rm{\hat D}}[J^{{\rm{ph,d}}} (\Omega )].
\end{split}	
\end{equation}
Inserting Eqs. (\ref{eq009}) to (\ref{eq011}) into the above equation, and expressing the effective interaction via the projection matrices according to Eq. (\ref{eq020}), we derive a concise matrix form of the TUFRG flow equations for the bosonic propagators \cite{ref35}:
\begin{equation}\label{eq024}
\begin{split}
\frac{{dP^\Omega  ({\bf{q}})}}{{d\Omega }} &= V^{{\rm{P}}(\Omega )} ({\bf{q}})\left[ {\frac{d}{{d\Omega }}\chi ^{{\rm{pp}}(\Omega )} ({\bf{q}})} \right]V^{{\rm{P}}(\Omega )} ({\bf{q}}),\hspace{1pc} 
\frac{{dC^\Omega  ({\bf{q}})}}{{d\Omega }} = V^{{\rm{C}}(\Omega )} ({\bf{q}})\left[ {\frac{d}{{d\Omega }}\chi ^{{\rm{ph}}(\Omega )} ({\bf{q}})} \right]V^{{\rm{C}}(\Omega )} ({\bf{q}}),\\
\frac{{dD^\Omega  ({\bf{q}})}}{{d\Omega }} &= [V^{{\rm{C}}(\Omega )} ({\bf{q}}) - V^{{\rm{D}}(\Omega )} ({\bf{q}})]\left[ {\frac{d}{{d\Omega }}\chi ^{{\rm{ph}}(\Omega )} ({\bf{q}})} \right]V^{{\rm{D}}(\Omega )} ({\bf{q}}) + V^{{\rm{D}}(\Omega )} ({\bf{q}})\left[ {\frac{d}{{d\Omega }}\chi ^{{\rm{ph}}(\Omega )} ({\bf{q}})} \right][V^{{\rm{C}}(\Omega )} ({\bf{q}}) - V^{{\rm{D}}(\Omega )} ({\bf{q}})].
\end{split}	
\end{equation}
Here the particle-particle and particle-hole susceptibility matrices are defined as
\begin{equation}\label{eq025}
\begin{split}
\chi _{o'_1 o'_2 m,o_1 o_2 n}^{{\rm{pp(}}\Omega {\rm{)}}} ({\bf{q}}) &\equiv  - \int {\frac{{d{\bf{k}}}}{{S_{{\rm{BZ}}} }}} f_m ({\bf{k}})f_n^* ({\bf{k}})\left[ {\frac{1}{{\beta \hbar ^2 }}\sum\limits_\omega  {G_{o'_1 o_1 }^{0,\Omega } ({\bf{k}} + {\bf{q}},\omega )G_{o'_2 o_2 }^{0,\Omega } ( - {\bf{k}}, - \omega )} } \right],\\
\chi _{o'_1 o'_2 m,o_1 o_2 n}^{{\rm{ph(}}\Omega {\rm{)}}} ({\bf{q}}) &\equiv  - \int {\frac{{d{\bf{k}}}}{{S_{{\rm{BZ}}} }}} f_m ({\bf{k}})f_n^* ({\bf{k}})\left[ {\frac{1}{{\beta \hbar ^2 }}\sum\limits_\omega  {G_{o'_1 o_1 }^{0,\Omega } ({\bf{k}} + {\bf{q}},\omega )G_{o_2 o'_2 }^{0,\Omega } ({\bf{k}},\omega )} } \right].
\end{split}	
\end{equation}
Combined with Eq. (\ref{eq021}), the flow equation (\ref{eq024}) constitutes a closed system of the equations for $P^\Omega  ,C^\Omega$, and $D^\Omega$.

Let us consider the universal symmetries for the bosonic propagators. In the FRG flow, the particle-hole symmetry (PHS) and the remnant of antisymmetry (RAS) of Grassmann variables, inherited from the initial bare interaction, are satisfied by the effective interaction \cite{ref30}:
\begin{equation}\label{eq026}
V_{o'_1 o'_2 ,o_1 o_2 }^\Omega  ({\bf{k'}}_1 ,{\bf{k'}}_2 ;{\bf{k}}_1 ,{\bf{k}}_2 ) = \left[ {V_{o_1 o_2 ,o'_1 o'_2 }^\Omega  ({\bf{k}}_1 ,{\bf{k}}_2 ;{\bf{k'}}_1 ,{\bf{k'}}_2 )} \right]^* \hspace{2pc} \textrm{(PHS)} ,
\end{equation}
\begin{equation}\label{eq027}
V_{o'_1 o'_2 ,o_1 o_2 }^\Omega  ({\bf{k'}}_1 ,{\bf{k'}}_2 ;{\bf{k}}_1 ,{\bf{k}}_2 ) = V_{o'_2 o'_1 ,o_2 o_1 }^\Omega  ({\bf{k'}}_2 ,{\bf{k'}}_1 ;{\bf{k}}_2 ,{\bf{k}}_1 )  \hspace{2pc} \textrm{(RAS)}.
\end{equation}
These relations should also be satisfied by each of the single-channel coupling functions. For example, the PHS and RAS in the particle-particle channel are expressed as
\begin{equation}\label{eq028}
\Phi _{o'_1 o'_2 ,o_1 o_2 }^{{\rm{pp}}(\Omega )} ({\bf{k'}}_1 ,{\bf{k'}}_2 ;{\bf{k}}_1 ,{\bf{k}}_2 ) = \left[ {\Phi _{o_1 o_2 ,o'_1 o'_2 }^{{\rm{pp}}(\Omega )} ({\bf{k}}_1 ,{\bf{k}}_2 ;{\bf{k'}}_1 ,{\bf{k'}}_2 )} \right]^* \hspace{2pc} \textrm{(PHS)},
\end{equation}
\begin{equation}\label{eq029}
\Phi _{o'_1 o'_2 ,o_1 o_2 }^{{\rm{pp}}(\Omega )} ({\bf{k'}}_1 ,{\bf{k'}}_2 ;{\bf{k}}_1 ,{\bf{k}}_2 ) = \Phi _{o'_2 o'_1 ,o_2 o_1 }^{{\rm{pp}}(\Omega )} ({\bf{k'}}_2 ,{\bf{k'}}_1 ;{\bf{k}}_2 ,{\bf{k}}_1 ) \hspace{2pc} \textrm{(RAS)}.
\end{equation}
Combining the first equation of Eq. (\ref{eq014}) with Eq. (\ref{eq028}), one can easily derive the relation $P_{o'_1 o'_2 m,o_1 o_2 n}^\Omega  ({\bf{q}}) = \left[ {P_{o_1 o_2 n,o'_1 o'_2 m}^\Omega  ({\bf{q}})} \right]^*$. It is extended to the other channels, leading to the following symmetry relation \cite{ref41}:
\begin{equation}\label{eq030}
X_{o'_1 o'_2 m,o_1 o_2 n}^\Omega  ({\bf{q}}) = \left[ {X_{o_1 o_2 n,o'_1 o'_2 m}^\Omega  ({\bf{q}})} \right]^*  \textrm{ with } X \in \{ P,C,D\}  \hspace{2pc} \textrm{(PHS)}.
\end{equation}
Inserting Eq. (\ref{eq029}) into the first equation of Eq. (\ref{eq014}), we obtain
\begin{equation}\nonumber
\begin{split}
P_{o'_1 o'_2 m,o_1 o_2 n}^\Omega  ({\bf{q}}) &= \frac{1}{{S_{{\rm{BZ}}}^2 }}\int {d{\bf{p}}} \int {d{\bf{p'}}} e^{i{\bf{R}}_m  \cdot {\bf{p}}} e^{ - i{\bf{R}}_n  \cdot {\bf{p'}}} \Phi _{o'_2 o'_1 ,o_2 o_1 }^{{\rm{pp}}(\Omega )} ( - {\bf{p}},{\bf{p}} + {\bf{q}}; - {\bf{p'}},{\bf{p'}} + {\bf{q}}) \\
 &= \frac{1}{{S_{{\rm{BZ}}}^2 }}\int {d{\bf{k}}} \int {d{\bf{k'}}} e^{i{\bf{R}}_m  \cdot ( - {\bf{k}} - {\bf{q}})} e^{ - i{\bf{R}}_n  \cdot ( - {\bf{k'}} - {\bf{q}})} \Phi _{o'_2 o'_1 ,o_2 o_1 }^{{\rm{pp}}(\Omega )} ({\bf{k}} + {\bf{q}}, - {\bf{k}};{\bf{k'}} + {\bf{q}}, - {\bf{k'}})\\
 &= e^{ - i{\bf{R}}_m  \cdot {\bf{q}}} e^{i{\bf{R}}_n  \cdot {\bf{q}}} \frac{1}{{S_{{\rm{BZ}}}^2 }}\int {d{\bf{k}}} \int {d{\bf{k'}}} e^{i( - {\bf{R}}_m ) \cdot {\bf{k}}} e^{ - i( - {\bf{R}}_n ) \cdot {\bf{k'}}} \Phi _{o'_2 o'_1 ,o_2 o_1 }^{{\rm{pp}}(\Omega )} ({\bf{k}} + {\bf{q}}, - {\bf{k}};{\bf{k'}} + {\bf{q}}, - {\bf{k'}})\\
 &= e^{ - i{\bf{R}}_m  \cdot {\bf{q}}} P_{o'_2 ,o'_1 , - {\bf{R}}_m ;o_2 ,o_1 , - {\bf{R}}_n }^\Omega  ({\bf{q}})e^{i{\bf{R}}_n  \cdot {\bf{q}}} .
\end{split}
\end{equation}
We can derive the RAS relations for $C^\Omega$ and $D^\Omega$ in a similar way. Thus, we have the RAS relations for three bosonic propagators: \cite{ref41}
\begin{equation}\label{eq031}
\begin{split}
&P_{o'_1 o'_2 m,o_1 o_2 n}^\Omega  ({\bf{q}})  = e^{ - i{\bf{R}}_m  \cdot {\bf{q}}} P_{o'_2 o'_1 \bar m,o_2 o_1 \bar n}^\Omega  ({\bf{q}})e^{i{\bf{R}}_n  \cdot {\bf{q}}},\\
&X_{o'_1 o'_2 m,o_1 o_2 n}^\Omega  ( - {\bf{q}}) = e^{i{\bf{R}}_m  \cdot {\bf{q}}} \left[ {X_{o'_2 o'_{\rm{1}} \bar m,o_{\rm{2}} o_1 \bar n}^\Omega  ({\bf{q}})} \right]^* e^{ - i{\bf{R}}_n  \cdot {\bf{q}}} \textrm{ with } X \in \{C,D\},  \hspace{2pc} \textrm{(RAS)},
\end{split}
\end{equation}
where $\bar m$ is the basis index associated with the Bravais vector $-{\bf{R}}_m$.

In the following, we elaborate on the four-point part of the action. From now on, we will denote $\Gamma ^{\Omega ,(4)}$ as $\Gamma ^\Omega$ for simplicity. Substituting Eq. (\ref{eq012}) into Eq. (\ref{eq007}), and then inserting Eq. (\ref{eq015}) into it, we obtain the following expression for $\Gamma ^\Omega$:
\begin{equation}\label{eq032}
\Gamma ^\Omega  [\psi ,\bar \psi ] = \Gamma ^{(0)} [\psi ,\bar \psi ] + \Gamma ^{{\rm{pp}}} [\psi ,\bar \psi ] + \Gamma ^{{\rm{ph,cr}}} [\psi ,\bar \psi ] + \Gamma ^{{\rm{ph,d}}} [\psi ,\bar \psi ],
\end{equation}
with $\Gamma ^{(0)},\Gamma ^{{\rm{pp}}},\Gamma ^{{\rm{ph,cr}}}$, and $\Gamma ^{{\rm{ph,d}}}$, defined by
\begin{equation}\label{eq033}
\begin{split}
\Gamma ^{(0)} [\psi ,\bar \psi ] = & S_{{\mathop{\rm int}} } [\psi ,\bar \psi ] = \frac{1}{{2N\beta \hbar ^2 }}\sum\limits_{o_1 , \cdots ,o_4 } {\sum\limits_{k_1 ,k_2 ,k_3 ,k_4 } {\sum\limits_{\sigma ,\sigma '} {V_{o_1 o_2 ,o_3 o_4 } ({\bf{k}}_1 ,{\bf{k}}_2 ;{\bf{k}}_3 ,{\bf{k}}_4 )} } } \\
& \times \delta _{k_1  + k_2 ,k_3  + k_4 } \bar \psi _\sigma  (k_1 ,o_1 )\bar \psi _{\sigma '} (k_2 ,o_2 )\psi _{\sigma '} (k_4 ,o_4 )\psi _\sigma  (k_3 ,o_3 ),
\end{split}
\end{equation}
\begin{equation}\label{eq034}
\begin{split}
\Gamma ^{{\rm{pp}}} [\psi ,\bar \psi ] \equiv& \frac{1}{{2N\beta \hbar ^2 }}\sum\limits_{o_1 , \cdots ,o_4 } {\sum\limits_{q,p,k} {\sum\limits_{m,n} {e^{ - i{\bf{R}}_m  \cdot {\bf{p}}} } P_{o_1 o_2 m,o_3 o_4 n}^\Omega  ({\bf{q}})e^{i{\bf{R}}_n  \cdot {\bf{k}}} } } \\
& \times \sum\limits_{\sigma ,\sigma '} {[\bar \psi _\sigma  (p + q,o_1 )\bar \psi _{\sigma '} ( - p,o_2 )]} [\psi _{\sigma '} ( - k,o_4 )\psi _\sigma  (k + q,o_3 )],
\end{split}
\end{equation}
\begin{equation}\label{eq035}
\begin{split}
\Gamma ^{{\rm{ph,cr}}} [\psi ,\bar \psi ] \equiv&  - \frac{1}{{2N\beta \hbar ^2 }}\sum\limits_{o_1 , \cdots ,o_4 } {\sum\limits_{q,p,k} {\sum\limits_{m,n} {e^{ - i{\bf{R}}_m  \cdot {\bf{p}}} } C_{o_1 o_2 m,o_3 o_4 n}^\Omega  ({\bf{q}})e^{i{\bf{R}}_n  \cdot {\bf{k}}} } } \\
&\times \sum\limits_{\sigma ,\sigma '} {[\bar \psi _\sigma  (p + q,o_1 )\psi _{\sigma '} (p,o_2 )]} [\bar \psi _{\sigma '} (k,o_4 )\psi _\sigma  (k + q,o_3 )],
\end{split}
\end{equation}
\begin{equation}\label{eq036}
\begin{split}
\Gamma ^{{\rm{ph,d}}} [\psi ,\bar \psi ] \equiv& \frac{1}{{2N\beta \hbar ^2 }}\sum\limits_{o_1 , \cdots ,o_4 } {\sum\limits_{q,p,k} {\sum\limits_{m,n} {e^{ - i{\bf{R}}_m  \cdot {\bf{p}}} } D_{o_1 o_2 m,o_3 o_4 n}^\Omega  ({\bf{q}})e^{i{\bf{R}}_n  \cdot {\bf{k}}} } }\\
&\times \sum\limits_{\sigma ,\sigma '} {[\bar \psi _\sigma  (p + q,o_1 )\psi _\sigma  (p,o_2 )]} [\bar \psi _{\sigma '} (k,o_4 )\psi _{\sigma '} (k + q,o_3 )].
\end{split}
\end{equation}
 Taking into account the relation,
\begin{equation}\nonumber
\begin{split}
&\sum\limits_{\sigma ,\sigma '} {[\bar \psi _\sigma  (p_1 ,o_1 )\bar \psi _{\sigma '} (p_2 ,o_2 )]} [\psi _{\sigma '} (p_4 ,o_4 )\psi _\sigma  (p_3 ,o_3 )] =
 \frac{{\rm{1}}}{2} \left\{ {\left[ {\sum\limits_\sigma  {\sigma \bar \psi _\sigma  (p_1 ,o_1 )\bar \psi _{ - \sigma } (p_2 ,o_2 )} } \right]} \right.\left[ {\sum\limits_{\sigma '} {{\sigma '} \psi _{ - \sigma '} (p_4 ,o_4 )\psi _{\sigma '} (p_3 ,o_3 )} } \right]\\
&+ \left[ { - \sum\limits_\sigma  {\sigma \bar \psi _\sigma  (p_1 ,o_1 )\bar \psi _\sigma  (p_2 ,o_2 )} } \right]\left[ { - \sum\limits_{\sigma '} {{\sigma '}\psi _{\sigma '} (p_4 ,o_4 )\psi _{\sigma '} (p_3 ,o_3 )} } \right]
+ \left[ {i\sum\limits_\sigma  {\bar \psi _\sigma  (p_1 ,o_1 )\bar \psi _\sigma  (p_2 ,o_2 )} } \right]\left[ { - i\sum\limits_{\sigma '} {\psi _{\sigma '} (p_4 ,o_4 )\psi _{\sigma '} (p_3 ,o_3 )} } \right]\\
& + \left[ {\sum\limits_\sigma  {\bar \psi _\sigma  (p_1 ,o_1 )\bar \psi _{ - \sigma } (p_2 ,o_2 )} } \right]\left. {\left[ {\sum\limits_{\sigma '} {\psi _{ - {\sigma '}} (p_4 ,o_4 )\psi _{\sigma '} (p_3 ,o_3 )} } \right]} \right\},
\end{split}
\end{equation}
we can decompose $\Gamma ^{{\rm{pp}}}$ into the spin-singlet $\Gamma ^{{\rm{sSC}}}$ and the spin-triplet $\Gamma ^{{\rm{tSC}}}$ parts.
\begin{equation}\label{eq037}
\Gamma ^{{\rm{pp}}} [\psi ,\bar \psi ] = \Gamma ^{{\rm{sSC}}} [\psi ,\bar \psi ] + \Gamma ^{{\rm{tSC}}} [\psi ,\bar \psi ],
\end{equation}
\begin{equation}\label{eq038}
\begin{split}
\Gamma ^{{\rm{sSC}}} [\psi ,\bar \psi ] \equiv&  - \frac{{\rm{1}}}{2}\frac{1}{{2N\beta \hbar ^2 }}\sum\limits_{o_1 , \cdots ,o_4 } {\sum\limits_{q,p,k} {\sum\limits_{m,n} {e^{ - i{\bf{R}}_m  \cdot {\bf{p}}} } \left( { - P_{o_1 o_2 m,o_3 o_4 n}^\Omega  ({\bf{q}})} \right)e^{i{\bf{R}}_n  \cdot {\bf{k}}} } }\\
& \times \left[ {\sum\limits_\sigma  {\sigma \psi _{ - \sigma } ( - p,o_2 )\psi _\sigma  (p + q,o_1 )} } \right]^* \left[ {\sum\limits_{\sigma '} {{\sigma '}\psi _{ - {\sigma '}} ( - k,o_4 )\psi _{\sigma '} (k + q,o_3 )} } \right],
\end{split}
\end{equation}
\begin{equation}\label{eq039}
\begin{split}
\Gamma ^{{\rm{tSC}}} [\psi ,\bar \psi ] \equiv&  - \frac{1}{2}\frac{1}{{2N\beta \hbar ^2 }}\sum\limits_{o_1 , \cdots ,o_4 } {\sum\limits_{q,p,k} {\sum\limits_{m,n} {e^{ - i{\bf{R}}_m  \cdot {\bf{p}}} } \left( { - P_{o_1 o_2 m,o_3 o_4 n}^\Omega  ({\bf{q}})} \right)e^{i{\bf{R}}_n  \cdot {\bf{k}}} } } \\
& \times \left\{ {\left[ { - \sum\limits_\sigma  {\sigma \psi _\sigma  ( - p,o_2 )\psi _\sigma  (p + q,o_1 )} } \right]^* } \right.\left[ { - \sum\limits_{\sigma '} {{\sigma '} \psi _{\sigma '} ( - k,o_4 )\psi _{\sigma '} (k + q,o_3 )} } \right]\\
&+ \left[ { - i\sum\limits_\sigma  {\psi _\sigma  ( - p,o_2 )\psi _\sigma  (p + q,o_1 )} } \right]^* \left[ { - i\sum\limits_{\sigma '} {\psi _{\sigma '} ( - k,o_4 )\psi _{\sigma '} (k + q,o_3 )} } \right]\\
&+ \left[ {\sum\limits_\sigma  {\psi _{ - \sigma } ( - p,o_2 )\psi _\sigma  (p + q,o_1 )} } \right]^* \left. {\left[ {\sum\limits_{\sigma '} {\psi _{ - {\sigma '}} ( - k,o_4 )\psi _{\sigma '} (k + q,o_3 )} } \right]} \right\}.
\end{split}
\end{equation}
In addition, by using the relation
\begin{equation}\nonumber
\begin{split}
\sum\limits_{\sigma ,\sigma '} {[\bar \psi _\sigma  (p + q,o_1 )\psi _{\sigma '} (p,o_2 )]} [\bar \psi _{\sigma '} (k,o_4 )\psi _\sigma  (k + q,o_3 )]
 = \frac{{\rm{1}}}{2}\left[ {\sum\limits_\sigma  {\bar \psi _\sigma  (p + q,o_1 )\psi _\sigma  (p,o_2 )} } \right]\left[ {\sum\limits_{\sigma '} {\bar \psi _{\sigma '} (k,o_4 )\psi _{\sigma '} (k + q,o_3 )} } \right]&\\
+ \frac{{\rm{1}}}{2}\left[ {\sum\limits_{\sigma ,\sigma '} {\bar \psi _\sigma  (p + q,o_1 )\vec \sigma _{\sigma \sigma '} \psi _{\sigma '} (p,o_2 )} } \right] \cdot \left[ {\sum\limits_{s,s'} {\bar \psi _s (k,o_4 )\vec \sigma _{ss'} \psi _{s'} (k + q,o_3 )} } \right]&,
\end{split}
\end{equation}
we can modify Eq. (\ref{eq035}) as
\begin{equation}\label{eq040}
\begin{split}
\Gamma ^{{\rm{ph,cr}}} [\psi ,\bar \psi ] =&  - \frac{1}{{2N\beta \hbar ^2 }}\sum\limits_{o_1 , \cdots ,o_4 } {\sum\limits_{q,p,k} {\sum\limits_{m,n} {e^{ - i{\bf{R}}_m  \cdot {\bf{p}}} } \frac{{\rm{1}}}{2}C_{o_1 o_2 m,o_3 o_4 n}^\Omega  ({\bf{q}})e^{i{\bf{R}}_n  \cdot {\bf{k}}} } } \\
 &\times \left\{ {\left[ {\sum\limits_{\sigma ,\sigma '} {\bar \psi _\sigma  (p + q,o_1 )\vec \sigma _{\sigma \sigma '} \psi _{\sigma '} (p,o_2 )} } \right] \cdot \left[ {\sum\limits_{s,s'} {\bar \psi _s (k,o_4 )\vec \sigma _{ss'} \psi _{s'} (k + q,o_3 )} } \right]} \right.\\
 &+ \left. {\left[ {\sum\limits_\sigma  {\bar \psi _\sigma  (p + q,o_1 )\psi _\sigma  (p,o_2 )} } \right]\left[ {\sum\limits_{\sigma '} {\bar \psi _{\sigma '} (k,o_4 )\psi _{\sigma '} (k + q,o_3 )} } \right]} \right\}.
\end{split}
\end{equation}
Substituting Eqs. (\ref{eq036}) to (\ref{eq040}) into Eq. (\ref{eq032}), we obtain the following representation for $\Gamma^\Omega$:
\begin{equation}\label{eq041}
\Gamma ^\Omega  [\psi ,\bar \psi ] = \Gamma ^{(0)} [\psi ,\bar \psi ] + \Gamma ^{{\rm{sSC}}} [\psi ,\bar \psi ] + \Gamma ^{{\rm{tSC}}} [\psi ,\bar \psi ] + \Gamma ^{{\rm{SPN}}} [\psi ,\bar \psi ] + \Gamma ^{{\rm{CHG}}} [\psi ,\bar \psi ],
\end{equation}
with $\Gamma ^{(0)}, \Gamma ^{{\rm{sSC}}}$, and $\Gamma ^{{\rm{tSC}}}$ defined in Eqs. (\ref{eq033}), (\ref{eq038}), and (\ref{eq039}), respectively, and with $\Gamma ^{{\rm{SPN}}}$ and $\Gamma ^{{\rm{SPN}}}$ defined as
\begin{equation}\label{eq042}
\begin{split}
\Gamma ^{{\rm{SPN}}} [\psi ,\bar \psi ] =&  - \frac{{\rm{1}}}{2}\frac{1}{{2N\beta \hbar ^2 }}\sum\limits_{o_1 , \cdots ,o_4 } {\sum\limits_{q,p,k} {\sum\limits_{m,n} {e^{ - i{\bf{R}}_m  \cdot {\bf{p}}} } C_{o_1 o_2 m,o_3 o_4 n}^\Omega  ({\bf{q}})e^{i{\bf{R}}_n  \cdot {\bf{k}}} } }\\
&\times \left[ {\sum\limits_{\sigma ,\sigma '} {\bar \psi _\sigma  (p,o_2 )\vec \sigma _{\sigma \sigma '} \psi _{\sigma '} (p + q,o_1 )} } \right]^*  \cdot \left[ {\sum\limits_{s,s'} {\bar \psi _s (k,o_4 )\vec \sigma _{ss'} \psi _{s'} (k + q,o_3 )} } \right],
\end{split}
\end{equation}
\begin{equation}\label{eq043}
\begin{split}
\Gamma ^{{\rm{CHG}}} [\psi ,\bar \psi ] =&  - \frac{{\rm{1}}}{2}\frac{1}{{2N\beta \hbar ^2 }}\sum\limits_{o_1 , \cdots ,o_4 } {\sum\limits_{q,p,k} {\sum\limits_{m,n} {e^{ - i{\bf{R}}_m  \cdot {\bf{p}}} } W_{o_1 o_2 m,o_3 o_4 n}^\Omega  ({\bf{q}})e^{i{\bf{R}}_n  \cdot {\bf{k}}} } }\\
&\times \left[ {\sum\limits_\sigma  {\bar \psi _\sigma  (p,o_2 )\psi _\sigma  (p + q,o_1 )} } \right]^* \left[ {\sum\limits_{\sigma '} {\bar \psi _{\sigma '} (k,o_4 )\psi _{\sigma '} (k + q,o_3 )} } \right].
\end{split}
\end{equation}
In the above equation, $W^\Omega  ({\bf{q}})=C^\Omega  ({\bf{q}}) - 2D^\Omega  ({\bf{q}})$ is the bosonic propagator in the charge channel. It is easy to verify that the flow equation (\ref{eq024}) can be rewritten as
\begin{equation}\label{eq044}
\begin{split}
\frac{{d[ - P^\Omega  ({\bf{q}})]}}{{d\Omega }} &= [ - V^{{\rm{P}}(\Omega )} ({\bf{q}})]\frac{{d[ - \chi ^{{\rm{pp}}(\Omega )} ({\bf{q}})]}}{{d\Omega }}[ - V^{{\rm{P}}(\Omega )} ({\bf{q}})],\\
\frac{{dC^\Omega  ({\bf{q}})}}{{d\Omega }} &= V^{{\rm{C}}(\Omega )} ({\bf{q}})\frac{{d\chi ^{{\rm{ph}}(\Omega )} ({\bf{q}})}}{{d\Omega }}V^{{\rm{C}}(\Omega )} ({\bf{q}}),\hspace{1pc} 
\frac{{dW^\Omega  ({\bf{q}})}}{{d\Omega }} = V^{{\rm{W}}(\Omega )} ({\bf{q}})\frac{{d\chi ^{{\rm{ph}}(\Omega )} ({\bf{q}})}}{{d\Omega }}V^{{\rm{W}}(\Omega )} ({\bf{q}}),
\end{split}
\end{equation}
where $W^\Omega  ({\bf{q}})$ and $V^{{\rm{W}}(\Omega )} ({\bf{q}})$ are defined as
\begin{equation}\label{eq045}
W^\Omega  ({\bf{q}}) \equiv C^\Omega  ({\bf{q}}) - 2D^\Omega  ({\bf{q}}),\hspace{1pc} 
V^{{\rm{W}}(\Omega )} ({\bf{q}}) \equiv V^{{\rm{C}}(\Omega )} ({\bf{q}}) - 2V^{{\rm{D}}(\Omega )} ({\bf{q}}).
\end{equation}

\end{widetext}

\section{TUFRG + MF approach}\label{sec3}

\subsection{TUFRG flow and RPA equations}\label{sec3A}
To access the symmetry-broken phases, we follow the idea of Wang, Eberlein, and Metzner \cite{ref33}, where the FRG flow of the vertex is exactly taken into account above the divergence scale ($\Omega  \ge \Omega _D$), while the contributions from the fluctuation channels are discarded at lower-energy scale ($\Omega  < \Omega _D$). Namely, we compute the bosonic propagators by integrating the TUFRG flow equation (\ref{eq044}) at high scale $\Omega  \ge \Omega _D$, but at low scale $\Omega  < \Omega _D$, we employ the following approximation for the projection matrices:
\begin{equation}\label{eq046}
\begin{split}
&V^{\rm{P}} (\Omega ) = V^{{\rm{P}},(0)}  + P^\Omega   + V^{{\rm{P}} \leftarrow {\rm{C}}} (\Omega ) + V^{{\rm{P}} \leftarrow {\rm{D}}} (\Omega ) \approx P^\Omega,\\
&V^{\rm{C}} (\Omega ) = V^{{\rm{C}},(0)}  + V^{{\rm{C}} \leftarrow {\rm{P}}} (\Omega ) + C^\Omega   + V^{{\rm{C}} \leftarrow {\rm{D}}} (\Omega ) \approx C^\Omega,\\
&V^{\rm{D}} (\Omega ) = V^{{\rm{D}},(0)}  + V^{{\rm{D}} \leftarrow {\rm{P}}} (\Omega ) + V^{{\rm{D}} \leftarrow {\rm{C}}} (\Omega ) + D^\Omega   \approx D^\Omega,\\
& \textrm{at } \Omega  < \Omega _D.
\end{split}
\end{equation}
This approximation is justified by the fact that, in the symmetry-broken regime, we focus only on the divergent parts of the effective interaction, and they are captured satisfactorily by Eq. (\ref{eq046}). Generally speaking, since the bare interaction is moderate, the initial projections, $V^{{\rm{P}},(0)} ,V^{{\rm{C}},(0)}$, and $V^{{\rm{D}},(0)}$ are not divergent. Moreover, the crossed contributions like, e.g., $V^{{\rm{C}} \leftarrow {\rm{P}}} (\Omega )$ and $V^{{\rm{D}} \leftarrow {\rm{P}}} (\Omega )$ do not have sharp peaks that are necessary for developing some orders, even though $P^{\Omega}$ has a peak. Under the approximation of Eq. (\ref{eq046}), the flow equation (\ref{eq044}) becomes
\begin{equation}\label{eq047}
\begin{split}
&\frac{{d \left[ - P^\Omega  ({\bf{q}}) \right]}}{{d\Omega }} \approx \left[ - P^\Omega  ({\bf{q}})\right]\frac{{d\left[ - \chi ^{{\rm{pp}}(\Omega )} ({\bf{q}})\right]}}{{d\Omega }} \left[ - P^\Omega  ({\bf{q}})\right], \\
&\frac{{dC^\Omega  ({\bf{q}})}}{{d\Omega }} \approx C^\Omega  ({\bf{q}})\frac{{d\chi ^{{\rm{ph}}(\Omega )} ({\bf{q}})}}{{d\Omega }}C^\Omega  ({\bf{q}}), \\
&\frac{{dW^\Omega  ({\bf{q}})}}{{d\Omega }} \approx W^\Omega  ({\bf{q}})\frac{{d\chi ^{{\rm{ph}}(\Omega )} ({\bf{q}})}}{{d\Omega }}W^\Omega  ({\bf{q}}), \\
&\textrm{at } \Omega  < \Omega _D,
\end{split}
\end{equation}
which has an exact solution
\begin{equation}\label{eq048}
\begin{split}
&\left[ - P^\Omega  ({\bf{q}})\right]^{ - 1}  - \left[ - P^{\Omega _D} ({\bf{q}})\right]^{ - 1}  =  - \chi ^{{\rm{pp}}(\Omega _D )} ({\bf{q}}) + \chi ^{{\rm{pp}}(\Omega )} ({\bf{q}}), \\
&\left[C^\Omega  ({\bf{q}})\right]^{ - 1}  - \left[C^{\Omega _D } ({\bf{q}})\right]^{ - 1}  = \chi ^{{\rm{ph}}(\Omega _D )} ({\bf{q}}) - \chi ^{{\rm{ph}}(\Omega )} ({\bf{q}}), \\
&\left[W^\Omega  ({\bf{q}})\right]^{ - 1}  - \left[W^{\Omega _D } ({\bf{q}})\right]^{ - 1}  = \chi ^{{\rm{ph}}(\Omega _D )} ({\bf{q}}) - \chi ^{{\rm{ph}}(\Omega )} ({\bf{q}}),\\
& \textrm{ at } \Omega  < \Omega _D.
\end{split}
\end{equation}

Now we introduce the irreducible bosonic propagators, $\tilde P,\tilde C$, and $\tilde W$, defined by
\begin{equation}\label{eq049}
\begin{split}
&\left[ - \tilde P({\bf{q}})\right]^{ - 1}  \equiv \left[ - P^{\Omega _D } ({\bf{q}})\right]^{ - 1}  - \chi ^{{\rm{pp}}(\Omega _D )} ({\bf{q}}),\\
&\left[\tilde C({\bf{q}})\right]^{ - 1}  \equiv \left[C^{\Omega _D } ({\bf{q}})\right]^{ - 1}  + \chi ^{{\rm{ph}}(\Omega _D )} ({\bf{q}}), \\
&\left[\tilde W({\bf{q}})\right]^{ - 1}  \equiv \left[W^{\Omega _D } ({\bf{q}})\right]^{ - 1}  + \chi ^{{\rm{ph}}(\Omega _D )} ({\bf{q}})\\
&\hspace{2pc}=\left[C^{\Omega _D } ({\bf{q}}) - 2D^{\Omega _D } ({\bf{q}})\right]^{ - 1}  + \chi ^{{\rm{ph}}(\Omega _D )} ({\bf{q}}).
\end{split}
\end{equation}
Then Eq. (\ref{eq048}) becomes the solution of the RPA matrix equation, which reads as
\begin{equation}\label{eq050}
\begin{split}
&\left[ - P^\Omega  ({\bf{q}})\right]^{ - 1}  = \left[ - \tilde P({\bf{q}})\right]^{ - 1}  + \chi ^{{\rm{pp}}(\Omega )} ({\bf{q}}),\\
&\left[C^\Omega  ({\bf{q}})\right]^{ - 1}  = \left[ \tilde C({\bf{q}})\right]^{ - 1}  - \chi ^{{\rm{ph}}(\Omega )} ({\bf{q}}), \\
&\left[W^\Omega  ({\bf{q}})\right]^{ - 1}  = \left[ \tilde W({\bf{q}})\right]^{ - 1}  - \chi ^{{\rm{ph}}(\Omega )} ({\bf{q}}),\\
&\textrm{ at } \Omega  < \Omega _D.
\end{split}
\end{equation}
Thus, in our approximation, where at high scale $\Omega  \ge \Omega _D $ the TUFRG is utilized and in the divergent regime ($\Omega  < \Omega _D$) the effects of the fluctuation channels are completely neglected, the bosonic propagators (at $\Omega  < \Omega _D$) can also be obtained purely by applying the RPA starting from the irreducible bosonic propagators.

\subsection{Singular modes of bosonic propagators}\label{sec3B}
The final purpose of the present paper is to derive the MF equation based on the irreducible singular modes. These modes are obtained by a linear combination of the singular modes of the bosonic propagators. Therefore, in this subsection, we first introduce the singular modes $\left| {\phi ^{{\rm{X}},\alpha } ({\bf{Q}})} \right\rangle$, and then, by using it, represent in detail the effective action $\Gamma ^{\Omega _D }$, the irreducible bosonic propagators $\tilde X({\bf{Q}})$, and the irreducible singular modes $\left| {\varphi ^{{\rm{X}},\alpha } ({\bf{Q}})} \right\rangle$. After that, we present the significant and universal symmetry conditions satisfied by $\left| {\varphi ^{{\rm{X}},\alpha } ({\bf{Q}})} \right\rangle$, which could simplify many of subsequent derivations.

Since the bosonic propagators are Hermitian matrices, they can be decomposed in terms of their eigenmodes:
\begin{equation}\nonumber
\begin{split}
 - P_{o_1 o_2 m,o_3 o_4 n}^\Omega  ({\bf{q}}) &= \sum\limits_\gamma  {\lambda ^{{\rm{P}},\gamma } ({\bf{q}})} \phi _{o_1 o_2 m}^{{\rm{P}},\gamma } ({\bf{q}})[\phi _{o_3 o_4 n}^{{\rm{P}},\gamma } ({\bf{q}})]^* ,\\
C_{o_1 o_2 m,o_3 o_4 n}^\Omega  ({\bf{q}}) &= \sum\limits_\gamma  {\lambda ^{{\rm{C}},\gamma } ({\bf{q}})} \phi _{o_1 o_2 m}^{{\rm{C}},\gamma } ({\bf{q}})[\phi _{o_3 o_4 n}^{{\rm{C}},\gamma } ({\bf{q}})]^* ,\\
W_{o_1 o_2 m,o_3 o_4 n}^\Omega  ({\bf{q}}) &= \sum\limits_\gamma  {\lambda ^{{\rm{W}},\gamma } ({\bf{q}})} \phi _{o_1 o_2 m}^{{\rm{W}},\gamma } ({\bf{q}})[\phi _{o_3 o_4 n}^{{\rm{W}},\gamma } ({\bf{q}})]^* .
\end{split}
\end{equation}
At the divergence scale, some propagators have strong divergence at particular transfer momenta ${\bf{Q}}_i$, and they can be approximated by the expansions in terms of several singular eigenmodes associated with dominant positive \footnote{As can be seen from Eq. (\ref{eq050}), in the RPA flow, due to the positivity of $\chi ^{{\rm{ph}}(\Omega )} $ and $- \chi ^{{\rm{pp}}(\Omega )}$, the positive eigenvalue will be amplified, while the negative eigenvalue weakened. In the TUFRG flow (\ref{eq044}) with the structure similar to the RPA, a similar behavior is exhibited when the scale approaches $\Omega _D $. Actually, in our experience to date, we have not found any dominant negative eigenvalue at the divergence scale.} eigenvalues
\begin{equation}\label{eq051}
\begin{split}
 - P_{o_1 o_2 m,o_3 o_4 n}^{\Omega _D } ({\bf{Q}}_i^{\rm{P}} ) &\approx \sum\limits_{\alpha  = 1}^{M_{{\rm{P}},i} } {\lambda ^{{\rm{P}},\alpha } ({\bf{Q}}_i^{\rm{P}} )} \\
&\times \phi _{o_1 o_2 m}^{{\rm{P}},\alpha } ({\bf{Q}}_i^{\rm{P}} )[\phi _{o_3 o_4 n}^{{\rm{P}},\alpha } ({\bf{Q}}_i^{\rm{P}} )]^* ,\\
C_{o_1 o_2 m,o_3 o_4 n}^{\Omega _D } ({\bf{Q}}_i^{\rm{C}} ) &\approx \sum\limits_{\alpha  = 1}^{M_{{\rm{C}},i} } {\lambda ^{{\rm{C}},\alpha } ({\bf{Q}}_i^{\rm{C}} )} \\
&\times \phi _{o_1 o_2 m}^{{\rm{C}},\alpha } ({\bf{Q}}_i^{\rm{C}} )[\phi _{o_3 o_4 n}^{{\rm{C}},\alpha } ({\bf{Q}}_i^{\rm{C}} )]^* ,\\
W_{o_1 o_2 m,o_3 o_4 n}^{\Omega _D } ({\bf{Q}}_i^{\rm{W}} ) &\approx \sum\limits_{\alpha  = 1}^{M_{{\rm{W}},i} } {\lambda ^{{\rm{W}},\alpha } ({\bf{Q}}_i^{\rm{W}} )}\\
&\times \phi _{o_1 o_2 m}^{{\rm{W}},\alpha } ({\bf{Q}}_i^{\rm{W}} )[\phi _{o_3 o_4 n}^{{\rm{W}},\alpha } ({\bf{Q}}_i^{\rm{W}} )]^* .
\end{split}
\end{equation}
Here, e.g., $\lambda ^{{\rm{P}},\alpha } ({\bf{Q}}_i^{\rm{P}} )$ is the $\alpha$-th largest positive eigenvalue of the matrix $- P^{\Omega _D } ({\bf{Q}}_i^{\rm{P}} )$, and $\phi _{o_1 o_2 m}^{{\rm{P}},\alpha } ({\bf{Q}}_i^{\rm{P}} )$ is an element of the corresponding orthonormal eigenvector (singular mode). Hence, for numerical implementation of the MF theory, we will retain only the divergent parts and take the following approximation for $\Gamma ^{\Omega _D } $:
\begin{equation}\label{eq052}
\begin{split}
\Gamma ^{\Omega _D } [\psi ,\bar \psi ] &\approx \Gamma ^{{\rm{sSC}}} [\psi ,\bar \psi ] + \Gamma ^{{\rm{tSC}}} [\psi ,\bar \psi ] \\
&+ \Gamma ^{{\rm{SPN}}} [\psi ,\bar \psi ] + \Gamma ^{{\rm{CHG}}} [\psi ,\bar \psi ] ,\\
\Gamma ^{{\rm{sSC}}} [\psi ,\bar \psi ] &\approx  - \frac{{\rm{1}}}{2}\frac{1}{{2N\beta \hbar ^2 }}\sum\limits_{i = 1}^{N_{\rm{P}} } {\sum\limits_{\alpha  = 1}^{M_{{\rm{P}},i} } {\lambda ^{{\rm{P}},\alpha } ({\bf{Q}}_i^{\rm{P}} )} }\\
&\times \sum\limits_\omega  {[A_\alpha ^{{\rm{sSC}}} ({\bf{Q}}_i^{\rm{P}} ,\omega )]^* A_\alpha ^{{\rm{sSC}}} ({\bf{Q}}_i^{\rm{P}} ,\omega )} ,\\
\Gamma ^{{\rm{tSC}}} [\psi ,\bar \psi ] &\approx  - \frac{{\rm{1}}}{2}\frac{1}{{2N\beta \hbar ^2 }}\sum\limits_{i = 1}^{N_{\rm{P}} } {\sum\limits_{\alpha  = 1}^{M_{{\rm{P}},i} } {\lambda ^{{\rm{P}},\alpha } ({\bf{Q}}_i^{\rm{P}} )} } \\
&\times\sum\limits_\omega  {[\vec A_\alpha ^{{\rm{tSC}}} ({\bf{Q}}_i^{\rm{P}} ,\omega )]^*  \cdot \vec A_\alpha ^{{\rm{tSC}}} ({\bf{Q}}_i^{\rm{P}} ,\omega )} ,\\
\Gamma ^{{\rm{SPN}}} [\psi ,\bar \psi ] &\approx  - \frac{{\rm{1}}}{2}\frac{1}{{2N\beta \hbar ^2 }}\sum\limits_{i = 1}^{N_{\rm{C}} } {\sum\limits_{\alpha  = 1}^{M_{{\rm{C}},i} } {\lambda ^{{\rm{C}},\alpha } ({\bf{Q}}_i^{\rm{C}} )} }\\
&\times \sum\limits_\omega  {[\vec A_\alpha ^{{\rm{SPN}}} ({\bf{Q}}_i^{\rm{C}} ,\omega )]^*  \cdot \vec A_\alpha ^{{\rm{SPN}}} ({\bf{Q}}_i^{\rm{C}} ,\omega )} ,\\
\Gamma ^{{\rm{CHG}}} [\psi ,\bar \psi ] &\approx  - \frac{{\rm{1}}}{2}\frac{1}{{2N\beta \hbar ^2 }}\sum\limits_{i = 1}^{N_{\rm{W}} } {\sum\limits_{\alpha  = 1}^{M_{{\rm{W}},i} } {\lambda ^{{\rm{W}},\alpha } ({\bf{Q}}_i^{\rm{W}} )} }\\
&\times \sum\limits_\omega  {[A_\alpha ^{{\rm{CHG}}} ({\bf{Q}}_i^{\rm{W}} ,\omega )]^* A_\alpha ^{{\rm{CHG}}} ({\bf{Q}}_i^{\rm{W}} ,\omega )} ,
\end{split}
\end{equation}
with the fermion bilinear in the X-channel, $A_\alpha ^{\rm{X}}$ (${\rm{X}} \in \{ {\rm{sSC, tSC, SPN, CHG}}\} $), defined as
\begin{equation}\label{eq053}
\begin{split}
&A_\alpha ^{{\rm{sSC}}} ({\bf{Q}}_i^{\rm{P}} ,\omega ) \equiv \sum\limits_{{\bf{p}},\omega _p } {\sum\limits_{o,o',m} {[\phi _{oo'm}^{{\rm{P}},\alpha } ({\bf{Q}}_i^{\rm{P}} )]^* e^{i{\bf{R}}_m  \cdot {\bf{p}}} } }\\
&\hspace{1pc}\times \sum\limits_\sigma  {\sigma \psi _{ - \sigma } ( - {\bf{p}}, - \omega _p ,o')\psi _\sigma  ({\bf{p}} + {\bf{Q}}_i^{\rm{P}} ,\omega _p  + \omega ,o)} ,\\
&\vec A_\alpha ^{{\rm{tSC}}} ({\bf{Q}}_i^{\rm{P}} ,\omega ) \equiv \left(A_{\alpha ,x}^{{\rm{tSC}}} ({\bf{Q}}_i^{\rm{P}} ,\omega ),A_{\alpha ,y}^{{\rm{tSC}}} ({\bf{Q}}_i^{\rm{P}} ,\omega ),A_{\alpha ,z}^{{\rm{tSC}}} ({\bf{Q}}_i^{\rm{P}} ,\omega )\right)\\
&\hspace{1pc} \equiv \sum\limits_{{\bf{p}},\omega _p } {\sum\limits_{o,o',m} {[\phi _{oo'm}^{{\rm{P}},\alpha } ({\bf{Q}}_i^{\rm{P}} )]^* e^{i{\bf{R}}_m  \cdot {\bf{p}}} } }\\
&\hspace{1.5pc}\times \left( { - \sum\limits_\sigma  {\sigma \psi _\sigma  ( - {\bf{p}}, - \omega _p ,o')\psi _\sigma  ({\bf{p}} + {\bf{Q}}_i^{\rm{P}} ,\omega _p  + \omega ,o)} ,} \right.\\
&\hspace{3.5pc} - i\sum\limits_\sigma  {\psi _\sigma  ( - {\bf{p}}, - \omega _p ,o')\psi _\sigma  ({\bf{p}} + {\bf{Q}}_i^{\rm{P}} ,\omega _p  + \omega ,o)} ,\\
&\hspace{3.5pc} \left. {\sum\limits_\sigma  {\psi _{ - \sigma } ( - {\bf{p}}, - \omega _p ,o')\psi _\sigma  ({\bf{p}} + {\bf{Q}}_i^{\rm{P}} ,\omega _p  + \omega ,o)} } \right),\\
&\vec A_\alpha ^{{\rm{SPN}}} ({\bf{Q}}_i^{\rm{C}} ,\omega ) \equiv \sum\limits_{{\bf{p}},\omega _p } {\sum\limits_{o,o',m} {[\phi _{oo'm}^{{\rm{C}},\alpha } ({\bf{Q}}_i^{\rm{C}} )]^* e^{i{\bf{R}}_m  \cdot {\bf{p}}} } }\\
&\hspace{1pc}\times \sum\limits_{\sigma ,\sigma '} {\bar \psi _\sigma  ({\bf{p}},\omega _p ,o')\vec \sigma _{\sigma \sigma '} \psi _{\sigma '} ({\bf{p}} + {\bf{Q}}_i^{\rm{C}} ,\omega _p  + \omega ,o)} ,\\
&A_\alpha ^{{\rm{CHG}}} ({\bf{Q}}_i^{\rm{W}} ,\omega ) \equiv \sum\limits_{{\bf{p}},\omega _p } {\sum\limits_{o,o',m} {[\phi _{oo'm}^{{\rm{W}},\alpha } ({\bf{Q}}_i^{\rm{W}} )]^* e^{i{\bf{R}}_m  \cdot {\bf{p}}} } }\\
&\hspace{1pc}\times \sum\limits_\sigma  {\bar \psi _\sigma  ({\bf{p}},\omega _p ,o')\psi _\sigma  ({\bf{p}} + {\bf{Q}}_i^{\rm{W}} ,\omega _p  + \omega ,o)} .
\end{split}
\end{equation}
One can rewrite briefly the bosonic propagators  in Eq. (\ref{eq051}) by introducing the notation $\left| {\phi ^\alpha  } \right\rangle \left\langle {\phi ^\beta  } \right|$ whose elements are defined as $\left( {\left| {\phi ^\alpha  } \right\rangle \left\langle {\phi ^\beta  } \right|} \right)_{o_1 o_2 m,o_3 o_4 n}  \equiv \phi _{o_1 o_2 m}^\alpha  \left( {\phi _{o_3 o_4 n}^\beta  } \right)^* $,
\begin{equation}\label{eq054}
\begin{split}
 - P^{\Omega _D } ({\bf{Q}}_i^{\rm{P}} ) =& \sum\limits_\alpha  {\lambda ^{{\rm{P}},\alpha } ({\bf{Q}}_i^{\rm{P}} )} \left| {\phi ^{{\rm{P}},\alpha } ({\bf{Q}}_i^{\rm{P}} )} \right\rangle \left\langle {\phi ^{{\rm{P}},\alpha } ({\bf{Q}}_i^{\rm{P}} )} \right|,\\
C^{\Omega _D } ({\bf{Q}}_i^{\rm{C}} ) =& \sum\limits_\alpha  {\lambda ^{{\rm{C}},\alpha } ({\bf{Q}}_i^{\rm{C}} )} \left| {\phi ^{{\rm{C}},\alpha } ({\bf{Q}}_i^{\rm{C}} )} \right\rangle \left\langle {\phi ^{{\rm{C}},\alpha } ({\bf{Q}}_i^{\rm{C}} )} \right|,\\
W^{\Omega _D } ({\bf{Q}}_i^{\rm{W}} ) =& \sum\limits_\alpha  {\lambda ^{{\rm{W}},\alpha } ({\bf{Q}}_i^{\rm{W}} )} \left| {\phi ^{{\rm{W}},\alpha } ({\bf{Q}}_i^{\rm{W}} )} \right\rangle \left\langle {\phi ^{{\rm{W}},\alpha } ({\bf{Q}}_i^{\rm{W}} )} \right|.
\end{split}
\end{equation}
Inserting Eq. (\ref{eq054}) into Eq. (\ref{eq049}), we can derive, e.g., the irreducible bosonic propagator in the pairing channel
\begin{equation}\nonumber
\begin{split}
&\left[ - \tilde P({\bf{Q}}_i^{\rm{P}} )\right] = \sqrt { - P^{\Omega _D} ({\bf{Q}}_i^{\rm{P}} )} \\
&\hspace{1pc} \times \left[ {1 + \sqrt { - P^{\Omega _D } ({\bf{Q}}_i^{\rm{P}} )} \left( { - \chi ^{{\rm{pp}}(\Omega _D )} ({\bf{Q}}_i^{\rm{P}} )} \right)\sqrt { - P^{\Omega _D } ({\bf{Q}}_i^{\rm{P}} )} } \right]^{ - 1} \\
&\hspace{1pc}\times \sqrt { - P^{\Omega _D } ({\bf{Q}}_i^{\rm{P}} )} 
\end{split}
\end{equation}
\begin{equation}\nonumber
\begin{split}
 &\hspace{1pc}= \left( {\sum\limits_\alpha  {\sqrt {\lambda ^{{\rm{P}},\alpha } ({\bf{Q}}_i^{\rm{P}} )} } \left| {\phi ^{{\rm{P}},\alpha } ({\bf{Q}}_i^{\rm{P}} )} \right\rangle \left\langle {\phi ^{{\rm{P}},\alpha } ({\bf{Q}}_i^{\rm{P}} )} \right|} \right) \\
&\hspace{1pc}\times \left[ {1 + \sqrt { - P^{\Omega _D } ({\bf{Q}}_i^{\rm{P}} )} \left( { - \chi ^{{\rm{pp}}(\Omega _D )} ({\bf{Q}}_i^{\rm{P}} )} \right) \sqrt { - P^{\Omega _D } ({\bf{Q}}_i^{\rm{P}} )} } \right]^{ - 1} \\
&\hspace{1pc} \times \left( {\sum\limits_{\alpha '} {\sqrt {\lambda ^{{\rm{P}},\alpha '} ({\bf{Q}}_i^{\rm{P}} )} } \left| {\phi ^{{\rm{P}},\alpha '} ({\bf{Q}}_i^{\rm{P}} )} \right\rangle \left\langle {\phi ^{{\rm{P}},\alpha '} ({\bf{Q}}_i^{\rm{P}} )} \right|} \right).
\end{split}
\end{equation}
The above equation can be rewritten as
\begin{equation}\nonumber
\left[ - \tilde P({\bf{Q}}_i^{\rm{P}} ) \right] = \sum\limits_\alpha  {\sum\limits_{\alpha '} {Y_{\alpha \alpha '}^{\rm{P}} ({\bf{Q}}_i^{\rm{P}} )} } \left| {\phi ^{{\rm{P}},\alpha } ({\bf{Q}}_i^{\rm{P}} )} \right\rangle \left\langle {\phi ^{{\rm{P}},\alpha '} ({\bf{Q}}_i^{\rm{P}} )} \right| ,
\end{equation}
with a $M_{{\rm{P}},i}  \times M_{{\rm{P}},i} $-matrix $Y^{\rm{P}} ({\bf{Q}}_i^{\rm{P}})$, defined as
\begin{equation}\nonumber
\begin{split}
&Y_{\alpha \alpha '}^{\rm{P}} ({\bf{Q}}_i^{\rm{P}} ) \equiv \sqrt {\lambda ^{{\rm{P}},\alpha } ({\bf{Q}}_i^{\rm{P}} )} \sqrt {\lambda ^{{\rm{P}},\alpha '} ({\bf{Q}}_i^{\rm{P}} )} \\
&\hspace{2pc} \times \left\langle {\phi ^{{\rm{P}},\alpha } ({\bf{Q}}_i^{\rm{P}} )} \right| \left[ {1 + \sqrt { - P^{\Omega _D } ({\bf{Q}}_i^{\rm{P}} )} } \right.\\
&\hspace{2pc} \times \left. {\left( { - \chi ^{{\rm{pp}}(\Omega _D )} ({\bf{Q}}_i^{\rm{P}} )} \right)\sqrt { - P^{\Omega _D } ({\bf{Q}}_i^{\rm{P}} )} } \right]^{ - 1} \left| {\phi ^{{\rm{P}},\alpha '} ({\bf{Q}}_i^{\rm{P}} )} \right\rangle .
\end{split}
\end{equation}
The inverse of $Y^{\rm{P}} ({\bf{Q}}_i^{\rm{P}} )$ is represented as
\begin{equation}\nonumber
\begin{split}
&\left[Y^{\rm{P}} ({\bf{Q}}_i^{\rm{P}} )^{ - 1} \right]_{\alpha \alpha '}  = \frac{1}{{\sqrt {\lambda ^{{\rm{P}},\alpha } ({\bf{Q}}_i^{\rm{P}} )} \sqrt {\lambda ^{{\rm{P}},\alpha '} ({\bf{Q}}_i^{\rm{P}} )} }}\\
&\hspace{2pc} \times \left( {\delta _{\alpha \alpha '}  + \sqrt {\lambda ^{{\rm{P}},\alpha } ({\bf{Q}}_i^{\rm{P}} )} \sqrt {\lambda ^{{\rm{P}},\alpha '} ({\bf{Q}}_i^{\rm{P}} )} Z_{\alpha \alpha '}^{\rm{P}} ({\bf{Q}}_i^{\rm{P}} )} \right)\\
&\hspace{2pc} = \left( {\frac{1}{{\lambda ^{{\rm{P}},\alpha } ({\bf{Q}}_i^{\rm{P}} )}}\delta _{\alpha \alpha '}  + Z_{\alpha \alpha '}^{\rm{P}} ({\bf{Q}}_i^{\rm{P}} )} \right),
\end{split}
\end{equation}
where $Z_{\alpha \alpha '}^{\rm{P}} ({\bf{Q}}_i^{\rm{P}} )$ is given by
\begin{equation}\nonumber
\begin{split}
Z_{\alpha \alpha '}^{\rm{P}} ({\bf{Q}}_i^{\rm{P}} )   \equiv & \left\langle {\phi ^{{\rm{P}},\alpha } ({\bf{Q}}_i^{\rm{P}} )} \right| \left[ - \chi ^{{\rm{pp}}(\Omega _D )} ({\bf{Q}}_i^{\rm{P}} ) \right] \left| {\phi ^{{\rm{P}},\alpha '} ({\bf{Q}}_i^{\rm{P}} )} \right\rangle \\
=& \sum\limits_{o_1 ,o_2 ,m} {\sum\limits_{o_3 ,o_4 ,n} {[\phi _{o_1 o_2 m}^{{\rm{P}},\alpha } ({\bf{Q}}_i^{\rm{P}} )]^* } }\\
&\times [ - \chi _{o_1 o_2 m,o_3 o_4 n}^{{\rm{pp}}(\Omega _D )} ({\bf{Q}}_i^{\rm{P}} )] \phi _{o_3 o_4 n}^{{\rm{P}},\alpha '  } ({\bf{Q}}_i^{\rm{P}} ).
\end{split}
\end{equation}
Thus, we have
\begin{equation}\label{eq055}
\begin{split}
&- \tilde P({\bf{Q}}_i^{\rm{P}} ) = \sum\limits_\alpha  {\sum\limits_{\alpha '} {Y_{\alpha \alpha '}^{\rm{P}} ({\bf{Q}}_i^{\rm{P}} )} } \left| {\phi ^{{\rm{P}},\alpha } ({\bf{Q}}_i^{\rm{P}} )} \right\rangle \left\langle {\phi ^{{\rm{P}},\alpha '} ({\bf{Q}}_i^{\rm{P}} )} \right|,\\
&\left[ {Y^{\rm{P}} ({\bf{Q}}_i^{\rm{P}} )^{ - 1} } \right]_{\alpha \alpha '}  \equiv \left( {\frac{1}{{\lambda ^{{\rm{P}},\alpha } ({\bf{Q}}_i^{\rm{P}} )}}\delta _{\alpha \alpha '} } \right.\\
&\hspace{3pc} + \left. {\left\langle {\phi ^{{\rm{P}},\alpha } ({\bf{Q}}_i^{\rm{P}} )} \right|\left[ { - \chi ^{{\rm{pp}}(\Omega _D )} ({\bf{Q}}_i^{\rm{P}} )} \right]\left| {\phi ^{{\rm{P}},\alpha '} ({\bf{Q}}_i^{\rm{P}} )} \right\rangle } \right).
\end{split}
\end{equation}
In a similar way, we can derive the expressions of the other two irreducible bosonic propagators
\begin{equation}\label{eq056}
\begin{split}
&\tilde C({\bf{Q}}_i^{\rm{C}} ) = \sum\limits_\alpha  {\sum\limits_{\alpha '} {Y_{\alpha \alpha '}^{\rm{C}} ({\bf{Q}}_i^{\rm{C}} )} } \left| {\phi ^{{\rm{C}},\alpha } ({\bf{Q}}_i^{\rm{C}} )} \right\rangle \left\langle {\phi ^{{\rm{C}},\alpha '} ({\bf{Q}}_i^{\rm{C}} )} \right|,\\
&\left[Y^{\rm{C}} ({\bf{Q}}_i^{\rm{C}} )^{ - 1} \right]_{\alpha \alpha '}  \equiv \left( {\frac{1}{{\lambda ^{{\rm{C}},\alpha } ({\bf{Q}}_i^{\rm{C}} )}}\delta _{\alpha \alpha '} } \right.\\
&\hspace{3pc} + \left. {\left\langle {\phi ^{{\rm{C}},\alpha } ({\bf{Q}}_i^{\rm{C}} )} \right|\chi ^{{\rm{ph}}(\Omega _D )} ({\bf{Q}}_i^{\rm{C}} )\left| {\phi ^{{\rm{C}},\alpha '} ({\bf{Q}}_i^{\rm{C}} )} \right\rangle } \right),
\end{split}
\end{equation}
\begin{equation}\label{eq057}
\begin{split}
&\tilde W({\bf{Q}}_i^{\rm{W}} ) = \sum\limits_\alpha  {\sum\limits_{\alpha '} {Y_{\alpha \alpha '}^{\rm{W}} ({\bf{Q}}_i^{\rm{W}} )} } \left| {\phi ^{{\rm{W}},\alpha } ({\bf{Q}}_i^{\rm{W}} )} \right\rangle \left\langle {\phi ^{{\rm{W}},\alpha '} ({\bf{Q}}_i^{\rm{W}} )} \right|,\\
&\left[Y^{\rm{W}} ({\bf{Q}}_i^{\rm{W}} )^{ - 1} \right]_{\alpha \alpha '}  \equiv \left( {\frac{1}{{\lambda ^{{\rm{W}},\alpha } ({\bf{Q}}_i^{\rm{W}} )}}\delta _{\alpha \alpha '} }\right.\\
&\hspace{3pc} + \left. {\left\langle {\phi ^{{\rm{W}},\alpha } ({\bf{Q}}_i^{\rm{W}} )} \right|\chi ^{{\rm{ph}}(\Omega _D )} ({\bf{Q}}_i^{\rm{W}} ) \left| {\phi ^{{\rm{W}},\alpha '} ({\bf{Q}}_i^{\rm{W}} )} \right\rangle } \right).
\end{split}
\end{equation}
In Eqs. (\ref{eq055}) to (\ref{eq057}), the matrices $Y^{\rm{X}} ({\bf{Q}}_i^{\rm{X}} )$ can be diagonalized as
\begin{equation}\label{eq058}
\begin{split}
&\left[Y^{\rm{X}} ({\bf{Q}}_i^{\rm{X}} )^{ - 1} \right]_{\alpha \alpha '}  = \sum\limits_{\beta  = 1}^{M_{{\rm{X}},i} } {\frac{1}{{\Lambda ^{{\rm{X}},\beta } ({\bf{Q}}_i^{\rm{X}} )}}} S_\alpha ^{{\rm{X}},\beta } ({\bf{Q}}_i^{\rm{X}} )[S_{\alpha '}^{{\rm{X}},\beta } ({\bf{Q}}_i^{\rm{X}} )]^*\\
&  \textrm{ with } {\rm{X}} \in \{ {\rm{P}},{\rm{C}},{\rm{W}}\} .
\end{split}
\end{equation}
Here $\Lambda ^{{\rm{X}},\beta } ({\bf{Q}}_i^{\rm{X}} )$ is the irreducible coupling constant for the $\beta$th mode in the X-channel, and ${\bf{S}}^{{\rm{X}},\beta } ({\bf{Q}}_i^{\rm{X}} ) = \left( {S_1^{{\rm{X}},\beta } ({\bf{Q}}_i^{\rm{X}} ),S_2^{{\rm{X}},\beta } ({\bf{Q}}_i^{\rm{X}} ), ... , S_{M_{{\rm{X}},i} }^{{\rm{X}},\beta } ({\bf{Q}}_i^{\rm{X}} )} \right)$ is its corresponding orthonormal eigenvector. One can rewrite Eqs. (\ref{eq055}) to (\ref{eq057}) as
\begin{equation}\label{eq059}
\begin{split}
 - \tilde P({\bf{Q}}_i^{\rm{P}} ) =& \sum\limits_{\alpha  = 1}^{M_{{\rm{P}},i} } {\Lambda ^{{\rm{P}},\alpha } ({\bf{Q}}_i^{\rm{P}} )} \left| {\varphi ^{{\rm{P}},\alpha } ({\bf{Q}}_i^{\rm{P}} )} \right\rangle \left\langle {\varphi ^{{\rm{P}},\alpha } ({\bf{Q}}_i^{\rm{P}} )} \right|,\\
\tilde C({\bf{Q}}_i^{\rm{C}} ) =& \sum\limits_{\alpha  = 1}^{M_{{\rm{C}},i} } {\Lambda ^{{\rm{C}},\alpha } ({\bf{Q}}_i^{\rm{C}} )} \left| {\varphi ^{{\rm{C}},\alpha } ({\bf{Q}}_i^{\rm{C}} )} \right\rangle \left\langle {\varphi ^{{\rm{C}},\alpha } ({\bf{Q}}_i^{\rm{C}} )} \right|,\\
\tilde W({\bf{Q}}_i^{\rm{W}} ) =& \sum\limits_{\alpha  = 1}^{M_{{\rm{W}},i} } {\Lambda ^{{\rm{W}},\alpha } ({\bf{Q}}_i^{\rm{W}} )} \left| {\varphi ^{{\rm{W}},\alpha } ({\bf{Q}}_i^{\rm{W}} )} \right\rangle \left\langle {\varphi ^{{\rm{W}},\alpha } ({\bf{Q}}_i^{\rm{W}} )} \right|,
\end{split}
\end{equation}
with the irreducible singular modes
\begin{equation}\label{eq060}
\begin{split}
\left| {\varphi ^{{\rm{X}},\alpha } ({\bf{Q}}_i^{\rm{X}} )} \right\rangle  \equiv& \sum\limits_{\beta  = 1}^{M_{{\rm{X}},i} } {S_\beta ^{{\rm{X}},\alpha } ({\bf{Q}}_i^{\rm{X}} )} \left| {\phi ^{{\rm{X}},\beta } ({\bf{Q}}_i^{\rm{X}} )} \right\rangle, \\
\textrm{ or } \varphi _{oo'm}^{{\rm{X}},\alpha } ({\bf{Q}}_i^{\rm{X}} ) \equiv& \sum\limits_{\beta  = 1}^{M_{{\rm{X}},i} } {S_\beta ^{{\rm{X}},\alpha } ({\bf{Q}}_i^{\rm{X}} )} \phi _{oo'm}^{{\rm{X}},\beta } ({\bf{Q}}_i^{\rm{X}} ).
\end{split}
\end{equation}

Now we consider the universal symmetry relations for the irreducible singular modes. The RAS relation (\ref{eq031}) should also be respected by the irreducible bosonic propagators
\begin{equation}\label{eq061}
\begin{split}
\tilde P_{o_1 o_2 m,o_3 o_4 n} ({\bf{Q}}) = e^{ - i{\bf{R}}_m  \cdot {\bf{Q}}} \tilde P_{o_2 o_1 \bar m,o_4 o_3 \bar n} ({\bf{Q}})e^{i{\bf{R}}_n  \cdot {\bf{Q}}},
\end{split}
\end{equation}
\begin{equation}\label{eq062}
\begin{split}
\tilde C_{o_1 o_2 m,o_3 o_4 n} ( - {\bf{Q}}) =& e^{i{\bf{R}}_m  \cdot {\bf{Q}}} [\tilde C_{o_2 o_{\rm{1}} \bar m,o_{\rm{4}} o_3 \bar n} ({\bf{Q}})]^* e^{ - i{\bf{R}}_n  \cdot {\bf{Q}}} , \\
\tilde W_{o_1 o_2 m,o_3 o_4 n} ( - {\bf{Q}}) =& e^{i{\bf{R}}_m  \cdot {\bf{Q}}} [\tilde W_{o_2 o_{\rm{1}} \bar m,o_{\rm{4}} o_3 \bar n} ({\bf{Q}})]^* e^{ - i{\bf{R}}_n  \cdot {\bf{Q}}} .
\end{split}
\end{equation}
From Eq. (\ref{eq061}) one can see that for any eigenvector $\left| {\varphi ^{{\rm{P}},\alpha } ({\bf{Q}})} \right\rangle $ of the matrix $\tilde P({\bf{Q}})$, its unitary transformation,
\begin{equation}\label{eq063}
\begin{split}
&\hat T^{{\rm{pp}}} :\left| {\varphi ^{{\rm{P}},\alpha } ({\bf{Q}})} \right\rangle  \to \left| {\tilde \varphi ^{{\rm{P}},\alpha } ({\bf{Q}})} \right\rangle ;\\
&\hspace{3pc} \tilde \varphi _{oo'm}^{{\rm{P}},\alpha } ({\bf{Q}}) \equiv \varphi _{o'o\bar m}^{{\rm{P}},\alpha } ({\bf{Q}})e^{ - i{\bf{R}}_m  \cdot {\bf{Q}}},
\end{split}
\end{equation}
gives also an eigenvector $\left| {\tilde \varphi ^{{\rm{P}},\alpha } ({\bf{Q}})} \right\rangle $ of $\tilde P({\bf{Q}})$, associated with the same eigenvalue. It means that the non-degenerate eigenvector $\left| {\varphi ^{{\rm{P}},\alpha } ({\bf{Q}})} \right\rangle$ should also be an eigenvector of the transformation $\hat T^{{\rm{pp}}}$. The relation $(\hat T^{{\rm{pp}}} )^2  = 1$ gives
\begin{equation}\label{eq064}
\begin{split}
\hat T^{{\rm{pp}}} \left| {\varphi ^{{\rm{P}},\alpha } ({\bf{Q}})} \right\rangle  =&  \pm \left| {\varphi ^{{\rm{P}},\alpha } ({\bf{Q}})} \right\rangle, \textrm{ or in more detail,}\\
\varphi _{oo'm}^{{\rm{P}},\alpha } ({\bf{Q}}) =&  \pm \varphi _{o'o\bar m}^{{\rm{P}},\alpha } ({\bf{Q}})e^{ - i{\bf{R}}_m  \cdot {\bf{Q}}} .
\end{split}
\end{equation}
For the degenerate eigenvectors, one can make new eigenvectors, which satisfy the above condition, by a suitable linear combination of the original vectors.

Similarly, it can be seen from Eq. (\ref{eq062}) that, for any eigenvector $\left| {\varphi ^{{\rm{X}},\alpha } ({\bf{Q}})} \right\rangle$ of the matrix $\tilde X({\bf{Q}})$ (${\rm{X}} \in \{ {\rm{C}},{\rm{W}}\}$) with the eigenvalue $\Lambda ^{{\rm{X}},\alpha } ({\bf{Q}})$, its antiunitary transformation
\begin{equation}\label{eq065}
\begin{split}
&\hat T^{{\rm{ph}}} :\left| {\varphi ^{{\rm{X}},\alpha } ({\bf{Q}})} \right\rangle  \to \left| {\tilde \varphi ^{{\rm{X}},\alpha } ( - {\bf{Q}})} \right\rangle ;\\
&\hspace{3pc} \tilde \varphi _{oo'm}^{{\rm{X}},\alpha } ( - {\bf{Q}}) \equiv [\varphi _{o'o\bar m}^{{\rm{X}},\alpha } ({\bf{Q}})]^* e^{i{\bf{R}}_m  \cdot {\bf{Q}}},
\end{split}
\end{equation}
presents an eigenvector $\left| {\tilde \varphi ^{{\rm{X}},\alpha } ( - {\bf{Q}})} \right\rangle $ of $\tilde X( - {\bf{Q}})$ with the same eigenvalue $\Lambda ^{{\rm{X}},\alpha } ({\bf{Q}})$. So, in the case of ${\bf{Q}} \notin \{ {\bf{G}}/2\}$ ($\{ {\bf{G}}\}$ is a group of all reciprocal vectors), we can take the eigenvectors of $\tilde X( - {\bf{Q}})$, using the ones of $\tilde X({\bf{Q}})$, by the relation
\begin{equation}\label{eq066}
\begin{split}
\left| {\varphi ^{{\rm{X}},\alpha } ( - {\bf{Q}})} \right\rangle  =& \hat T^{{\rm{ph}}} \left| {\varphi ^{{\rm{X}},\alpha } ({\bf{Q}})} \right\rangle , \textrm{ or in more detail,} \\
\varphi _{oo'm}^{{\rm{X}},\alpha } ( - {\bf{Q}}) =& [\varphi _{o'o\bar m}^{{\rm{X}},\alpha } ({\bf{Q}})]^* e^{i{\bf{R}}_m  \cdot {\bf{Q}}} , \textrm{ for } {\bf{Q}} \notin \{ {\bf{G}}/2\} .
\end{split}
\end{equation}
For ${\bf{Q}} \in \{ {\bf{G}}/2\}$, the wave vector ${\bf{Q}}$ is physically equivalent to $ - {\bf{Q}}$. In this case, we can construct, by multiplying some phase factors and/or taking suitable linear combinations of degenerate eigenvectors, a new set of eigenvectors of $\tilde X({\bf{Q}})$ that respects the following condition:
\begin{equation}\label{eq067}
\begin{split}
\hat T^{{\rm{ph}}} \left| {\varphi ^{{\rm{X}},\alpha } ({\bf{Q}})} \right\rangle  =& \left| {\varphi ^{{\rm{X}},\alpha } ({\bf{Q}})} \right\rangle ,\\
\textrm{ or } \varphi _{oo'm}^{{\rm{X}},\alpha } ({\bf{Q}}) =& [\varphi _{o'o\bar m}^{{\rm{X}},\alpha } ({\bf{Q}})]^* e^{i{\bf{R}}_m  \cdot {\bf{Q}}} ,\\
\textrm{ for } {\bf{Q}} \in \{ {\bf{G}}/2\}& .
\end{split}
\end{equation}
The constraints of Eqs. (\ref{eq064}), (\ref{eq066}), and (\ref{eq067}) serve as necessary conditions satisfied by the irreducible singular modes. These constraints should also be respected by the singular modes $ \left| {\phi ^{{\rm{X}},\alpha } ({\bf{Q}})} \right\rangle$.

\subsection{Irreducible action as an input for MF treatment}\label{sec3C}
In this paper, we derive the MF equation, based on the saddle-point approximation in the field-theoretical framework. This needs a specific form of the input action. Here we present a detailed expression of the irreducible action. Then, we discuss the relation between the RPA, the MF theory, and the saddle-point approximation. Specifically, we focus on their equivalence in the critical conditions. Finally, on the base of it, we justify our novel TUFRG + MF approach, namely, we explain why the irreducible action should be the interaction part of the input action in our MF theory.

The irreducible bosonic propagators in Eq. (\ref{eq059}) have a structure similar to that of the bosonic propagators in Eq. (\ref{eq054}), and they are thus associated with the following irreducible action:
\begin{equation}\label{eq068}
\begin{split}
\tilde \Gamma [\psi ,\bar \psi ] =& \tilde \Gamma ^{{\rm{sSC}}} [\psi ,\bar \psi ] + \tilde \Gamma ^{{\rm{tSC}}} [\psi ,\bar \psi ]\\
& + \tilde \Gamma ^{{\rm{SPN}}} [\psi ,\bar \psi ] + \tilde \Gamma ^{{\rm{CHG}}} [\psi ,\bar \psi ], \\
\tilde \Gamma ^{{\rm{sSC}}} [\psi ,\bar \psi ] \equiv&  - \frac{{\rm{1}}}{2}\frac{1}{{2N\beta \hbar ^2 }}\sum\limits_{i = 1}^{N_{\rm{P}} } {\sum\limits_{\alpha  = 1}^{M_{{\rm{P}},i} } {\Lambda ^{{\rm{P}},\alpha } ({\bf{Q}}_i^{\rm{P}} )} } \\
&\times \sum\limits_\omega  {[O_\alpha ^{{\rm{sSC}}} ({\bf{Q}}_i^{\rm{P}} ,\omega )]^* O_\alpha ^{{\rm{sSC}}} ({\bf{Q}}_i^{\rm{P}} ,\omega )}, \\
\tilde \Gamma ^{{\rm{tSC}}} [\psi ,\bar \psi ] \equiv&  - \frac{{\rm{1}}}{2}\frac{1}{{2N\beta \hbar ^2 }}\sum\limits_{i = 1}^{N_{\rm{P}} } {\sum\limits_{\alpha  = 1}^{M_{{\rm{P}},i} } {\Lambda ^{{\rm{P}},\alpha } ({\bf{Q}}_i^{\rm{P}} )} } \\
&\times \sum\limits_\omega  {[\vec O_\alpha ^{{\rm{tSC}}} ({\bf{Q}}_i^{\rm{P}} ,\omega )]^*  \cdot \vec O_\alpha ^{{\rm{tSC}}} ({\bf{Q}}_i^{\rm{P}} ,\omega )}, \\
\tilde \Gamma ^{{\rm{SPN}}} [\psi ,\bar \psi ] \equiv&  - \frac{{\rm{1}}}{2}\frac{1}{{2N\beta \hbar ^2 }}\sum\limits_{i = 1}^{N_{\rm{C}} } {\sum\limits_{\alpha  = 1}^{M_{{\rm{C}},i} } {\Lambda ^{{\rm{C}},\alpha } ({\bf{Q}}_i^{\rm{C}} )} } \\
&\times  \sum\limits_\omega  {[\vec O_\alpha ^{{\rm{SPN}}} ({\bf{Q}}_i^{\rm{C}} ,\omega )]^*  \cdot \vec O_\alpha ^{{\rm{SPN}}} ({\bf{Q}}_i^{\rm{C}} ,\omega )},\\
\tilde \Gamma ^{{\rm{CHG}}} [\psi ,\bar \psi ] \equiv&  - \frac{{\rm{1}}}{2}\frac{1}{{2N\beta \hbar ^2 }}\sum\limits_{i = 1}^{N_{\rm{W}} } {\sum\limits_{\alpha  = 1}^{M_{{\rm{W}},i} } {\Lambda ^{{\rm{W}},\alpha } ({\bf{Q}}_i^{\rm{W}} )} } \\
&\times \sum\limits_\omega  {[O_\alpha ^{{\rm{CHG}}} ({\bf{Q}}_i^{\rm{W}} ,\omega )]^* O_\alpha ^{{\rm{CHG}}} ({\bf{Q}}_i^{\rm{W}} ,\omega )} .
\end{split}
\end{equation}
Here the fermion bilinear in the X-channel, $O_\alpha ^{\rm{X}}$ (${\rm{X}} \in \{ {\rm{sSC, tSC, SPN, CHG}}\}$) is defined by
\begin{equation}\nonumber
\begin{split}
&O_\alpha ^{{\rm{sSC}}} ({\bf{Q}}_i^{\rm{P}} ,\omega ) \equiv \sum\limits_{{\bf{p}},\omega _p } {\sum\limits_{o,o',m} {[\varphi _{oo'm}^{{\rm{P}},\alpha } ({\bf{Q}}_i^{\rm{P}} )]^* e^{i{\bf{R}}_m  \cdot {\bf{p}}} } } \\
&\hspace{2pc} \times \sum\limits_\sigma  {\sigma \psi _{ - \sigma } ( - {\bf{p}}, - \omega _p ,o')\psi _\sigma  ({\bf{p}} + {\bf{Q}}_i^{\rm{P}} ,\omega _p  + \omega ,o)}, \\
&\vec O_\alpha ^{{\rm{tSC}}} ({\bf{Q}}_i^{\rm{P}} ,\omega ) = \left(O_{\alpha ,x}^{{\rm{tSC}}} ({\bf{Q}}_i^{\rm{P}} ,\omega ),O_{\alpha ,y}^{{\rm{tSC}}} ({\bf{Q}}_i^{\rm{P}} ,\omega ),O_{\alpha ,z}^{{\rm{tSC}}} ({\bf{Q}}_i^{\rm{P}} ,\omega )\right) \\
&\hspace{2pc} \equiv \sum\limits_{{\bf{p}},\omega _p } {\sum\limits_{o,o',m} {[\varphi _{oo'm}^{{\rm{P}},\alpha } ({\bf{Q}}_i^{\rm{P}} )]^* e^{i{\bf{R}}_m  \cdot {\bf{p}}} } } \\
&\hspace{2pc}\times \left( { - \sum\limits_\sigma  {\sigma \psi _\sigma  ( - {\bf{p}}, - \omega _p ,o')\psi _\sigma  ({\bf{p}} + {\bf{Q}}_i^{\rm{P}} ,\omega _p  + \omega ,o)} ,} \right.\\
&\hspace{3pc} - i\sum\limits_\sigma  {\psi _\sigma  ( - {\bf{p}}, - \omega _p ,o')\psi _\sigma  ({\bf{p}} + {\bf{Q}}_i^{\rm{P}} ,\omega _p  + \omega ,o)} ,
\end{split}
\end{equation}
\begin{equation}\label{eq069}
\begin{split}
&\hspace{3pc} \left. {\sum\limits_\sigma  {\psi _{ - \sigma } ( - {\bf{p}}, - \omega _p ,o')\psi _\sigma  ({\bf{p}} + {\bf{Q}}_i^{\rm{P}} ,\omega _p  + \omega ,o)} } \right),\\
&\vec O_\alpha ^{{\rm{SPN}}} ({\bf{Q}}_i^{\rm{C}} ,\omega ) \equiv \sum\limits_{{\bf{p}},\omega _p } {\sum\limits_{o,o',m} {[\varphi _{oo'm}^{{\rm{C}},\alpha } ({\bf{Q}}_i^{\rm{C}} )]^* e^{i{\bf{R}}_m  \cdot {\bf{p}}} } }\\
&\hspace{2pc}\times \sum\limits_{\sigma ,\sigma '} {\bar \psi _\sigma  ({\bf{p}},\omega _p ,o')\vec \sigma _{\sigma \sigma '} \psi _{\sigma '} ({\bf{p}} + {\bf{Q}}_i^{\rm{C}} ,\omega _p  + \omega ,o)} ,\\
&O_\alpha ^{{\rm{CHG}}} ({\bf{Q}}_i^{\rm{W}} ,\omega ) \equiv \sum\limits_{{\bf{p}},\omega _p } {\sum\limits_{o,o',m} {[\varphi _{oo'm}^{{\rm{W}},\alpha } ({\bf{Q}}_i^{\rm{W}} )]^* e^{i{\bf{R}}_m  \cdot {\bf{p}}} } } \\
&\hspace{2pc}\times \sum\limits_\sigma  {\bar \psi _\sigma  ({\bf{p}},\omega _p ,o')\psi _\sigma  ({\bf{p}} + {\bf{Q}}_i^{\rm{W}} ,\omega _p  + \omega ,o)} .
\end{split}
\end{equation}

It is easy to verify that the constraint of Eq. (\ref{eq064}) yields the following relation:
\begin{equation}\label{eq070}
\begin{split}
&O_\alpha ^{{\rm{sSC}}} ({\bf{Q}}_i^{\rm{P}} ,\omega ) = 0, \\
&\hspace{2pc} \textrm{ if } \varphi _{oo'm}^{{\rm{P}},\alpha } ({\bf{Q}}_i^{\rm{P}} ) =  - \varphi _{o'o\bar m}^{{\rm{P}},\alpha } ({\bf{Q}}_i^{\rm{P}} )e^{ - i{\bf{R}}_m  \cdot {\bf{Q}}_i^{\rm{P}} } ,\\
&\vec O_\alpha ^{{\rm{tSC}}} ({\bf{Q}}_i^{\rm{P}} ,\omega ) = (0,0,0), \\
&\hspace{2pc} \textrm{ if } \varphi _{oo'm}^{{\rm{P}},\alpha } ({\bf{Q}}_i^{\rm{P}} ) =  + \varphi _{o'o\bar m}^{{\rm{P}},\alpha } ({\bf{Q}}_i^{\rm{P}} )e^{ - i{\bf{R}}_m  \cdot {\bf{Q}}_i^{\rm{P}} } .
\end{split}
\end{equation}
This means that the irreducible singular modes in the pairing channel are divided into two groups, namely, the spin-singlet [$\varphi _{oo'm}^{{\rm{sSC}},\alpha } ({\bf{Q}}_i^{\rm{P}} ) =  + e^{ - i{\bf{R}}_m  \cdot {\bf{Q}}_i^{\rm{P}} } \varphi _{o'o\bar m}^{{\rm{sSC}},\alpha } ({\bf{Q}}_i^{\rm{P}} )$] and spin-triplet [$\varphi _{oo'm}^{{\rm{tSC}},\alpha } ({\bf{Q}}_i^{\rm{P}} ) =  - e^{ - i{\bf{R}}_m  \cdot {\bf{Q}}_i^{\rm{P}} } \varphi _{o'o\bar m}^{{\rm{tSC}},\alpha } ({\bf{Q}}_i^{\rm{P}} )$] modes. Also, starting from Eqs. (\ref{eq066}) and (\ref{eq067}), one can easily derive
\begin{equation}\label{eq071}
\begin{split}
\vec O_\alpha ^{{\rm{SPN}}} ({\bf{Q}}_i^{\rm{C}} ,\omega ) =& [\vec O_\alpha ^{{\rm{SPN}}} ( - {\bf{Q}}_i^{\rm{C}} , - \omega )]^* ,\\
O_\alpha ^{{\rm{CHG}}} ({\bf{Q}}_i^{\rm{W}} ,\omega ) =& [O_\alpha ^{{\rm{CHG}}} ( - {\bf{Q}}_i^{\rm{W}} , - \omega )]^* .
\end{split}
\end{equation}

Let us consider the relationship between the RPA, the MF theory, and the so-called saddle-point approximation in the path-integral formalism. The RPA starting from $\Gamma ^{(0)} [\psi ,\bar \psi ]$ yields the critical condition equivalent to the MF theory (see Appendix \ref{appendA}). Concretely speaking, one can represent $\Gamma ^{(0)} [\psi ,\bar \psi ]$, by using $V^{{\rm{P}},(0)}$, in the form of the pairing channel, and then use the ladder approximation (RPA in pairing channel) to obtain $V^{{\rm{P}},{\rm{RPA}}}$ (corresponding to $\Gamma ^{{\rm{P,RPA}}} [\psi ,\bar \psi ]$). If the resulting $V^{{\rm{P}},{\rm{RPA}}} ({\bf{Q}})$ has any eigenvector with infinite eigenvalue at some particular ${\bf{Q}}$, the system is said to be at critical point between the disordered (metal) and ordered (superconductor) phases. This is the critical condition in the RPA.

In the MF theory, one can introduce the mean-field decomposition of the interaction Hamiltonian (corresponding to $\Gamma ^{(0)} [\psi ,\bar \psi ]$) in the pairing channel, and calculate the superconducting order parameter (energy gap) by applying the self-consistency condition. If the order parameter starts being different from zero (the critical condition in the MF theory), the system can be considered to have a transition from metallic to superconducting phase. As will be discussed in Appendix \ref{appendA}, the RPA and the MF theory have identical critical conditions, implying a sort of the equivalence between them.

\begin{figure}[h!]
	\begin{center}
		\includegraphics[width=8.6cm]{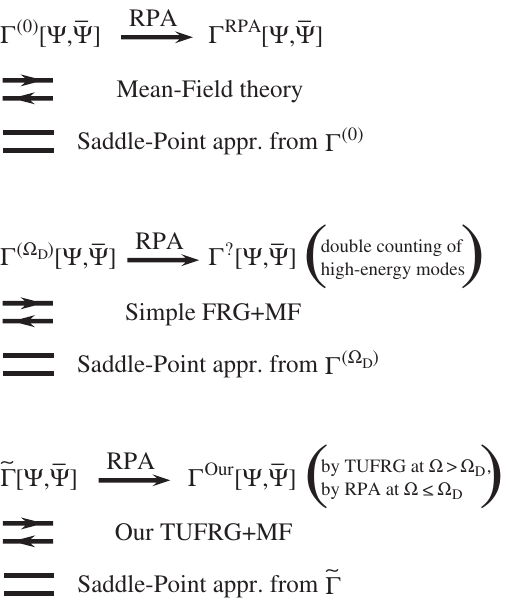}
	\end{center}
	\caption{Relationship between the conventional MF theory, the simple FRG + MF approach, and our TUFRG + MF approach. A pair of opposite arrows indicates the agreement between their critical conditions, while an equal sign means the physical equivalence.}
	\label{fig1}
\end{figure}

We can also obtain the result identical to the MF theory within the path-integral formalism, starting from the interaction part $\Gamma ^{(0)} [\psi ,\bar \psi ]$ of the action. Namely, introducing, via a Hubbard-Stratonovich transformation, the auxiliary variable associated with superconducting order parameters, and employing the saddle-point approximation, one can arrive at the result that is completely identical to the MF approximation. In this sense, the above three methods can be considered to be physically equivalent. However, the MF theory has an advantage, when compared with the RPA, that it can address the ordered phases of the system.

Now we return to the irreducible action $\tilde \Gamma [\psi ,\bar \psi ]$ presented in Eq. (\ref{eq068}). In the multichannel RPA, it evolves into $\Gamma ^{\Omega _D }$ of Eq. (\ref{eq052}) at the divergence scale. Namely, the pairing part of the action changes from $\tilde \Gamma ^{{\rm{sSC}}}  + \tilde \Gamma ^{{\rm{tSC}}}$ of Eq. (\ref{eq068}) to $\Gamma ^{{\rm{sSC}}}  + \Gamma ^{{\rm{tSC}}}$ of Eq. (\ref{eq052}) in the RPA flow of the pairing channel, while the spin and charge parts transform into $\Gamma ^{{\rm{SPN}}} 
$ and $\Gamma ^{{\rm{CHG}}}$, starting from $\tilde \Gamma ^{{\rm{SPN}}}$ and $\tilde \Gamma ^{{\rm{CHG}}}$, in the RPA flows of the spin and charge channels, respectively. Moreover, in our approach (using the TUFRG at $\Omega  \ge \Omega _D$, while using the RPA at $\Omega  < \Omega _D$), the resulting low-energy-scale action $\Gamma ^{\Omega  < \Omega _D }$ is exactly the same as the one that is obtained by the multichannel RPA starting from $\tilde \Gamma$ given in Eq. (\ref{eq068}). Taking into account the equivalence between the RPA and MF theory, we arrive at the conclusion that it would be reasonable to choose the interaction Hamiltonian corresponding to $\tilde \Gamma [\psi ,\bar \psi ]$ as an input for our MF calculation.

Our TUFRG + MF approach is advantageous when compared with the conventional MF theory and the simple FRG + MF approach. As a biased method, the MF theory neglects the interplay between different channels and emphasize a particular channel. When it is applied to the case of coexisting orders, the bare interaction should be split into the parts of multiple channels, which leads to the Fierz ambiguity in the multichannel MF calculation and may cause a certain bias. In the simple FRG + MF scheme, the resulting effective interaction at the divergence scale, which was obtained by the FRG flow, is directly inserted into the MF calculation. This approach has a drawback that it counts doubly the contributions from high-energy modes and generally overemphasizes the ordering tendencies. The relationship between these three methods is schematically shown in Fig. \ref{fig1}.

\begin{widetext}
\subsection{Hubbard-Stratonovich transformation and saddle-point approximation}\label{sec3D}
In the following, we will derive the MF equation for our TUFRG + MF approach by resorting to the saddle-point approximation of statistical field theory. Following the discussion in the previous subsection, we use $\tilde \Gamma$, instead of $\Gamma ^{(0)}$ or $\Gamma ^{\Omega _D } $, as the interaction part of the action. Therefore, in our TUFRG + MF scheme, we consider the partition function represented as follows:
\begin{equation}\label{eq072}
\begin{split}
\Xi  =& \int {D\bar \psi {\kern 1pt} D\psi } \exp \{  - S_0 [\psi ,\bar \psi ] - \tilde \Gamma [\psi ,\bar \psi ]\} = \int {D\bar \psi {\kern 1pt} D\psi } \exp \{  - S_0 [\psi ,\bar \psi ]\} \\
& \times \exp \{  - \tilde \Gamma ^{{\rm{sSC}}} [\psi ,\bar \psi ]\}  \times \exp \{  - \tilde \Gamma ^{{\rm{tSC}}} [\psi ,\bar \psi ]\} 
\times \exp \{  - \tilde \Gamma ^{{\rm{SPN}}} [\psi ,\bar \psi ]\}  \times \exp \{  - \tilde \Gamma ^{{\rm{CHG}}} [\psi ,\bar \psi ]\} ,
\end{split}
\end{equation}
with $S_0$ and $\tilde \Gamma$ given in Eqs. (\ref{eq003}) and (\ref{eq068}), respectively. We can use the Hubbard-Stratonovich transformation to decompose the fermionic quartic terms in $\tilde \Gamma$ into the fermion bilinears.

For example, let us consider the term containing $\tilde \Gamma ^{{\rm{sSC}}}$ in Eq. (\ref{eq072}). It can be expressed in a factorized form.
\begin{equation}\label{eq073}
\begin{split}
\exp \{  - \tilde \Gamma ^{{\rm{sSC}}} [\psi ,\bar \psi ]\}  = \prod\limits_{i = 1}^{N_{\rm{P}} } {\prod\limits_{\alpha  = {\rm{1}}}^{M_{{\rm{sSC}},i} } {\prod\limits_\omega  {\exp \left\{ {\frac{{\rm{1}}}{2}\frac{1}{{2N\beta \hbar ^2 }}\Lambda ^{{\rm{sSC}},\alpha } ({\bf{Q}}_i^{\rm{P}} )[O_\alpha ^{{\rm{sSC}}} ({\bf{Q}}_i^{\rm{P}} ,\omega )]^* O_\alpha ^{{\rm{sSC}}} ({\bf{Q}}_i^{\rm{P}} ,\omega )} \right\}} } } 
\end{split}
\end{equation}
We can perform the Hubbard-Stratonovich transformation for each factor of Eq. (\ref{eq073}) in the following way:
\begin{equation}\nonumber
\begin{split}
&\exp \left\{ {\frac{{\rm{1}}}{2}\frac{1}{{2N\beta \hbar ^2 }}\Lambda ^{{\rm{sSC}},\alpha } ({\bf{Q}}_i^{\rm{P}} )[O_\alpha ^{{\rm{sSC}}} ({\bf{Q}}_i^{\rm{P}} ,\omega )]^* O_\alpha ^{{\rm{sSC}}} ({\bf{Q}}_i^{\rm{P}} ,\omega )} \right\}
 = \exp \{ \eta [O_\alpha ^{{\rm{sSC}}} ({\bf{Q}}_i^{\rm{P}} ,\omega )]^* O_\alpha ^{{\rm{sSC}}} ({\bf{Q}}_i^{\rm{P}} ,\omega )\} \\
&\hspace{12pc} \times \frac{{\int {d\delta ^* } \int {d\delta } \exp \left\{ { - \frac{1}{\eta }} \right.\left[ {\frac{\delta }{\hbar } - \eta O_\alpha ^{{\rm{sSC}}} ({\bf{Q}}_i^{\rm{P}} ,\omega )} \right]^* \left. {\left[ {\frac{\delta }{\hbar } - \eta O_\alpha ^{{\rm{sSC}}} ({\bf{Q}}_i^{\rm{P}} ,\omega )} \right]} \right\}}}{{\int {d\delta ^* } \int {d\delta } \exp \left\{ { - \frac{1}{\eta }\left( {\frac{\delta }{\hbar }} \right)^* \left( {\frac{\delta }{\hbar }} \right)} \right\}}}\\
&\hspace{2pc} = \frac{{\int {d\delta ^* } \int {d\delta } \exp \left\{ { - \frac{1}{{\eta \hbar ^2 }}\delta ^* \delta  + \frac{{\delta ^* }}{\hbar }O_\alpha ^{{\rm{sSC}}} ({\bf{Q}}_i^{\rm{P}} ,\omega ) + \frac{\delta }{\hbar }[O_\alpha ^{{\rm{sSC}}} ({\bf{Q}}_i^{\rm{P}} ,\omega )]^* } \right\}}}{{\int {d\delta ^* } \int {d\delta } \exp \left\{ { - \frac{1}{{\eta \hbar ^2 }}\delta ^* \delta } \right\}}}\\
&\hspace{2pc} = c\int {d\delta ^* } \int {d\delta } \exp \left\{ { - \frac{{4N\beta }}{{\Lambda ^{{\rm{sSC}},\alpha } ({\bf{Q}}_i^{\rm{P}} )}}\delta ^* \delta  + \frac{{\delta ^* }}{\hbar }O_\alpha ^{{\rm{sSC}}} ({\bf{Q}}_i^{\rm{P}} ,\omega ) + \frac{\delta }{\hbar }[O_\alpha ^{{\rm{sSC}}} ({\bf{Q}}_i^{\rm{P}} ,\omega )]^* } \right\},
\end{split}
\end{equation}
with a complex variable $\delta$, a real constant $c$, and $\eta  = \frac{{\Lambda ^{{\rm{sSC}},\alpha } ({\bf{Q}}_i^{\rm{P}} )}}{{4N\beta \hbar ^2 }}$. Inserting the above equation into Eq. (\ref{eq073}), we obtain the following relation:
\begin{equation}\label{eq074}
\begin{split}
\exp \left\{  - \tilde \Gamma ^{{\rm{sSC}}} [\psi ,\bar \psi ]\right\}  =& C_{{\rm{sSC}}} \int {D\delta _{{\rm{sSC}}}^* } \int {D\delta _{{\rm{sSC}}} } \exp \left\{ { - S_{{\rm{HS}}}^{{\rm{sSC}}} [\psi ,\bar \psi ;\delta _{{\rm{sSC}}}^* ,\delta _{{\rm{sSC}}} ]} \right\},\\
S_{{\rm{HS}}}^{{\rm{sSC}}} [\psi ,\bar \psi ;\delta _{{\rm{sSC}}}^* ,\delta _{{\rm{sSC}}} ] \equiv& \sum\limits_{i = 1}^{N_{\rm{P}} } {\sum\limits_{\alpha  = 1}^{M_{{\rm{sSC}},i} } {\sum\limits_\omega  {\left\{ {\frac{{4N\beta }}{{\Lambda ^{{\rm{sSC}},\alpha } ({\bf{Q}}_i^{\rm{P}} )}}} \right.[\delta _\alpha ^{{\rm{sSC}}} ({\bf{Q}}_i^{\rm{P}} ,\omega )]^* \delta _\alpha ^{{\rm{sSC}}} ({\bf{Q}}_i^{\rm{P}} ,\omega )} } } \\
& - \frac{1}{\hbar }[\delta _\alpha ^{{\rm{sSC}}} ({\bf{Q}}_i^{\rm{P}} ,\omega )]^* O_\alpha ^{{\rm{sSC}}} ({\bf{Q}}_i^{\rm{P}} ,\omega ) - \left. {\frac{1}{\hbar }\delta _\alpha ^{{\rm{sSC}}} ({\bf{Q}}_i^{\rm{P}} ,\omega )[O_\alpha ^{{\rm{sSC}}} ({\bf{Q}}_i^{\rm{P}} ,\omega )]^* } \right\}.
\end{split}
\end{equation}
In a similar way, after a tedious calculation, we derive the following equations:
\begin{equation}\label{eq075}
\begin{split}
\exp \left\{  - \tilde \Gamma ^{{\rm{tSC}}} [\psi ,\bar \psi ] \right\}  =& C_{{\rm{tSC}}} \int {D\vec \delta _{{\rm{tSC}}}^* } \int {D\vec \delta _{{\rm{tSC}}} } \exp \left\{ {  - S_{{\rm{HS}}}^{{\rm{tSC}}} [\psi ,\bar \psi ;\vec \delta _{{\rm{tSC}}}^* ,\vec \delta _{{\rm{tSC}}} ]} \right\} ,\\
S_{{\rm{HS}}}^{{\rm{tSC}}} [\psi ,\bar \psi ;\vec \delta _{{\rm{tSC}}}^* ,\vec \delta _{{\rm{tSC}}} ] \equiv& \sum\limits_{i = 1}^{N_{\rm{P}} } {\sum\limits_{\alpha  = 1}^{M_{{\rm{tSC}},i} } {\sum\limits_\omega  {\left\{ {\frac{{4N\beta }}{{\Lambda ^{{\rm{tSC}},\alpha } ({\bf{Q}}_i^{\rm{P}} )}}} \right.[\vec \delta _\alpha ^{{\rm{tSC}}} ({\bf{Q}}_i^{\rm{P}} ,\omega )]^*  \cdot \vec \delta _\alpha ^{{\rm{tSC}}} ({\bf{Q}}_i^{\rm{P}} ,\omega )} } } \\
& - \frac{1}{\hbar }[\vec \delta _\alpha ^{{\rm{tSC}}} ({\bf{Q}}_i^{\rm{P}} ,\omega )]^*  \cdot \vec O_\alpha ^{{\rm{tSC}}} ({\bf{Q}}_i^{\rm{P}} ,\omega ) - \left. {\frac{1}{\hbar }\vec \delta _\alpha ^{{\rm{tSC}}} ({\bf{Q}}_i^{\rm{P}} ,\omega ) \cdot [\vec O_\alpha ^{{\rm{tSC}}} ({\bf{Q}}_i^{\rm{P}} ,\omega )]^* } \right\},
\end{split}
\end{equation}
\begin{equation}\label{eq076}
\begin{split}
\exp \left \{  - \tilde \Gamma ^{{\rm{SPN}}} [\psi ,\bar \psi ] \right\}  =& C_{{\rm{SPN}}} \int {D\vec \delta _{{\rm{SPN}}}^* } \int {D\vec \delta _{{\rm{SPN}}} } \exp \left \{  - S_{{\rm{HS}}}^{{\rm{SPN}}} [\psi ,\bar \psi ;\vec \delta _{{\rm{SPN}}}^* ,\vec \delta _{{\rm{SPN}}} ] \right\} ,\\
S_{{\rm{HS}}}^{{\rm{SPN}}} [\psi ,\bar \psi ;\vec \delta _{{\rm{SPN}}}^* ,\vec \delta _{{\rm{SPN}}} ] \equiv& \sum\limits_{i = 1}^{N_{\rm{C}} } {\sum\limits_{\alpha  = 1}^{M_{{\rm{C}},i} } {\sum\limits_\omega  {\left\{ {\frac{{4N\beta }}{{\Lambda ^{{\rm{C}},\alpha } ({\bf{Q}}_i^{\rm{C}} )}}} \right.[\vec \delta _\alpha ^{{\rm{SPN}}} ({\bf{Q}}_i^{\rm{C}} ,\omega )]^*  \cdot \vec \delta _\alpha ^{{\rm{SPN}}} ({\bf{Q}}_i^{\rm{C}} ,\omega )} } } \\
& - \frac{1}{\hbar }[\vec \delta _\alpha ^{{\rm{SPN}}} ({\bf{Q}}_i^{\rm{C}} ,\omega )]^*  \cdot \vec O_\alpha ^{{\rm{SPN}}} ({\bf{Q}}_i^{\rm{C}} ,\omega ) - \left. {\frac{1}{\hbar }\vec \delta _\alpha ^{{\rm{SPN}}} ({\bf{Q}}_i^{\rm{C}} ,\omega ) \cdot [\vec O_\alpha ^{{\rm{SPN}}} ({\bf{Q}}_i^{\rm{C}} ,\omega )]^* } \right\},\\
\textrm{ with a constraint } &\vec \delta _\alpha ^{{\rm{SPN}}} ({\bf{Q}}_i^{\rm{C}} ,\omega ) = [\vec \delta _\alpha ^{{\rm{SPN}}} ( - {\bf{Q}}_i^{\rm{C}} , - \omega )]^* ,
\end{split}
\end{equation}
\begin{equation}\label{eq077}
\begin{split}
\exp \left\{  - \tilde \Gamma ^{{\rm{CHG}}} [\psi ,\bar \psi ] \right\}  =& C_{{\rm{CHG}}} \int {D\delta _{{\rm{CHG}}}^* } \int {D\delta _{{\rm{CHG}}} } \exp \left\{  - S_{{\rm{HS}}}^{{\rm{CHG}}} [\psi ,\bar \psi ;\delta _{{\rm{CHG}}}^* ,\delta _{{\rm{CHG}}} ] \right\} ,\\
S_{{\rm{HS}}}^{{\rm{CHG}}} [\psi ,\bar \psi ;\delta _{{\rm{CHG}}}^* ,\delta _{{\rm{CHG}}} ] \equiv& \sum\limits_{i = 1}^{N_{\rm{W}} } {\sum\limits_{\alpha  = 1}^{M_{{\rm{W}},i} } {\sum\limits_\omega  {\left\{ {\frac{{4N\beta }}{{\Lambda ^{{\rm{W}},\alpha } ({\bf{Q}}_i^{\rm{W}} )}}} \right.[\delta _\alpha ^{{\rm{CHG}}} ({\bf{Q}}_i^{\rm{W}} ,\omega )]^* \delta _\alpha ^{{\rm{CHG}}} ({\bf{Q}}_i^{\rm{W}} ,\omega )} } } \\
& - \frac{1}{\hbar }[\delta _\alpha ^{{\rm{CHG}}} ({\bf{Q}}_i^{\rm{W}} ,\omega )]^* O_\alpha ^{{\rm{CHG}}} ({\bf{Q}}_i^{\rm{W}} ,\omega ) - \left. {\frac{1}{\hbar }\delta _\alpha ^{{\rm{CHG}}} ({\bf{Q}}_i^{\rm{W}} ,\omega )[O_\alpha ^{{\rm{CHG}}} ({\bf{Q}}_i^{\rm{W}} ,\omega )]^* } \right\},\\
\textrm{ with a constraint } &\delta _\alpha ^{{\rm{CHG}}} ({\bf{Q}}_i^{\rm{W}} ,\omega ) = [\delta _\alpha ^{{\rm{CHG}}} ( - {\bf{Q}}_i^{\rm{W}} , - \omega )]^* .
\end{split}
\end{equation}

Near the critical points, the electronic instabilities are dominated by $\omega  = 0$ components of the auxiliary variables $\delta _\alpha ^{\rm{X}} ({\bf{Q}},\omega )$ (for a detailed discussion, see pages 89--91 of Ref. [\onlinecite{ref58}]). Therefore, in what follows we will only consider these components and eliminate the $\omega$-dependence of $\delta _\alpha ^{\rm{X}} ({\bf{Q}},\omega )$ and $O_\alpha ^{\rm{X}} ({\bf{Q}},\omega )$, with implicit definition of $\delta _\alpha ^{\rm{X}} ({\bf{Q}}) \equiv \delta _\alpha ^{\rm{X}} ({\bf{Q}},\omega  = 0)$ and $O_\alpha ^{\rm{X}} ({\bf{Q}}) \equiv O_\alpha ^{\rm{X}} ({\bf{Q}},\omega  = 0)$.

Inserting Eqs. (\ref{eq074}) to (\ref{eq077}) into Eq. (\ref{eq072}), we have
\begin{equation}\label{eq078}
\begin{split}
\Xi  =& C\int {D\{ \delta ^* \} } \int {D\{ \delta \} } \left( {\int {D\bar \psi {\kern 1pt} D\psi } \exp \left\{  - S_0 [\psi ,\bar \psi ] - S_{{\rm{HS}}} [\psi ,\bar \psi ;\{ \delta ^* \} ,\{ \delta \} ] \right\} } \right)\\
 =& C\int {D\{ \delta ^* \} } \int {D\{ \delta \} } \exp \left\{  - \beta \Omega ^{{\rm{HS}}} [\{ \delta ^* \} ,\{ \delta \} ] \right\} ,
\end{split}
\end{equation}
with a definition of
\begin{equation}\label{eq079}
\begin{split}
S_{{\rm{HS}}} [\psi ,\bar \psi ;\{ \delta ^* \} ,\{ \delta \} ] \equiv& S_{{\rm{HS}}}^{{\rm{sSC}}} [\psi ,\bar \psi ;\delta _{{\rm{sSC}}}^* ,\delta _{{\rm{sSC}}} ] + S_{{\rm{HS}}}^{{\rm{tSC}}} [\psi ,\bar \psi ;\vec \delta _{{\rm{tSC}}}^* ,\vec \delta _{{\rm{tSC}}} ]\\
& + S_{{\rm{HS}}}^{{\rm{SPN}}} [\psi ,\bar \psi ;\vec \delta _{{\rm{SPN}}}^* ,\vec \delta _{{\rm{SPN}}} ] + S_{{\rm{HS}}}^{{\rm{CHG}}} [\psi ,\bar \psi ;\delta _{{\rm{CHG}}}^* ,\delta _{{\rm{CHG}}} ] ,
\end{split}
\end{equation}
\begin{equation}\label{eq080}
\begin{split}
\Omega ^{{\rm{HS}}} [\{ \delta ^* \} ,\{ \delta \} ] \equiv  - \frac{1}{\beta }\ln \left( {\int {D\bar \psi {\kern 1pt} D\psi } \exp \left\{ { - S_0 [\psi ,\bar \psi ] - S_{{\rm{HS}}} [\psi ,\bar \psi ;\{ \delta ^* \} ,\{ \delta \} ]} \right\}} \right),
\end{split}
\end{equation}
and with an abbreviation $\{ \delta \}  \equiv (\delta _{{\rm{sSC}}} ,\vec \delta _{{\rm{tSC}}} ,\vec \delta _{{\rm{SPN}}} ,\delta _{{\rm{CHG}}} )$. Equation (\ref{eq078}) indicates that the probability of the auxiliary variables to take the values $(\delta _{{\rm{sSC}}} ,\vec \delta _{{\rm{tSC}}} ,\vec \delta _{{\rm{SPN}}} ,\delta _{{\rm{CHG}}} )$ is proportional to $e^{ - \beta \Omega ^{{\rm{HS}}} [\{ \delta ^* \} ,\{ \delta \} ]}$. Therefore, the value of $\{ \delta \}$ with the maximal provability corresponds to the minimum of the Hubbard-Stratonovich thermodynamic potential $\Omega ^{{\rm{HS}}} [\{ \delta ^* \} ,\{ \delta \} ]$.

In the integration of Eq. (\ref{eq078}), one usually employs the saddle-point approximation. Within this approximation the result of the integration is approximated to be the maximal value of the integrand. Namely, the saddle-point approximation is represented as
\begin{equation}\label{eq081}
\begin{split}
\Xi  \approx \exp \left\{  - \beta \Omega ^{{\rm{HS}}} [\{ \Delta ^* \} ,\{ \Delta \} ] \right\}  = \int {D\bar \psi  D\psi } \exp \left\{  - S_0 [\psi ,\bar \psi ] - S_{{\rm{HS}}} [\psi ,\bar \psi ;\{ \Delta ^* \} ,\{ \Delta \} ] \right \}.
\end{split}
\end{equation}
Here the mean-field parameters $\{ \Delta \}  \equiv (\Delta _{{\rm{sSC}}} ,\vec \Delta _{{\rm{tSC}}} ,\vec \Delta _{{\rm{SPN}}} ,\Delta _{{\rm{CHG}}} )$ are determined by the minimization of the effective thermodynamic potential:
\begin{equation}\label{eq082}
\begin{split}
\Omega ^{{\rm{HS}}} [\{ \Delta ^* \} ,\{ \Delta \} ] = \min \left\{ {\Omega ^{{\rm{HS}}} [\{ \delta ^* \} ,\{ \delta \} ]} \right\}.
\end{split}
\end{equation}
Equation (\ref{eq081}) implies that the interacting system in the saddle-point approximation turns into the noninteracting system in the external field determined by the parameters $\{ \Delta \}$. Combining the minimization condition (\ref{eq082}) and Eq. (\ref{eq080}), we obtain the following self-consistency condition:
\begin{equation}\label{eq083}
\begin{split}
\Delta _\alpha ^{{\rm{sSC}}} ({\bf{Q}}_i^{\rm{P}} ) =& \frac{{\Lambda ^{{\rm{sSC}},\alpha } ({\bf{Q}}_i^{\rm{P}} )}}{{4N\beta \hbar }}\left\langle {O_\alpha ^{{\rm{sSC}}} ({\bf{Q}}_i^{\rm{P}} )} \right\rangle _\Delta  , 
\vec \Delta _\alpha ^{{\rm{tSC}}} ({\bf{Q}}_i^{\rm{P}} ) = \frac{{\Lambda ^{{\rm{tSC}},\alpha } ({\bf{Q}}_i^{\rm{P}} )}}{{4N\beta \hbar }}\left\langle {\vec O_\alpha ^{{\rm{tSC}}} ({\bf{Q}}_i^{\rm{P}} )} \right\rangle _\Delta  ,\\
\vec \Delta _\alpha ^{{\rm{SPN}}} ({\bf{Q}}_i^{\rm{C}} ) =& \frac{{\Lambda ^{{\rm{C}},\alpha } ({\bf{Q}}_i^{\rm{C}} )}}{{4N\beta \hbar }}\left\langle {\vec O_\alpha ^{{\rm{SPN}}} ({\bf{Q}}_i^{\rm{C}} )} \right\rangle _\Delta  ,
\Delta _\alpha ^{{\rm{CHG}}} ({\bf{Q}}_i^{\rm{W}} ) = \frac{{\Lambda ^{{\rm{W}},\alpha } ({\bf{Q}}_i^{\rm{W}} )}}{{4N\beta \hbar }}\left\langle {O_\alpha ^{{\rm{CHG}}} ({\bf{Q}}_i^{\rm{W}} )} \right\rangle _\Delta  ,
\end{split}
\end{equation}
with the mean value defined by
\begin{equation}\label{eq084}
\begin{split}
\left\langle A \right\rangle _\Delta   \equiv \frac{{\int {D\bar \psi {\kern 1pt} D\psi } \exp \left\{  - S_0 [\psi ,\bar \psi ] - S_{{\rm{HS}}} [\psi ,\bar \psi ;\{ \Delta ^* \} ,\{ \Delta \}] \right\} A}}{{\int {D\bar \psi {\kern 1pt} D\psi } \exp \left\{  - S_0 [\psi ,\bar \psi ] - S_{{\rm{HS}}} [\psi ,\bar \psi ;\{ \Delta ^* \} ,\{ \Delta \} ] \right\} }}.
\end{split}
\end{equation}
Thus, in the saddle-point approximation, the system is described by the approximate action:
\begin{equation}\label{eq085}
\begin{split}
S_{{\rm{MF}}}  = S_0 [\psi ,\bar \psi ] + S_{{\rm{HS}}} [\psi ,\bar \psi ;\{ \Delta ^* \} ,\{ \Delta \} ],
\end{split}
\end{equation}
with $S_0$ and $S_{{\rm{HS}}}$, represented by Eqs. (\ref{eq003}) and (\ref{eq079}), respectively.

Now we return to the operator formalism. The action (\ref{eq085}) is equivalent to the following mean-field Hamiltonian:
\begin{equation}\label{eq086}
\begin{split}
&\hat H_{{\rm{MF}}}  = \hat H_0  + \hat H_{{\rm{sSC}}}  + \hat H_{{\rm{tSC}}}  + \hat H_{{\rm{SPN}}}  + \hat H_{{\rm{CHG}}}, 
\hat H_0  = \sum\limits_{o,o'} {\sum\limits_{{\bf{k}},\sigma } {\hat c_{{\bf{k}}o\sigma }^ \dagger  (H_{oo'}^0 ({\bf{k}}) - \mu \delta _{oo'} )\hat c_{{\bf{k}}o'\sigma } } } ,\\
&\hat H_{{\rm{sSC}}}  = \sum\limits_{i = 1}^{N_{\rm{P}} } {\sum\limits_{\alpha  = 1}^{M_{{\rm{sSC}},i} } {\left( {\frac{{4N}}{{\Lambda ^{{\rm{sSC}},\alpha } ({\bf{Q}}_i^{\rm{P}} )}}[\Delta _\alpha ^{{\rm{sSC}}} ({\bf{Q}}_i^{\rm{P}} )]^* \Delta _\alpha ^{{\rm{sSC}}} ({\bf{Q}}_i^{\rm{P}} ) - [\Delta _\alpha ^{{\rm{sSC}}} ({\bf{Q}}_i^{\rm{P}} )]^* \hat O_\alpha ^{{\rm{sSC}}} ({\bf{Q}}_i^{\rm{P}} ) - \Delta _\alpha ^{{\rm{sSC}}} ({\bf{Q}}_i^{\rm{P}} )[\hat O_\alpha ^{{\rm{sSC}}} ({\bf{Q}}_i^{\rm{P}} )]^ \dagger  } \right)} } ,\\
&\hat H_{{\rm{tSC}}}  = \sum\limits_{i = 1}^{N_{\rm{P}} } {\sum\limits_{\alpha  = 1}^{M_{{\rm{tSC}},i} } {\left( {\frac{{4N}}{{\Lambda ^{{\rm{tSC}},\alpha } ({\bf{Q}}_i^{\rm{P}} )}}[\vec \Delta _\alpha ^{{\rm{tSC}}} ({\bf{Q}}_i^{\rm{P}} )]^*  \cdot \vec \Delta _\alpha ^{{\rm{tSC}}} ({\bf{Q}}_i^{\rm{P}} ) - [\vec \Delta _\alpha ^{{\rm{tSC}}} ({\bf{Q}}_i^{\rm{P}} )]^*  \cdot \hat {\vec O}_\alpha ^{{\rm{tSC}}} ({\bf{Q}}_i^{\rm{P}} ) - \vec \Delta _\alpha ^{{\rm{tSC}}} ({\bf{Q}}_i^{\rm{P}} ) \cdot [\hat {\vec O}_\alpha ^{{\rm{tSC}}} ({\bf{Q}}_i^{\rm{P}} )]^ \dagger  } \right)} } ,\\
&\hat H_{{\rm{SPN}}}  = \sum\limits_{i = 1}^{N_{\rm{C}} } {\sum\limits_{\alpha  = 1}^{M_{{\rm{C}},i} } {\left( {\frac{{4N}}{{\Lambda ^{{\rm{C}},\alpha } ({\bf{Q}}_i^{\rm{C}} )}}[\vec \Delta _\alpha ^{{\rm{SPN}}} ({\bf{Q}}_i^{\rm{C}} ,\omega )]^*  \cdot \vec \Delta _\alpha ^{{\rm{SPN}}} ({\bf{Q}}_i^{\rm{C}} ,\omega ) - [\vec \Delta _\alpha ^{{\rm{SPN}}} ({\bf{Q}}_i^{\rm{C}} )]^*  \cdot \hat {\vec O}_\alpha ^{{\rm{SPN}}} ({\bf{Q}}_i^{\rm{C}} ) - \vec \Delta _\alpha ^{{\rm{SPN}}} ({\bf{Q}}_i^{\rm{C}} ) \cdot [\hat {\vec O}_\alpha ^{{\rm{SPN}}} ({\bf{Q}}_i^{\rm{C}} )]^ \dagger  } \right)} } ,\\
&\hat H_{{\rm{CHG}}}  = \sum\limits_{i = 1}^{N_{\rm{W}} } {\sum\limits_{\alpha  = 1}^{M_{{\rm{W}},i} } {\left( {\frac{{4N}}{{\Lambda ^{{\rm{W}},\alpha } ({\bf{Q}}_i^{\rm{W}} )}}[\Delta _\alpha ^{{\rm{CHG}}} ({\bf{Q}}_i^{\rm{W}} )]^* \Delta _\alpha ^{{\rm{CHG}}} ({\bf{Q}}_i^{\rm{W}} ) - [\Delta _\alpha ^{{\rm{CHG}}} ({\bf{Q}}_i^{\rm{W}} )]^* [\hat {\vec O}_\alpha ^{{\rm{CHG}}} ({\bf{Q}}_i^{\rm{W}} ) - \Delta _\alpha ^{{\rm{CHG}}} ({\bf{Q}}_i^{\rm{W}} )[\hat {\vec O}_\alpha ^{{\rm{CHG}}} ({\bf{Q}}_i^{\rm{W}} )]^ \dagger  } \right)} } ,\\
&\hspace{2pc} \textrm{ with constraints } \vec \Delta _\alpha ^{{\rm{SPN}}} ({\bf{Q}}_i^{\rm{C}} ) = [\vec \Delta _\alpha ^{{\rm{SPN}}} ( - {\bf{Q}}_i^{\rm{C}} )]^* ,\Delta _\alpha ^{{\rm{CHG}}} ({\bf{Q}}_i^{\rm{W}} ) = [\Delta _\alpha ^{{\rm{CHG}}} ( - {\bf{Q}}_i^{\rm{W}} )]^* .
\end{split}
\end{equation}
Here the operators $\hat O_\alpha ^{\rm{X}}$ are expressed as
\begin{equation}\label{eq087}
\begin{split}
\hat O_\alpha ^{{\rm{sSC}}} ({\bf{Q}}_i^{\rm{P}} ) =& \sum\limits_{\bf{k}} {\sum\limits_{o,o',m} {[\varphi _{oo'm}^{{\rm{P}},\alpha } ({\bf{Q}}_i^{\rm{P}} )]^* e^{i{\bf{R}}_m  \cdot {\bf{k}}} } } \sum\limits_\sigma  {\sigma \hat c_{ - {\bf{k}},o', - \sigma } \hat c_{{\bf{k}} + {\bf{Q}}_i^{\rm{P}} ,o,\sigma } } ,\\
\hat {\vec O}_\alpha ^{{\rm{tSC}}} ({\bf{Q}}_i^{\rm{P}} ) =& (\hat O_{\alpha ,x}^{{\rm{tSC}}} ({\bf{Q}}_i^{\rm{P}} ),\hat O_{\alpha ,y}^{{\rm{tSC}}} ({\bf{Q}}_i^{\rm{P}} ),\hat O_{\alpha ,z}^{{\rm{tSC}}} ({\bf{Q}}_i^{\rm{P}} )) = \sum\limits_{\bf{k}} {\sum\limits_{o,o',m} {[\varphi _{oo'm}^{{\rm{P}},\alpha } ({\bf{Q}}_i^{\rm{P}} )]^* e^{i{\bf{R}}_m  \cdot {\bf{k}}} } } \\
&\times  \left. { \left( { - \sum\limits_\sigma  {\sigma \hat c_{ - {\bf{k}},o',\sigma } \hat c_{{\bf{k}} + {\bf{Q}}_i^{\rm{P}} ,o,\sigma } } ,} \right. - i\sum\limits_\sigma  {\hat c_{ - {\bf{k}},o',\sigma } \hat c_{{\bf{k}} + {\bf{Q}}_i^{\rm{P}} ,o,\sigma } } ,\sum\limits_\sigma  {\hat c_{ - {\bf{k}},o', - \sigma } \hat c_{{\bf{k}} + {\bf{Q}}_i^{\rm{P}} ,o,\sigma } } } \right),\\
\hat {\vec O}_\alpha ^{{\rm{SPN}}} ({\bf{Q}}_i^{\rm{C}} ) =& \sum\limits_{\bf{k}} {\sum\limits_{o,o',m} {[\varphi _{oo'm}^{{\rm{C}},\alpha } ({\bf{Q}}_i^{\rm{C}} )]^* e^{i{\bf{R}}_m  \cdot {\bf{k}}} } } \sum\limits_{\sigma ,\sigma '} {\hat c_{{\bf{k}},o',\sigma }^ \dagger  \vec \sigma _{\sigma \sigma '} \hat c_{{\bf{k}} + {\bf{Q}}_i^{\rm{C}} ,o,\sigma '} } ,\\
\hat O_\alpha ^{{\rm{CHG}}} ({\bf{Q}}_i^{\rm{W}} ) =& \sum\limits_{\bf{k}} {\sum\limits_{o,o',m} {[\varphi _{oo'm}^{{\rm{W}},\alpha } ({\bf{Q}}_i^{\rm{W}} )]^* e^{i{\bf{R}}_m  \cdot {\bf{k}}} } } \sum\limits_\sigma  {\hat c_{{\bf{k}},o',\sigma }^ \dagger  \hat c_{{\bf{k}} + {\bf{Q}}_i^{\rm{W}} ,o,\sigma } } ,
\end{split}
\end{equation}
with the symmetry relation
\begin{equation}\label{eq088}
\begin{split}
\hat {\vec O}_\alpha ^{{\rm{SPN}}} ( - {\bf{Q}}_i^{\rm{C}} ) = [\hat {\vec O}_\alpha ^{{\rm{SPN}}} ({\bf{Q}}_i^{\rm{C}} )]^ \dagger  ,\hat O_\alpha ^{{\rm{CHG}}} ( - {\bf{Q}}_i^{\rm{W}} ) = [\hat O_\alpha ^{{\rm{CHG}}} ({\bf{Q}}_i^{\rm{W}} )]^ \dagger  .
\end{split}
\end{equation}
In addition, inserting the relations
\begin{equation}\nonumber
\begin{split}
&\frac{1}{{\beta \hbar }}\left\langle {\sum\limits_\omega  {\bar \psi (\alpha ,\omega )\psi (\beta ,\omega )} } \right\rangle _\Delta   = \left\langle {\hat c_\alpha ^ \dagger  \hat c_\beta  } \right\rangle _{{\rm{MF}}} ,\\
&\frac{1}{{\beta \hbar }}\left\langle {\sum\limits_\omega  {\psi (\alpha ,\omega )\psi (\beta , - \omega )} } \right\rangle _\Delta   = \left\langle {\hat c_\alpha  \hat c_\beta  } \right\rangle _{{\rm{MF}}} ,
\frac{1}{{\beta \hbar }}\left\langle {\sum\limits_\omega  {\bar \psi (\alpha ,\omega )\bar \psi (\beta , - \omega )} } \right\rangle _\Delta   = \left\langle {\hat c_\alpha ^ \dagger  \hat c_\beta ^ \dagger  } \right\rangle _{{\rm{MF}}} ,
\end{split}
\end{equation}
into Eq. (\ref{eq083}), we have the self-consistency condition in the operator formalism:
\begin{equation}\label{eq089}
\begin{split}
\Delta _\alpha ^{{\rm{sSC}}} ({\bf{Q}}_i^{\rm{P}} ) =& \frac{{\Lambda ^{{\rm{sSC}},\alpha } ({\bf{Q}}_i^{\rm{P}} )}}{{4N}}\left\langle {\hat O_\alpha ^{{\rm{sSC}}} ({\bf{Q}}_i^{\rm{P}} )} \right\rangle _{{\rm{MF}}} ,
\vec \Delta _\alpha ^{{\rm{tSC}}} ({\bf{Q}}_i^{\rm{P}} ) = \frac{{\Lambda ^{{\rm{tSC}},\alpha } ({\bf{Q}}_i^{\rm{P}} )}}{{4N}}\left\langle {\hat {\vec O}_\alpha ^{{\rm{tSC}}} ({\bf{Q}}_i^{\rm{P}} )} \right\rangle _{{\rm{MF}}} ,\\
\vec \Delta _\alpha ^{{\rm{SPN}}} ({\bf{Q}}_i^{\rm{C}} ) =& \frac{{\Lambda ^{{\rm{C}},\alpha } ({\bf{Q}}_i^{\rm{C}} )}}{{4N}}\left\langle {\hat {\vec O}_\alpha ^{{\rm{SPN}}} ({\bf{Q}}_i^{\rm{C}} )} \right\rangle _{{\rm{MF}}} ,
\Delta _\alpha ^{{\rm{CHG}}} ({\bf{Q}}_i^{\rm{W}} ) = \frac{{\Lambda ^{{\rm{W}},\alpha } ({\bf{Q}}_i^{\rm{W}} )}}{{4N}}\left\langle {\hat O_\alpha ^{{\rm{CHG}}} ({\bf{Q}}_i^{\rm{W}} )} \right\rangle _{{\rm{MF}}} .
\end{split}
\end{equation}
Here $\left\langle {\hat A} \right\rangle _{{\rm{MF}}}$ is the mean value of the operator $\hat A$ of the system described by $\hat H_{{\rm{MF}}}$, in the grand canonical ensemble.

If the total number of electrons in the system is taken to be constant, the chemical potential $\mu$ of the MF Hamiltonian should be a function of the variables $\{ \delta \}$. In this case, according to the knowledge of statistical mechanics, the minimization condition of Eq. (\ref{eq082}) translates into the minimization of the free energy of the MF Hamiltonian (\ref{eq086}). At the same time, the self-consistency condition (\ref{eq089}) will also coincide with the minimization condition of the free energy. At zero temperature, the free energy becomes the ground-state energy. Therefore, the self-consistency condition is achieved by the minimization of the ground-state energy of the MF Hamiltonian.
\end{widetext}

\section{Coexistence phase of chiral SC and chiral SDW} \label{sec4}

This section presents the first application of our TUFRG + MF approach to competing or coexisting orders. Actually, our approach has already been applied, with poor elucidation and derivation, to the simple case with a unique singular mode in the spin channel \cite{ref43}. A similar scheme has been used to describe the SC phases in the Rashba-Hubbard model \cite{ref48}, but it has not addressed competing orders in distinct channels. Furthermore, its detailed derivation and reasonable validation are not presented. A similar approach has also been utilized in the context of the singular-mode FRG to determine the SC gap of strontium ruthenate \cite{ref59}, but again without any derivation and justification. In this section, we use our TUFRG + MF approach to analyze competing chiral $d$-wave SC \cite{ref60, ref61, ref62, ref63} and chiral SDW \cite{ref63, ref64, ref65} orders which have been predicted to be possible in graphene near van Hove filling. In some FRG studies \cite{ref36, ref42}, it was concluded that the chiral SDW phase is generated right around van Hove filling, while the chiral $d$-wave SC emerges slightly away from it. Here we will focus our attention on the value of both order parameters, $\Delta _{d_{1,2} }^{{\rm{sSC}}} ({\bf{Q}} = 0)$ and $\vec \Delta ^{{\rm{SPN}}} ({\bf{Q}} = {\bf{M}}_{1,2,3} )$, as a function of doping. In the following we will denote these parameters as $\Delta _{1,2}^{{\rm{sSC}}}  = \Delta _{d_{1,2} }^{{\rm{sSC}}} ({\bf{Q}} = 0)$ and $\vec \Delta ^{{\rm{SPN}}} ({\bf{M}}_{1,2,3} ) = \vec \Delta ^{{\rm{SPN}}} ({\bf{Q}} = {\bf{M}}_{1,2,3} )$.

\begin{figure}[h!]
	\begin{center}
		\includegraphics[width=8.6cm]{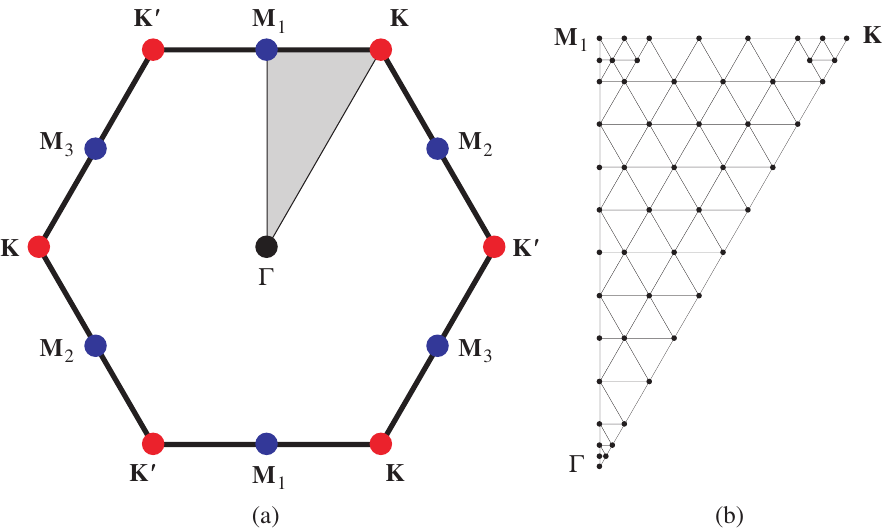}
	\end{center}
	\caption{(a) Brillouin zone and high-symmetry points of the honeycomb lattice. The gray triangle is the irreducible region of the Brillouin zone. (b) Mesh of sampling points for transfer momenta within the irreducible region. Only the bosonic propagators with these transfer momenta are numerically calculated at each step of the TUFRG, while others are generated by using the point-group symmetry relations \cite{ref41, ref42}.}
	\label{fig2}
\end{figure}

Here we study the honeycomb lattice at zero temperature that is doped close to van Hove filling. As mentioned above, for the system at zero temperature, we determine the order parameters by minimizing the ground-state energy of the MF Hamiltonian, with respect to the order parameters. The honeycomb lattice is described by the following Hubbard model:
\begin{equation}\label{eq090}
\begin{split}
\hat H =& \hat H_0  + \hat H_{{\mathop{\rm int}} } ,\\
\hat H_0  =&  - t\sum\limits_{\left\langle {iA,jB} \right\rangle ,\sigma } {(\hat c_{iA\sigma }^ \dagger  \hat c_{jB\sigma } }  + {\rm{H}}{\rm{.c}}{\rm{.}})\\
& - t' \sum\limits_{\left\langle {\left\langle {io,jo} \right\rangle } \right\rangle ,o,\sigma } {(\hat c_{io\sigma }^ \dagger  \hat c_{jo\sigma } }  + {\rm{H}}{\rm{.c}}{\rm{.}}) - \mu \sum\limits_{i,o,\sigma } {\hat c_{io\sigma }^ \dagger  \hat c_{io\sigma } },\\
\hat H_{{\mathop{\rm int}} }  =& U\sum\limits_{i,0} {\hat n_{io \uparrow } \hat n_{io \downarrow } }.
\end{split}
\end{equation}
Here $t$ and $t'$ are the hopping amplitudes between nearest and next-nearest neighbors, $\mu$ is the chemical potential, while $\left\langle {iA,jB} \right\rangle$ and $\left\langle {\left\langle {io,jo} \right\rangle } \right\rangle$ denote nearest-neighbor and next-nearest-neighbor bonds. The doping level is defined by $\delta  = n_e  - 1$ with $n_e$ being the number of electrons per site. The chemical potential and the doping level have the values of $\mu _{{\rm{VH}}}  = t + 2t',\delta _{{\rm{VH}}}  = 0.25$ at van Hove filling. We set the parameters as $t = 2.8{\rm{eV}},t' = 0.1{\rm{eV}},U = 10.08{\rm{eV}}$, which were used in our previous work \cite{ref42}. We investigate the range of doping corresponding to $n_e = 1.25$--$1.29$. Figure \ref{fig2} shows the Brillouin zone (BZ) of the honeycomb lattice and sampling points for transfer momenta within the irreducible region of BZ.

\begin{table}[h!]
	\caption{The ratios of two computation times elapsed for the MF and TUFRG calculations for several values of $n_e$.}
	\begin{center}
		\begin{tabular}{|c|c|}
			\hline
			$n_e$ & $t_{{\rm{MF}}} /t_{{\rm{TUFRG}}}$\\ 
			\hline
			1.25 & 0.126\\
			1.26 & 0.721\\
			1.27 & 0.407\\
			1.28 & 0.264\\
			1.29 & 0.263\\
			\hline
		\end{tabular}
	\end{center}
	\label{tab01}
\end{table}

\begin{figure*}
	\begin{center}
		\includegraphics[width=15.2cm]{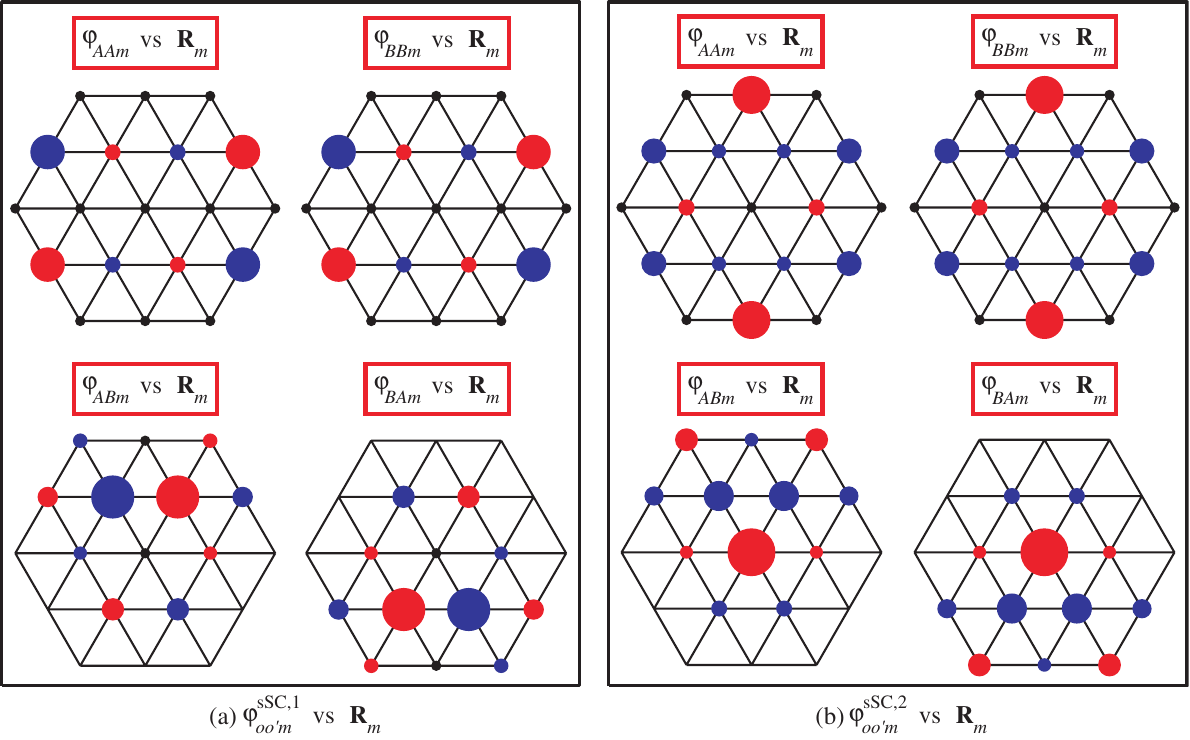}
	\end{center}
	\caption{Values of the two-fold degenerate spin-singlet irreducible singular modes (a) $\varphi _{oo'm}^{{\rm{sSC}},1}$ and (b) $\varphi _{oo'm}^{{\rm{sSC}},2}$ for $n_e  = 1.286$ ($\delta  = 0.286$). The red and blue circles indicate the positive and negative values, respectively, and the absolute values $\left| {\varphi _{oo'm} } \right|$ are encoded by the radius of the circles. The small dots denote the sites ${\bf{R}}_m$ having negligible $\varphi _{oo'm}$, while the empty sites are eliminated by the filtering process \cite{ref42}.}
	\label{fig3}
\end{figure*}

\begin{figure*}
	\begin{center}
		\includegraphics[width=15.2cm]{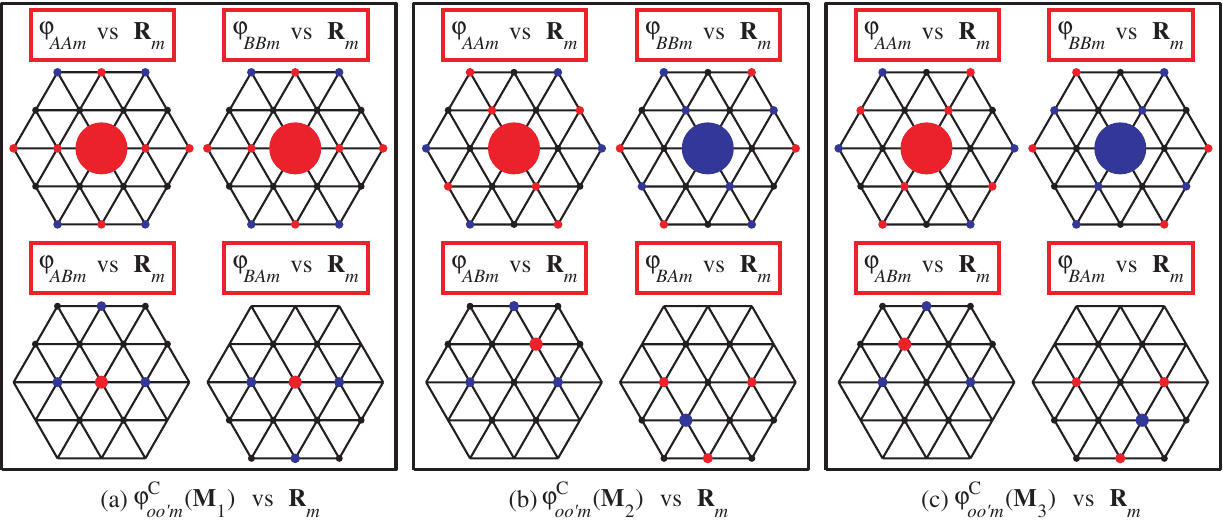}
	\end{center}
	\caption{Values of the irreducible singular modes in the spin channel (a) $\varphi _{oo'm}^{\rm{C}} ({\bf{M}}_1)$, (b) $\varphi _{oo'm}^{\rm{C}} ({\bf{M}}_2)$, and (c) $\varphi _{oo'm}^{\rm{C}} ({\bf{M}}_3)$ for $n_e  = 1.252$ ($\delta  = 0.252$). The red and blue circles, the radius of them, the small dots, and the empty sites have the same meanings as in Fig. \ref{fig3}.}
	\label{fig4}
\end{figure*}

The TUFRG calculation is performed until the maximum absolute value of the elements of $P^\Omega  ({\bf{q}}), C^\Omega  ({\bf{q}}),$ or $W^\Omega  ({\bf{q}})$ exceeds a certain threshold value $S = 10E_{{\rm{BW}}}$ ($E_{{\rm{BW}}}$ is the band width). This process is mostly time consuming. After that, one obtains the singular modes of the bosonic propagators, and then, extracts the irreducible singular modes from those. This calculation is done quickly, so that this stage needs very little computation time. Finally, the MF calculation is performed. This process is in essence a minimization of the free energy (ground-state energy at zero temperature) of the MF Hamiltonian. It needs also much computation time. We show in Table \ref{tab01} the ratios of two computation times elapsed for the MF ($t_{{\rm{MF}}}$) and the TUFRG ($t_{{\rm{TUFRG}}}$) calculations, for several values of $n_e$.

For the parameter sets considered in the present work, the chiral $d$-wave SC and the chiral SDW constitute main ingredients of the resulting phase diagram. Following the process given in Sec.~\ref{sec3B}, the irreducible coupling constants and singular modes are extracted from the bosonic propagators in the pairing and spin channels that were obtained by the TUFRG flow. Then we perform the MF calculation, with these coupling constants and singular modes, following the procedure described in Sec.~\ref{sec3D}. At this stage the system is depicted by the following MF Hamiltonian:
\begin{equation}\label{eq091}
\begin{split}
\hat H_{{\rm{MF}}}  =& \hat H_0  + \hat H_{{\rm{sSC}}}  + \hat H_{{\rm{SPN}}} ,\\
\hat H_0  =& \sum\limits_{{\bf{k}},\sigma } {\left\{ {[ - tF({\bf{k}})]\hat c_{{\bf{k}}A\sigma }^ \dagger  \hat c_{{\bf{k}}B\sigma } } \right.} \\
&+ \left. {[ - tF^* ({\bf{k}})]\hat c_{{\bf{k}}B\sigma }^ \dagger  \hat c_{{\bf{k}}A\sigma } } \right\} \\
&+ \sum\limits_{{\bf{k}},o,\sigma } {[ - t'g({\bf{k}}) - \mu ]\hat c_{{\bf{k}}o\sigma }^ \dagger  \hat c_{{\bf{k}}o\sigma } } ,\\
\hat H_{{\rm{sSC}}}  =& \sum\limits_{\alpha  = 1,2} {\left( {\frac{{4N}}{{\Lambda ^{{\rm{sSC}}} }}} \right.} (\Delta _\alpha ^{{\rm{sSC}}} )^* \Delta _\alpha ^{{\rm{sSC}}}  \\ 
&\left. { - (\Delta _\alpha ^{{\rm{sSC}}} )^* \hat O_\alpha ^{{\rm{sSC}}}  - \Delta _\alpha ^{{\rm{sSC}}} (\hat O_\alpha ^{{\rm{sSC}}} )^ \dagger  } \right) \\ 
=& \sum\limits_{\alpha  = 1,2} {\left( {\frac{{4N}}{{\Lambda ^{{\rm{sSC}}} }}(\Delta _\alpha ^{{\rm{sSC}}} )^* \Delta _\alpha ^{{\rm{sSC}}} } \right.}  \\ 
 &- \sum\limits_{\bf{k}} {\sum\limits_{o,o',m} {(\varphi _{oo'm}^{{\rm{sSC}},\alpha } )^* e^{i{\bf{R}}_m  \cdot {\bf{k}}} } }  \\ 
 & \times (\Delta _\alpha ^{{\rm{sSC}}} )^* \sum\limits_\sigma  {\sigma \hat c_{ - {\bf{k}},o', - \sigma } \hat c_{{\bf{k}},o,\sigma } }  \\ 
& - \sum\limits_{\bf{k}} {\sum\limits_{o,o',m} {\varphi _{oo'm}^{{\rm{sSC}},\alpha } e^{ - i{\bf{R}}_m  \cdot {\bf{k}}} } }  \\ 
& \times \left. {\Delta _\alpha ^{{\rm{sSC}}} \sum\limits_\sigma  {\sigma \hat c_{{\bf{k}},o,\sigma }^ \dagger  \hat c_{ - {\bf{k}},o', - \sigma }^ \dagger  } } \right) ,\\ 
 \hat H_{{\rm{SPN}}}  =& \sum\limits_{i = 1}^3 {\left( {\frac{{4N}}{{\Lambda ^{\rm{C}} }}} \right.} \vec \Delta ^{{\rm{SPN}}} ({\bf{M}}_i ) \cdot \vec \Delta ^{{\rm{SPN}}} ({\bf{M}}_i ) \\ 
&- 2\left. {\vec \Delta ^{{\rm{SPN}}} ({\bf{M}}_i ) \cdot \hat {\vec O}^{{\rm{SPN}}} ({\bf{M}}_i )} \right) \\ 
 =& \sum\limits_{i = 1}^3 {\left( {\frac{{4N}}{{\Lambda ^{\rm{C}} }}\vec \Delta ^{{\rm{SPN}}} ({\bf{M}}_i ) \cdot \vec \Delta ^{{\rm{SPN}}} ({\bf{M}}_i )} \right.} \\
&- 2\sum\limits_{\bf{k}} {\sum\limits_{o,o',m} {[\varphi _{oo'm}^{\rm{C}} ({\bf{M}}_i )]^* e^{i{\bf{R}}_m  \cdot {\bf{k}}} } }  \\ 
& \times \vec \Delta ^{{\rm{SPN}}} ({\bf{M}}_i ) \cdot \left. {\sum\limits_{\sigma ,\sigma '} {\hat c_{{\bf{k}},o',\sigma }^ \dagger  \vec \sigma _{\sigma \sigma '} \hat c_{{\bf{k}} + {\bf{M}}_i ,o,\sigma '} } } \right).
\end{split}
\end{equation}
Here we used the relations, $\vec \Delta ^{{\rm{SPN}}} ({\bf{Q}}) = [\vec \Delta ^{{\rm{SPN}}} ({\bf{Q}})]^* $ and $\hat {\vec O}^{{\rm{SPN}}} ({\bf{Q}}) = [\hat {\vec O}^{{\rm{SPN}}} ({\bf{Q}})]^ \dagger $, valid for ${\bf{Q}} = {\bf{M}}_{1,2,3}  \in \{ {\bf{G}}/2\}$. The functions $F({\bf{k}})$ and $g({\bf{k}})$ are defined by
\begin{equation}\label{eq092}
\begin{split}
F({\bf{k}}) \equiv& 1 + 2\cos \left( {\frac{1}{2}k_x a} \right)e^{ - i\frac{{\sqrt 3 }}{2}k_y a} ,\\
g({\bf{k}}) \equiv& 2\left[ {\cos (k_x a) + 2\cos \left( {\frac{1}{2}k_x a} \right)\cos \left( {\frac{{\sqrt 3 }}{2}k_y a} \right)} \right].
\end{split}
\end{equation}

As an example, the irreducible singular modes in the pairing channel for $n_e  = 1.286$ ($\delta  = 0.286$) and in the spin channel for $n_e  = 1.252$ ($\delta  = 0.252$) are shown schematically in Figs. \ref{fig3} and \ref{fig4}. In Fig. \ref{fig4}, large circles centered at the origin demonstrate that the resulting spin order is a kind of SDW.

\begin{figure}[h!]
	\begin{center}
		\includegraphics[width=8.6cm]{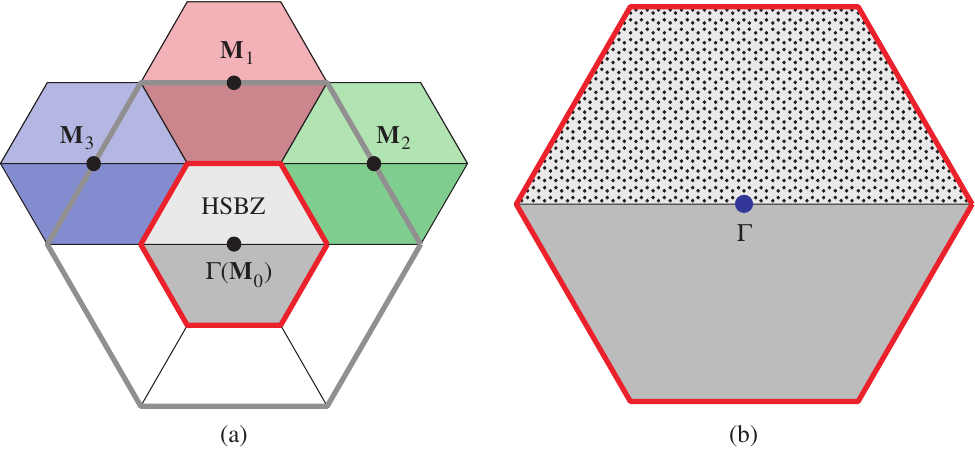}
	\end{center}
	\caption{(a) Small BZ reduced by the spin order (small hexagon surrounded by a red border) and the original BZ (large hexagon with a gray border). The spin order couples a momentum in the small BZ with the ones in three little hexagons above it, while the superconducting order links the momenta in a half of the small BZ (HSBZ) to the ones within another half region below it. (b) Sampling points for the momenta in the HSBZ, used in the MF calculation. The HSBZ has an area smaller by a factor of 8 than that of the original BZ.}
	\label{fig5}
\end{figure}

In the presence of this SDW, the unit cell gets enlarged, while the BZ reduced. The BZ of the system is shown in Fig. \ref{fig5}(a), compared with the BZ of the honeycomb lattice. Moreover, due to the SC order, a pair of the states associated with momenta ${\bf{k}}$ and $-{\bf{k}}$ are coupled with each other, and therefore, the region of independent momentum becomes half of small BZ (HSBZ). The sampling momentum points in it, employed in the MF calculation, are shown in Fig. \ref{fig5}(b).

\begin{figure}[h!]
	\begin{center}
		\includegraphics[width=8.6cm]{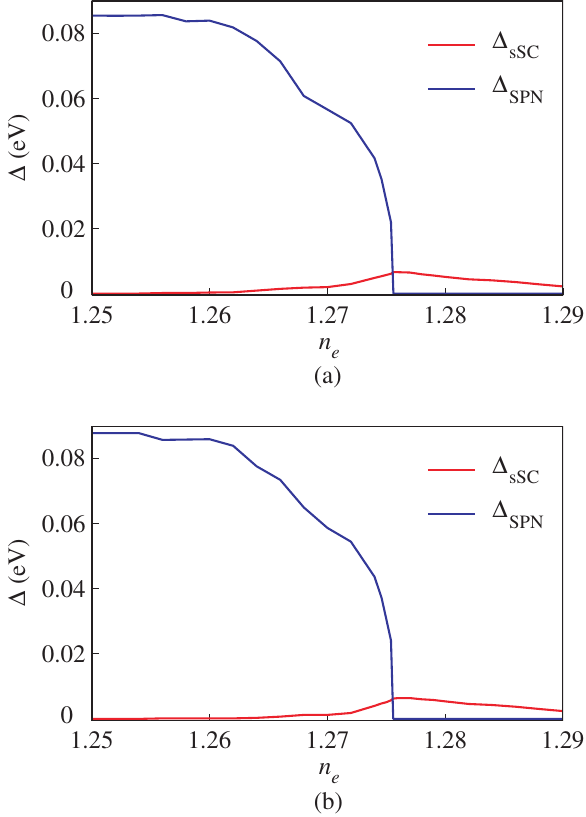}
	\end{center}
	\caption{Order parameters $\Delta _{{\rm{sSC}}}$ and $\Delta _{{\rm{SPN}}}$ as a function of $n_e$ obtained by the TUFRG + MF calculations starting from the bosonic propagators (a) at the divergence scale $\Omega _D$ and (b) just before $\Omega _D$, i.e., at $\Omega  = \Omega _D  + \Delta \Omega$. Two plots are nearly identical, demonstrating the robustness of the results of our TUFRG + MF scheme.}
	\label{fig6}
\end{figure}

One can express the Hamiltonian (\ref{eq091}) in terms of ${\bf{\hat X}}_{\bf{k}}  \equiv ({\rm{\hat C}}_{\bf{k}} ,{\rm{\hat C}}_{{\bf{k}} + {\bf{M}}_1 } ,{\rm{\hat C}}_{{\bf{k}} + {\bf{M}}_2 } ,{\rm{\hat C}}_{{\bf{k}} + {\bf{M}}_3 }, {\rm{\hat C}}_{ - {\bf{k}}}^ \dagger  ,{\rm{\hat C}}_{ - {\bf{k}} + {\bf{M}}_1 }^ \dagger  ,{\rm{\hat C}}_{ - {\bf{k}} + {\bf{M}}_2 }^ \dagger$, ${\rm{\hat C}}_{ - {\bf{k}} + {\bf{M}}_3 }^ \dagger  )^T$ with ${\rm{\hat C}}_{\bf{k}}  \equiv (\hat c_{{\bf{k}},A, \uparrow } ,\hat c_{{\bf{k}},A, \downarrow } ,\hat c_{{\bf{k}},B, \uparrow }, \hat c_{{\bf{k}},B, \downarrow } )$. Diagonalizing it, we can obtain 32 eigenstates per momentum in the HSBZ. From the eigenstates for all the sampling points, one can calculate the ground-state energy $E_G$ as a function of $\Delta _1^{{\rm{sSC}}} ,\Delta _2^{{\rm{sSC}}} ,\vec \Delta ^{{\rm{SPN}}} ({\bf{M}}_1 ),\vec \Delta ^{{\rm{SPN}}} ({\bf{M}}_2 )$, and $\vec \Delta ^{{\rm{SPN}}} ({\bf{M}}_3 )$. Finally, we can determine the order parameters by minimization of $E_G$. In our calculation the resulting order parameters have the form of
\begin{equation}\label{eq093}
\begin{split}
&\Delta _1^{{\rm{sSC}}}  = \Delta _{{\rm{sSC}}} ,\Delta _2^{{\rm{sSC}}}  = i\Delta _{{\rm{sSC}}} ,\\
&\vec \Delta ^{{\rm{SPN}}} ({\bf{M}}_1 ) =  - \Delta _{{\rm{SPN}}} {\bf{e}}_z ,\\
&\vec \Delta ^{{\rm{SPN}}} ({\bf{M}}_2 ) =  - \Delta _{{\rm{SPN}}} {\bf{e}}_x ,\\
&\vec \Delta ^{{\rm{SPN}}} ({\bf{M}}_3 ) =  - \Delta _{{\rm{SPN}}} {\bf{e}}_y ,
\end{split}
\end{equation}
which indicate the chiral SC and the chiral SDW orders.

The order parameters $\Delta _{{\rm{sSC}}}$ and $\Delta _{{\rm{SPN}}}$ as a function of the electron density are shown in Fig. \ref{fig6}. For comparison, we calculated the order parameters using our TUFRG + MF scheme, in two ways. In the first calculation we obtained the irreducible bosonic propagators from the TUFRG result of the bosonic propagators at the divergence scale $\Omega _D$, and then plugged them into the MF calculation. In the second calculation the irreducible bosonic propagators are extracted from the TUFRG result just before entering the divergent regime, namely, from the result at the scale slight larger than $\Omega _D$. Comparative analysis of the plots in Figs. \ref{fig6}(a) and \ref{fig6}(b) demonstrates the robustness of the results of our TUFRG + MF scheme. The plots are characterized by weak and persistent chiral SC order and strong, but suddenly disappearing chiral SDW. These features are very similar to the amplitudes of antiferromagnetic and SC gaps in the ground state of the 2D Hubbard model \cite{ref33}. Near $n_e  = 1.27$, there is an extended region where the chiral SC and chiral SDW orders coexist. In this region the SC order is much smaller than the SDW order. Hence the SC order can be thought of as a secondary order within the chiral SDW phase having the unit cell four times larger than the original one. Since the Fermi surface shrinks to two points, the SC order disappears at van Hove filling. At $n_e  \approx 1.275$ the SDW order drops suddenly and the plot of $\Delta _{{\rm{sSC}}}$ exhibits a kink, which implies that the two order parameters compete with each other (see the discussion in page 3 of Ref. [\onlinecite{ref33}]). The drop of the chiral SDW order is related with the generalized Stoner criterion for this instability and the Fermi surface structure away from the nesting.

\section{Conclusion} \label{sec5}
In the present work, we proposed an approach for combining efficiently the TUFRG and the MF theory, extending the efficient FRG + MF scheme \cite{ref33} developed by Wang, Eberlein, and Metzner. Following the FRG + MF, fluctuation effects from other channels were neglected in the symmetry-broken regime of the TUFRG flow, yielding the RPA flow of the bosonic propagators. The irreducible bosonic propagators were defined as the initial values of these RPA flow equations. We retained only the dominantly divergent parts of the propagators at the divergence scale, and determined the singular eigenmodes of them. The irreducible bosonic propagators and their eigenmodes (irreducible singular modes) are obtained by resolving inversely the RPA matrix equations. The singular and irreducible singular modes have to satisfy the universal RAS symmetry relations derived from the antisymmetry of Grassmann variables. The MF equation based on the irreducible singular modes was derived by introducing the Hubbard-Stratonovich transformation and employing the saddle-point approximation, in the framework of the path-integral formalism. Details of our TUFRG + MF algorithm are described below.

First, the TUFRG flow equation (\ref{eq044}) is integrated until the largest element of some bosonic propagator exceeds a certain value at the divergence scale $\Omega_D$. Second, the singular eigenvalues and eigenmodes are found by a diagonalization [Eq. (\ref{eq051})] of the resulting propagators, $P^{\Omega _D }, C^{\Omega _D }$, and $W^{\Omega _D}$. Third, the irreducible bosonic propagators [Eqs. (\ref{eq055}) to (\ref{eq057})] and the irreducible singular modes [Eqs. (\ref{eq058}) and (\ref{eq060})] are determined. The irreducible singular modes in the pairing channel should respect the condition (\ref{eq064}) and are divided into the spin-singlet and spin-triplet modes. The modes in the spin and charge channels should satisfy the constraints (\ref{eq066}) and (\ref{eq067}). Finally, the irreducible coupling constants $\Lambda ^{{\rm{X}},\alpha } ({\bf{Q}}_i^{\rm{X}} )$ and singular modes $\left| {\varphi ^{{\rm{X}},\alpha } ({\bf{Q}}_i^{\rm{X}} )} \right\rangle$ are inserted into the MF equation with the Hamiltonian given by Eq. (\ref{eq086}). This equation should be combined with the self-consistency condition (\ref{eq089}), leading to final results of the order parameters.

Our novel scheme was applied to a quantitatively reasonable analysis of the competing chiral $d$-wave SC and chiral SDW orders, predicted near van Hove filling of the honeycomb lattice. The plot of the magnitudes of both order parameters are obtained as a function of the electron density. Comparative analysis of the plots, which were obtained from the TUFRG results at different scales, indicates the robustness of the results of our TUFRG + MF scheme. The plots are characterized by weak and durable chiral SC order and strong, but suddenly dropped chiral SDW, and these features are similar to the previous work where the antiferromagnetic and SC gaps are discussed for the ground state of the 2D Hubbard model. This calculation result shows that our TUFRG + MF approach can elevate the power of the TUFRG to a quantitatively reasonable level and extend its application to the study of the coexisting orders.

Finally, we give a brief comment on the comparison of our approach with the renormalized MF theory \cite{ref26}. When using the sharp momentum cutoff regulator, the latter will be physically reasonable. So, it would be interesting to compare these two MF approaches, while using the momentum cutoff regulator in both of them. In Appendix \ref{appendB}, we derived the critical conditions of both approaches. According to our result, the ordering tendencies are a bit more enhanced in our TUFRG + MF than in the renormalized MF. However, in the special case of the SC order in the single-band systems, the two critical conditions become identical.

\section*{Acknowledgments}
The author thanks Prof. Kwang-Il Kim, Prof. Guang-Shan Tian, and Prof. Hak-Chol Pak for their great educational efforts.

\appendix

\begin{widetext}
	
\section{Equivalence of critical conditions in RPA and MF theory} \label{appendA}

We prove the equivalence of two critical conditions in the RPA and the MF theory under the assumption of the positivity of the initial matrices. First, we consider the critical conditions of the RPA. The RPA flow equations in the pairing, spin, and charge channels are, respectively [see Eq. (\ref{eq050})]:
\begin{equation}\label{eqA01}
\begin{split}
&\left[ { - P^\Omega  ({\bf{q}})} \right]^{ - 1}  = \left[ { - P^{(0)} ({\bf{q}})} \right]^{ - 1}  + \chi ^{{\rm{pp}}(\Omega )} ({\bf{q}}),\\
&\left[ {C^\Omega  ({\bf{q}})} \right]^{ - 1}  = \left[ {C^{(0)} ({\bf{q}})} \right]^{ - 1}  - \chi ^{{\rm{ph}}(\Omega )} ({\bf{q}}),
\left[ {W^\Omega  ({\bf{q}})} \right]^{ - 1}  = \left[ {W^{(0)} ({\bf{q}})} \right]^{ - 1}  - \chi ^{{\rm{ph}}(\Omega )} ({\bf{q}}).
\end{split}
\end{equation}
Here $P^{(0)} ,C^{(0)},$ and $W^{(0)}$ are the initial values of $P^\Omega  ,C^\Omega,$ and $W^\Omega$. In our TUFRG + MF they are the irreducible bosonic propagators, $\tilde P,\tilde C,$ and $\tilde W$, while in the conventional MF theory they would become the initial projection matrices, $V^{{\rm{P}},(0)} ,V^{{\rm{C}},(0)},$ and $V^{{\rm{W,}}(0)}  \equiv V^{{\rm{C,}}(0)}  - 2V^{{\rm{D,}}(0)}$. They should be Hermitian [Eq. (\ref{eq030})] and satisfy the RAS relations [Eq. (\ref{eq031})]:
\begin{equation}\label{eqA02}
\begin{split}
&P_{o'_1 o'_2 m,o_1 o_2 n}^{(0)} ({\bf{q}}) = e^{ - i{\bf{R}}_m  \cdot {\bf{q}}} P_{o'_2 o'_1 \bar m,o_2 o_1 \bar n}^{(0)}({\bf{q}})e^{i{\bf{R}}_n  \cdot {\bf{q}}}, \\
&X_{o'_1 o'_2 m,o_1 o_2 n}^{(0)} ( - {\bf{q}}) = e^{i{\bf{R}}_m  \cdot {\bf{q}}} \left[ {X_{o'_2 o'_{\rm{1}} \bar m,o_{\rm{2}} o_1 \bar n}^{(0)} ({\bf{q}})} \right]^* e^{ - i{\bf{R}}_n  \cdot {\bf{q}}} \textrm{ with } X \in \{ C,W\},\hspace{1pc} \textrm{(RAS)}.
\end{split}
\end{equation}
For simplicity, we assume that $- P^{(0)} ,C^{(0)},$ and $W^{(0)}$ are positive matrices like $\tilde P,\tilde C,$ and $\tilde W$. In this case, the initial matrices, $- P^{(0)} ,C^{(0)},$ and $W^{(0)}$, can be decomposed in terms of their eigenmodes associated with non-zero positive eigenvalues as done in Eq. (\ref{eq059}).
\begin{equation}\label{eqA03}
\begin{split}
\left[ { - P^{(0)} ({\bf{q}})} \right] =& \sum\limits_{\alpha  = 1}^{M_{\rm{P}} } {\Lambda ^{{\rm{P}},\alpha } ({\bf{q}})} \left| {\varphi ^{{\rm{P}},\alpha } ({\bf{q}})} \right\rangle \left\langle {\varphi ^{{\rm{P}},\alpha } ({\bf{q}})} \right|, \textrm{ with } \Lambda ^{{\rm{P}},\alpha } ({\bf{q}}) > 0,\\
C^{(0)} ({\bf{q}}) =& \sum\limits_{\alpha  = 1}^{M_{\rm{C}} } {\Lambda ^{{\rm{C}},\alpha } ({\bf{q}})} \left| {\varphi ^{{\rm{C}},\alpha } ({\bf{q}})} \right\rangle \left\langle {\varphi ^{{\rm{C}},\alpha } ({\bf{q}})} \right|, \textrm{ with } \Lambda ^{{\rm{C}},\alpha } ({\bf{q}}) > 0,\\
W^{(0)} ({\bf{q}}) =& \sum\limits_{\alpha  = 1}^{M_{\rm{W}} } {\Lambda ^{{\rm{W}},\alpha } ({\bf{q}})} \left| {\varphi ^{{\rm{W}},\alpha } ({\bf{q}})} \right\rangle \left\langle {\varphi ^{{\rm{W}},\alpha } ({\bf{q}})} \right|, \textrm{ with } \Lambda ^{{\rm{W}},\alpha } ({\bf{q}}) > 0.
\end{split}
\end{equation}
Here the eigenmodes have to respect the constraints of Eqs. (\ref{eq064}), (\ref{eq066}), and (\ref{eq067}).
\begin{equation}\label{eqA04}
\varphi _{oo'm}^{{\rm{P}},\alpha } ({\bf{q}}) =  \pm \varphi _{o'o\bar m}^{{\rm{P}},\alpha } ({\bf{q}})e^{ - i{\bf{R}}_m  \cdot {\bf{q}}} ,
\end{equation}
\begin{equation}\label{eqA05}
\begin{split}
\varphi _{oo'm}^{{\rm{X}},\alpha } ( - {\bf{q}}) = [\varphi _{o'o\bar m}^{{\rm{X}},\alpha } ({\bf{q}})]^* e^{i{\bf{R}}_m  \cdot {\bf{q}}}  \textrm{ with } {\rm{X}} \in \{ {\rm{C}}, {\rm{W}}\} .
\end{split}
\end{equation}
The eigenmodes of $P^{(0)}$ are divided into two sets, i.e., the spin-singlet $\left\{ {\left| {\varphi ^{{\rm{sSC}},\alpha } } \right\rangle } \right\}$ and the spin-triplet $\left\{ {\left| {\varphi ^{{\rm{tSC}},\alpha } } \right\rangle } \right\}$ modes, which satisfy
\begin{equation}\label{eqA06}
\begin{split}
\varphi _{oo'm}^{{\rm{sSC}},\alpha } ({\bf{q}}) =  + \varphi _{o'o\bar m}^{{\rm{sSC}},\alpha } ({\bf{q}})e^{ - i{\bf{R}}_m  \cdot {\bf{q}}} ,
\varphi _{oo'm}^{{\rm{tSC}},\alpha } ({\bf{q}}) =  - \varphi _{o'o\bar m}^{{\rm{tSC}},\alpha } ({\bf{q}})e^{ - i{\bf{R}}_m  \cdot {\bf{q}}} .
\end{split}
\end{equation}
Due to the positivity of $ - P^{(0)} ,C^{(0)},$ and $W^{(0)}$, we can introduce
\begin{equation}\label{eqA07}
\begin{split}
&\sqrt { - P^{(0)} ({\bf{q}})}  = \sum\limits_{\alpha  = 1}^{M_{\rm{P}} } {\sqrt {\Lambda ^{{\rm{P}},\alpha } ({\bf{q}})} } \left| {\varphi ^{{\rm{P}},\alpha } ({\bf{q}})} \right\rangle \left\langle {\varphi ^{{\rm{P}},\alpha } ({\bf{q}})} \right|,\\
&\sqrt {C^{(0)} ({\bf{q}})}  = \sum\limits_{\alpha  = 1}^{M_{\rm{C}} } {\sqrt {\Lambda ^{{\rm{C}},\alpha } ({\bf{q}})} } \left| {\varphi ^{{\rm{C}},\alpha } ({\bf{q}})} \right\rangle \left\langle {\varphi ^{{\rm{C}},\alpha } ({\bf{q}})} \right|,
\sqrt {W^{(0)} ({\bf{q}})}  = \sum\limits_{\alpha  = 1}^{M_{\rm{W}} } {\sqrt {\Lambda ^{{\rm{W}},\alpha } ({\bf{q}})} } \left| {\varphi ^{{\rm{W}},\alpha } ({\bf{q}})} \right\rangle \left\langle {\varphi ^{{\rm{W}},\alpha } ({\bf{q}})} \right|,
\end{split}
\end{equation}
and rewrite Eq. (\ref{eqA01}) as follows:
\begin{equation}\label{eqA08}
\begin{split}
- P^\Omega  ({\bf{q}}) =& \sqrt { - P^{(0)} ({\bf{q}})} \left( {1 + \sqrt { - P^{(0)} ({\bf{q}})} \chi ^{{\rm{pp}}(\Omega )} ({\bf{q}})\sqrt { - P^{(0)} ({\bf{q}})} } \right)^{ - 1} \sqrt { - P^{(0)} ({\bf{q}})} ,\\
C^\Omega  ({\bf{q}}) =& \sqrt {C^{(0)} ({\bf{q}})} \left( {1 - \sqrt {C^{(0)} ({\bf{q}})} \chi ^{{\rm{ph}}(\Omega )} ({\bf{q}})\sqrt {C^{(0)} ({\bf{q}})} } \right)^{ - 1} \sqrt {C^{(0)} ({\bf{q}})} ,\\
W^\Omega  ({\bf{q}}) =& \sqrt {W^{(0)} ({\bf{q}})} \left( {1 - \sqrt {W^{(0)} ({\bf{q}})} \chi ^{{\rm{ph}}(\Omega )} ({\bf{q}})\sqrt {W^{(0)} ({\bf{q}})} } \right)^{ - 1} \sqrt {W^{(0)} ({\bf{q}})} .
\end{split}
\end{equation}
Inserting Eq. (\ref{eqA07}) into Eq. (\ref{eqA08}) and setting $\Omega$ as $\Omega  = 0$, we get the final results of the RPA flows.
\begin{equation}\label{eqA09}
\begin{split}
 - P^{\Omega  = 0} ({\bf{q}}) =& \sum\limits_{\alpha  = 1}^{M_{\rm{P}} } {\sum\limits_{\beta  = 1}^{M_{\rm{P}} } {\left[ {\sqrt {\Lambda ^{\rm{P}} ({\bf{q}})} \left( {1 + \sqrt {\Lambda ^{\rm{P}} ({\bf{q}})} X^{{\rm{pp}}} ({\bf{q}})\sqrt {\Lambda ^{\rm{P}} ({\bf{q}})} } \right)^{ - 1} \sqrt {\Lambda ^{\rm{P}} ({\bf{q}})} } \right]_{\alpha \beta } } } \left| {\varphi ^{{\rm{P}},\alpha } ({\bf{q}})} \right\rangle \left\langle {\varphi ^{{\rm{P}},\beta } ({\bf{q}})} \right|,\\
C^{\Omega  = 0} ({\bf{q}}) =& \sum\limits_{\alpha  = 1}^{M_{\rm{C}} } {\sum\limits_{\beta  = 1}^{M_{\rm{C}} } {\left[ {\sqrt {\Lambda ^{\rm{C}} ({\bf{q}})} \left( {1 - \sqrt {\Lambda ^{\rm{C}} ({\bf{q}})} X^{{\rm{ph,C}}} ({\bf{q}})\sqrt {\Lambda ^{\rm{C}} ({\bf{q}})} } \right)^{ - 1} \sqrt {\Lambda ^{\rm{C}} ({\bf{q}})} } \right]_{\alpha \beta } } } \left| {\varphi ^{{\rm{C}},\alpha } ({\bf{q}})} \right\rangle \left\langle {\varphi ^{{\rm{C}},\beta } ({\bf{q}})} \right|,\\
W^{\Omega  = 0} ({\bf{q}}) =& \sum\limits_{\alpha  = 1}^{M_{\rm{W}} } {\sum\limits_{\beta  = 1}^{M_{\rm{W}} } {\left[ {\sqrt {\Lambda ^{\rm{W}} ({\bf{q}})} \left( {1 - \sqrt {\Lambda ^{\rm{W}} ({\bf{q}})} X^{{\rm{ph,W}}} ({\bf{q}})\sqrt {\Lambda ^{\rm{W}} ({\bf{q}})} } \right)^{ - 1} \sqrt {\Lambda ^{\rm{W}} ({\bf{q}})} } \right]_{\alpha \beta } } } \left| {\varphi ^{{\rm{W}},\alpha } ({\bf{q}})} \right\rangle \left\langle {\varphi ^{{\rm{W}},\beta } ({\bf{q}})} \right|.
\end{split}
\end{equation}
Here three diagonal matrices, $\Lambda ^{\rm{P}} ({\bf{q}}),\Lambda ^{\rm{C}} ({\bf{q}}),$ and $\Lambda ^{\rm{W}} ({\bf{q}})$, are given by
\begin{equation}\label{eqA10}
\begin{split}
\Lambda _{\alpha \beta }^{\rm{P}} ({\bf{q}}) \equiv \Lambda ^{{\rm{P}},\alpha } ({\bf{q}})\delta _{\alpha \beta } ,\Lambda _{\alpha \beta }^{\rm{C}} ({\bf{q}}) \equiv \Lambda ^{{\rm{C}},\alpha } ({\bf{q}})\delta _{\alpha \beta } ,\Lambda _{\alpha \beta }^{\rm{W}} ({\bf{q}}) \equiv \Lambda ^{{\rm{W}},\alpha } ({\bf{q}})\delta _{\alpha \beta } ,
\end{split}
\end{equation}
and three matrices, $X^{{\rm{pp}}} ({\bf{q}}),X^{{\rm{ph,C}}} ({\bf{q}}),$ and $X^{{\rm{ph,W}}} ({\bf{q}})$, are defined by
\begin{equation}\label{eqA11}
\begin{split}
&X_{\alpha \beta }^{{\rm{pp}}} ({\bf{q}}) \equiv \left\langle {\varphi ^{{\rm{P}},\alpha } ({\bf{q}})} \right|\chi ^{{\rm{pp}}(\Omega  = 0)} ({\bf{q}})\left| {\varphi ^{{\rm{P}},\beta } ({\bf{q}})} \right\rangle ,\\
&X_{\alpha \beta }^{{\rm{ph,C}}} ({\bf{q}}) \equiv \left\langle {\varphi ^{{\rm{C}},\alpha } ({\bf{q}})} \right|\chi ^{{\rm{ph}}(\Omega  = 0)} ({\bf{q}})\left| {\varphi ^{{\rm{C}},\beta } ({\bf{q}})} \right\rangle ,X_{\alpha \beta }^{{\rm{ph,W}}} ({\bf{q}}) \equiv \left\langle {\varphi ^{{\rm{W}},\alpha } ({\bf{q}})} \right|\chi ^{{\rm{ph}}(\Omega  = 0)} ({\bf{q}})\left| {\varphi ^{{\rm{W}},\beta } ({\bf{q}})} \right\rangle .
\end{split}
\end{equation}

The critical condition of the RPA in the pairing channel is given by the requirement that the matrix $-P^{\Omega  = 0} ({\bf{q}})$ should have an eigenvector associated with infinite eigenvalue. It can be represented by the following equation:
\begin{equation}\nonumber
\begin{split}
\sum\limits_{\beta  = 1}^{M_{\rm{P}} } {\left[ {\sqrt {\Lambda ^{\rm{P}} ({\bf{q}})} \left( {1 + \sqrt {\Lambda ^{\rm{P}} ({\bf{q}})} X^{{\rm{pp}}} ({\bf{q}})\sqrt {\Lambda ^{\rm{P}} ({\bf{q}})} } \right)^{ - 1} \sqrt {\Lambda ^{\rm{P}} ({\bf{q}})} } \right]_{\alpha \beta } } C_\beta   = \mathop {\lim }\limits_{\lambda  \to \infty } \lambda C_\alpha ,
\end{split}
\end{equation}
or equivalently,
\begin{equation}\nonumber
\begin{split}
\sum\limits_{\beta  = 1}^{M_{\rm{P}} } {\left[ {\left( {\sqrt {\Lambda ^{\rm{P}} ({\bf{q}})} } \right)^{ - 1} \left( {1 + \sqrt {\Lambda ^{\rm{P}} ({\bf{q}})} X^{{\rm{pp}}} ({\bf{q}})\sqrt {\Lambda ^{\rm{P}} ({\bf{q}})} } \right)\left( {\sqrt {\Lambda ^{\rm{P}} ({\bf{q}})} } \right)^{ - 1} } \right]_{\alpha \beta } } C_\beta   = \mathop {\lim }\limits_{\lambda  \to \infty } \frac{1}{\lambda }C_\alpha   = 0.
\end{split}
\end{equation}
Due to the finiteness of the diagonal matrix $\sqrt {\Lambda ^{\rm{P}} ({\bf{q}})}$, this condition is satisfied only if
\begin{equation}\nonumber
\begin{split}
\sum\limits_{\beta  = 1}^{M_{\rm{P}} } {\left( {1 + \sqrt {\Lambda ^{\rm{P}} ({\bf{q}})} X^{{\rm{pp}}} ({\bf{q}})\sqrt {\Lambda ^{\rm{P}} ({\bf{q}})} } \right)_{\alpha \beta } } \left( {\frac{{C_\beta  }}{{\sqrt {\Lambda ^{{\rm{P}},\beta } ({\bf{q}})} }}} \right) = 0.
\end{split}
\end{equation}
This indicates that the matrix $\sqrt {\Lambda ^{\rm{P}} ({\bf{q}})} X^{{\rm{pp}}} ({\bf{q}})\sqrt {\Lambda ^{\rm{P}} ({\bf{q}})}$ should have the eigenvalue of $-1$. Thus we have the critical condition at ${\bf{q}} = {\bf{Q}}_i^{\rm{P}}$,
\begin{equation}\label{eqA12}
\begin{split}
&\sum\limits_{\beta  = 1}^{M_{{\rm{P}},i} } {\left\{ {\sqrt {\Lambda ^{{\rm{P,}}\alpha } ({\bf{Q}}_i^{\rm{P}} )} } \right.} \left[ { - X_{\alpha \beta }^{{\rm{pp}}} ({\bf{Q}}_i^{\rm{P}} )} \right]\left. {\sqrt {\Lambda ^{{\rm{P,}}\beta } ({\bf{Q}}_i^{\rm{P}} )} } \right\}Y_\beta   = Y_\alpha ,\\
&\textrm{(Critical condition of the RPA in the pairing channel)}.
\end{split}
\end{equation}
In a similar way, we can obtain the critical conditions of the RPA in the spin and charge channels, which read
\begin{equation}\label{eqA13}
\begin{split}
&\sum\limits_{\beta  = 1}^{M_{{\rm{C}},i} } {\left\{ {\sqrt {\Lambda ^{{\rm{C,}}\alpha } ({\bf{Q}}_i^{\rm{C}} )} } \right.} X_{\alpha \beta }^{{\rm{ph,C}}} ({\bf{Q}}_i^{\rm{C}} )\left. {\sqrt {\Lambda ^{{\rm{C,}}\beta } ({\bf{Q}}_i^{\rm{C}} )} } \right\}Y_\beta   = Y_\alpha,\\
&\textrm{(Critical condition of the RPA in the spin channel)},
\end{split}
\end{equation}
\begin{equation}\label{eqA14}
\begin{split}
&\sum\limits_{\beta  = 1}^{M_{{\rm{W}},i} } {\left\{ {\sqrt {\Lambda ^{{\rm{W,}}\alpha } ({\bf{Q}}_i^{\rm{W}} )} } \right.} X_{\alpha \beta }^{{\rm{ph,W}}} ({\bf{Q}}_i^{\rm{W}} )\left. {\sqrt {\Lambda ^{{\rm{W,}}\beta } ({\bf{Q}}_i^{\rm{W}} )} } \right\}Y_\beta   = Y_\alpha,\\
&\textrm{(Critical condition of the RPA in the charge channel)}.
\end{split}
\end{equation}

The critical condition (\ref{eqA12}) can be expressed more concisely. Starting from the definition of $\chi ^{{\rm{pp}}(\Omega )} ({\bf{q}})$ [Eq. (\ref{eq025})], one can easily derive following relation:
\begin{equation}\label{eqA15}
\begin{split}
\chi _{o'_1 o'_2 m,o_1 o_2 n}^{{\rm{pp}}(\Omega )} ({\bf{q}}) = e^{ - i{\bf{R}}_m  \cdot {\bf{q}}} \chi _{o'_2 o'_1 \bar m,o_2 o_1 \bar n}^{{\rm{pp}}(\Omega )} ({\bf{q}})e^{i{\bf{R}}_n  \cdot {\bf{q}}} .
\end{split}
\end{equation}
Combining Eqs. (\ref{eqA06}) and (\ref{eqA15}), we obtain
\begin{equation}\label{eqA16}
\begin{split}
\left\langle {\varphi ^{{\rm{sSC}},\alpha } ({\bf{q}})} \right|\chi ^{{\rm{pp}}(\Omega )} ({\bf{q}})\left| {\varphi ^{{\rm{tSC}},\beta } ({\bf{q}})} \right\rangle  = \left\langle {\varphi ^{{\rm{tSC}},\alpha } ({\bf{q}})} \right|\chi ^{{\rm{pp}}(\Omega )} ({\bf{q}})\left| {\varphi ^{{\rm{sSC}},\beta } ({\bf{q}})} \right\rangle  = 0,
\end{split}
\end{equation}
which demonstrates that each of the two sets, $\left\{ {\left| {\varphi ^{{\rm{sSC}},\alpha } } \right\rangle } \right\}$ and $\left\{ {\left| {\varphi ^{{\rm{tSC}},\alpha } } \right\rangle } \right\}$, becomes the invariant subspace with respect to the matrix $X^{{\rm{pp}}} ({\bf{q}})$. Therefore, the critical condition (\ref{eqA12}) is divided into two separate conditions for the spin-singlet and spin-triplet pairing channels
\begin{equation}\label{eqA17}
\begin{split}
&\sum\limits_{\beta  = 1}^{M_{{\rm{sSC}},i} } {\left\{ {\sqrt {\Lambda ^{{\rm{sSC,}}\alpha } ({\bf{Q}}_i^{\rm{P}} )} } \right.} \left[ { - X_{\alpha \beta }^{{\rm{pp,sSC}}} ({\bf{Q}}_i^{\rm{P}} )} \right]\left. {\sqrt {\Lambda ^{{\rm{sSC,}}\beta } ({\bf{Q}}_i^{\rm{P}} )} } \right\}Y_\beta   = Y_\alpha ,\\
&\textrm{(Critical condition of the RPA in the spin-singlet pairing channel)},
\end{split}
\end{equation}
\begin{equation}\label{eqA18}
\begin{split}
&\sum\limits_{\beta  = 1}^{M_{{\rm{tSC}},i} } {\left\{ {\sqrt {\Lambda ^{{\rm{tSC,}}\alpha } ({\bf{Q}}_i^{\rm{P}} )} } \right.} \left[ { - X_{\alpha \beta }^{{\rm{pp,tSC}}} ({\bf{Q}}_i^{\rm{P}} )} \right]\left. {\sqrt {\Lambda ^{{\rm{tSC,}}\beta } ({\bf{Q}}_i^{\rm{P}} )} } \right\}Y_\beta   = Y_\alpha ,\\
&\textrm{(Critical condition of the RPA in the spin-triplet pairing channel)}.
\end{split}
\end{equation}

Second, we consider the critical conditions of the MF theory. All discussions in Sec.~\ref{sec3} are valid here, except for the irreducible bosonic propagators, $\tilde P,\tilde C,$ and $\tilde W$, replaced with $P^{(0)} ,C^{(0)},$ and $W^{(0)}$. The self-consistency conditions are given by [see Eq. (\ref{eq083})]
\begin{equation}\label{eqA19}
\begin{split}
&\Delta _\alpha ^{{\rm{sSC}}} ({\bf{Q}}_i^{\rm{P}} ) = \frac{{\Lambda ^{{\rm{sSC}},\alpha } ({\bf{Q}}_i^{\rm{P}} )}}{{4N\beta \hbar }}\left\langle {O_\alpha ^{{\rm{sSC}}} ({\bf{Q}}_i^{\rm{P}} )} \right\rangle _\Delta ,
\vec \Delta _\alpha ^{{\rm{tSC}}} ({\bf{Q}}_i^{\rm{P}} ) = \frac{{\Lambda ^{{\rm{tSC}},\alpha } ({\bf{Q}}_i^{\rm{P}} )}}{{4N\beta \hbar }}\left\langle {\vec O_\alpha ^{{\rm{tSC}}} ({\bf{Q}}_i^{\rm{P}} )} \right\rangle _\Delta ,\\
&\vec \Delta _\alpha ^{{\rm{SPN}}} ({\bf{Q}}_i^{\rm{C}} ) = \frac{{\Lambda ^{{\rm{C}},\alpha } ({\bf{Q}}_i^{\rm{C}} )}}{{4N\beta \hbar }}\left\langle {\vec O_\alpha ^{{\rm{SPN}}} ({\bf{Q}}_i^{\rm{C}} )} \right\rangle _\Delta ,
\Delta _\alpha ^{{\rm{CHG}}} ({\bf{Q}}_i^{\rm{W}} ) = \frac{{\Lambda ^{{\rm{W}},\alpha } ({\bf{Q}}_i^{\rm{W}} )}}{{4N\beta \hbar }}\left\langle {O_\alpha ^{{\rm{CHG}}} ({\bf{Q}}_i^{\rm{W}} )} \right\rangle _\Delta.
\end{split}
\end{equation}
Here the mean values are defined as
\begin{equation}\label{eqA20}
\begin{split}
\left\langle A \right\rangle _\Delta   \equiv \frac{{\int {D\bar \psi {\kern 1pt} D\psi } \exp \{  - S_0 [\psi ,\bar \psi ] - S_{{\rm{HS}}} [\psi ,\bar \psi ;\{ \Delta ^* \} ,\{ \Delta \} ]\} A}}{{\int {D\bar \psi {\kern 1pt} D\psi } \exp \{  - S_0 [\psi ,\bar \psi ] - S_{{\rm{HS}}} [\psi ,\bar \psi ;\{ \Delta ^* \} ,\{ \Delta \} ]\} }},
\end{split}
\end{equation}
with $S_0$ given in Eq. (\ref{eq003}), and $S_{{\rm{HS}}}$ defined by Eqs. (\ref{eq074}) to (\ref{eq077}) and (\ref{eq079}). In addition, the fermion bilinears in four channels, $O_\alpha ^{{\rm{sSC}}} ({\bf{q}}),\vec O_\alpha ^{{\rm{tSC}}} ({\bf{q}}),\vec O_\alpha ^{{\rm{SPN}}} ({\bf{q}}),$ and $O_\alpha ^{{\rm{CHG}}} ({\bf{q}})$ are defined by Eq. (\ref{eq069}). In the limit of $\left\{ {\Delta ^* } \right\} \to 0$ and $\left\{ \Delta  \right\} \to 0$, the additional action $S_{{\rm{HS}}} $ vanishes and we have
\begin{equation}\nonumber
\exp \left\{ { - S_{{\rm{HS}}} [\psi ,\bar \psi ;\{ \Delta ^* \} ,\{ \Delta \} ]} \right\} \approx 1 - S_{{\rm{HS}}} [\psi ,\bar \psi ;\{ \Delta ^* \} ,\{ \Delta \} ],
\end{equation}
from which the following equation is obtained:
\begin{equation}\nonumber
\begin{split}
\left\langle A \right\rangle _\Delta   \approx \frac{{\left\langle A \right\rangle _0  - \left\langle {AS_{{\rm{HS}}} [\psi ,\bar \psi ;\{ \Delta ^* \} ,\{ \Delta \} ]} \right\rangle _0 }}{{1 - \left\langle {S_{{\rm{HS}}} [\psi ,\bar \psi ;\{ \Delta ^* \} ,\{ \Delta \} ]} \right\rangle _0 }}
 = \left\langle A \right\rangle _0  + \frac{{\left\langle A \right\rangle _0 \left\langle {S_{{\rm{HS}}} [\psi ,\bar \psi ;\{ \Delta ^* \} ,\{ \Delta \} ]} \right\rangle _0  - \left\langle {AS_{{\rm{HS}}} [\psi ,\bar \psi ;\{ \Delta ^* \} ,\{ \Delta \} ]} \right\rangle _0 }}{{1 - \left\langle {S_{{\rm{HS}}} [\psi ,\bar \psi ;\{ \Delta ^* \} ,\{ \Delta \} ]} \right\rangle _0 }}.
\end{split}
\end{equation}
Here we introduced a notation $\left\langle A \right\rangle _0  \equiv \frac{{\int {D\bar \psi D \psi } \exp \left( {-S_0 } \right)A}}{{\int {D\bar \psi D \psi } \exp \left( {-S_0 } \right)}}$. Since $S_{{\rm{HS}}}$ has the order of magnitude of $\{ \Delta ^* \}$ and $\{ \Delta \}$, the above equation becomes
\begin{equation}\label{eqA21}
\begin{split}
\left\langle A \right\rangle _\Delta   = \left\langle A \right\rangle _0  + \left\langle A \right\rangle _0 \left\langle {S_{{\rm{HS}}} [\psi ,\bar \psi ;\{ \Delta ^* \} ,\{ \Delta \} ]} \right\rangle _0  - \left\langle {AS_{{\rm{HS}}} [\psi ,\bar \psi ;\{ \Delta ^* \} ,\{ \Delta \} ]} \right\rangle _0  + O\left( {\left\{ {\left| \Delta  \right|^2 } \right\}} \right).
\end{split}
\end{equation}

Let us consider the mean value $\left\langle {O_\alpha ^{{\rm{sSC}}} ({\bf{Q}}_i^{\rm{P}} )} \right\rangle _\Delta$. Because of $\left\langle {O_\alpha ^{{\rm{sSC}}} ({\bf{Q}}_i^{\rm{P}} )} \right\rangle _0  = 0$, we have
\begin{equation}\label{eqA22}
\begin{split}
\left\langle {O_\alpha ^{{\rm{sSC}}} ({\bf{Q}}_i^{\rm{P}} )} \right\rangle _\Delta   \approx &  - \left\langle {O_\alpha ^{{\rm{sSC}}} ({\bf{Q}}_i^{\rm{P}} )S_{{\rm{HS}}} [\psi ,\bar \psi ;\{ \Delta ^* \} ,\{ \Delta \} ]} \right\rangle _0 
 =  - \left\langle {O_\alpha ^{{\rm{sSC}}} ({\bf{Q}}_i^{\rm{P}} )} \right.\left( {S_{{\rm{HS}}}^{{\rm{sSC}}} [\psi ,\bar \psi ;\Delta _{{\rm{sSC}}}^* ,\Delta _{{\rm{sSC}}} ]} \right. \\
&+ S_{{\rm{HS}}}^{{\rm{tSC}}} [\psi ,\bar \psi ;\vec \Delta _{{\rm{tSC}}}^* ,\vec \Delta _{{\rm{tSC}}} ]
+ S_{{\rm{HS}}}^{{\rm{SPN}}} [\psi ,\bar \psi ;\vec \Delta _{{\rm{SPN}}}^* ,\vec \Delta _{{\rm{SPN}}} ] + \left. {\left. {S_{{\rm{HS}}}^{{\rm{CHG}}} [\psi ,\bar \psi ;\Delta _{{\rm{CHG}}}^* ,\Delta _{{\rm{CHG}}} ]} \right)} \right\rangle _0 .
\end{split}
\end{equation}
Inserting Eq. (\ref{eq074})  \footnote{Note that only the $\omega  = 0$ components are involved in the summations of Eqs. (\ref{eq074}) to (\ref{eq077}) [see the discussion below Eq. (\ref{eq077})].} into the above equation, one can rewrite the first term in it as follows:
\begin{equation}\label{eqA23}
\begin{split}
&\left\langle {O_\alpha ^{{\rm{sSC}}} ({\bf{Q}}_i^{\rm{P}} )S_{{\rm{HS}}}^{{\rm{sSC}}} [\psi ,\bar \psi ;\Delta _{{\rm{sSC}}}^* ,\Delta _{{\rm{sSC}}} ]} \right\rangle _0 \\
& \hspace{1pc}  = - \frac{1}{\hbar }\sum\limits_{j = 1}^{N_{\rm{P}} } {\sum\limits_{\beta  = 1}^{M_{{\rm{sSC}},j} } {\left\{ {\left[ {\Delta _\beta ^{{\rm{sSC}}} ({\bf{Q}}_j^{\rm{P}} )} \right]^* } \right.} } \left\langle {O_\alpha ^{{\rm{sSC}}} ({\bf{Q}}_i^{\rm{P}} )O_\beta ^{{\rm{sSC}}} ({\bf{Q}}_j^{\rm{P}} )} \right\rangle _0  + \left. {\Delta _\beta ^{{\rm{sSC}}} ({\bf{Q}}_j^{\rm{P}} )\left\langle {O_\alpha ^{{\rm{sSC}}} ({\bf{Q}}_i^{\rm{P}} )\left[ {O_\beta ^{{\rm{sSC}}} ({\bf{Q}}_j^{\rm{P}} )} \right]^* } \right\rangle _0 } \right\}\\
& \hspace{1pc}  =  - \frac{1}{\hbar }\sum\limits_{\beta  = 1}^{M_{{\rm{sSC}},i} } {\Delta _\beta ^{{\rm{sSC}}} ({\bf{Q}}_i^{\rm{P}} )} \left\langle {O_\alpha ^{{\rm{sSC}}} ({\bf{Q}}_i^{\rm{P}} )\left[ {O_\beta ^{{\rm{sSC}}} ({\bf{Q}}_i^{\rm{P}} )} \right]^* } \right\rangle _0.
\end{split}
\end{equation}
Here we used the relations of
\begin{equation}\nonumber
\left\langle {O_\alpha ^{{\rm{sSC}}} ({\bf{Q}}_i^{\rm{P}} )O_\beta ^{{\rm{sSC}}} ({\bf{Q}}_j^{\rm{P}} )} \right\rangle _0  = 0,\left\langle {O_\alpha ^{{\rm{sSC}}} ({\bf{Q}}_i^{\rm{P}} )\left[ {O_\beta ^{{\rm{sSC}}} ({\bf{Q}}_j^{\rm{P}} )} \right]^* } \right\rangle _0  = \delta _{ij} \left\langle {O_\alpha ^{{\rm{sSC}}} ({\bf{Q}}_i^{\rm{P}} )\left[ {O_\beta ^{{\rm{sSC}}} ({\bf{Q}}_i^{\rm{P}} )} \right]^* } \right\rangle _0 .
\end{equation}
Inserting the first equation in Eq. (\ref{eq069}) into Eq. (\ref{eqA23}), we have
\begin{equation}\label{eqA24}
\begin{split}
&\left\langle {O_\alpha ^{{\rm{sSC}}} ({\bf{Q}}_i^{\rm{P}} )S_{{\rm{HS}}}^{{\rm{sSC}}} [\psi ,\bar \psi ;\Delta _{{\rm{sSC}}}^* ,\Delta _{{\rm{sSC}}} ]} \right\rangle _0  =  - \frac{1}{\hbar }\sum\limits_{\beta  = 1}^{M_{{\rm{sSC}},i} } {\Delta _\beta ^{{\rm{sSC}}} ({\bf{Q}}_i^{\rm{P}} )} \left\langle {O_\alpha ^{{\rm{sSC}}} ({\bf{Q}}_i^{\rm{P}} )\left[ {O_\beta ^{{\rm{sSC}}} ({\bf{Q}}_i^{\rm{P}} )} \right]^* } \right\rangle _0 \\
&=  - \frac{1}{\hbar }\sum\limits_{\beta  = 1}^{M_{{\rm{sSC}},i} } {\Delta _\beta ^{{\rm{sSC}}} ({\bf{Q}}_i^{\rm{P}} )} \sum\limits_{{\bf{p}},\omega _p } {\sum\limits_{o,o',m} {[\varphi _{oo'm}^{{\rm{sSC}},\alpha } ({\bf{Q}}_i^{\rm{P}} )]^* e^{i{\bf{R}}_m  \cdot {\bf{p}}} } } \sum\limits_{{\bf{k}},\omega _k } {\sum\limits_{\tilde o,\tilde o',n} {\varphi _{\tilde o\tilde o'n}^{{\rm{sSC}},\beta } ({\bf{Q}}_i^{\rm{P}} )e^{ - i{\bf{R}}_n  \cdot {\bf{k}}} } } \\
& \hspace{1pc} \times \sum\limits_{\sigma ,\sigma '} {\sigma \sigma '} \left\langle {\psi _{ - \sigma } ( - {\bf{p}}, - \omega _p ,o')\psi _\sigma  ({\bf{p}} + {\bf{Q}}_i^{\rm{P}} ,\omega _p ,o)\bar \psi _{\sigma '} ({\bf{k}} + {\bf{Q}}_i^{\rm{P}} ,\omega _k ,\tilde o)\bar \psi _{ - \sigma '} ( - {\bf{k}}, - \omega _k ,\tilde o')} \right\rangle _0 .
\end{split}
\end{equation}
We can use Wick's theorem \cite{ref53} to evaluate the mean value in the above equation
\begin{equation}\nonumber
\begin{split}
&\sum\limits_{\sigma ,\sigma '} {\sigma \sigma '} \left\langle {\psi _{ - \sigma } ( - {\bf{p}}, - \omega _p ,o')\psi _\sigma  ({\bf{p}} + {\bf{Q}}_i^{\rm{P}} ,\omega _p ,o)\bar \psi _{\sigma '} ({\bf{k}} + {\bf{Q}}_i^{\rm{P}} ,\omega _k ,\tilde o)\bar \psi _{ - \sigma '} ( - {\bf{k}}, - \omega _k ,\tilde o')} \right\rangle _0 \\
& = \sum\limits_{\sigma ,\sigma '} {\sigma \sigma '} \left\langle {\psi _{ - \sigma } ( - {\bf{p}}, - \omega _p ,o')\bar \psi _{ - \sigma '} ( - {\bf{k}}, - \omega _k ,\tilde o')} \right\rangle _0 \left\langle {\psi _\sigma  ({\bf{p}} + {\bf{Q}}_i^{\rm{P}} ,\omega _p ,o)\bar \psi _{\sigma '} ({\bf{k}} + {\bf{Q}}_i^{\rm{P}} ,\omega _k ,\tilde o)} \right\rangle _0 \\
& + \sum\limits_{\sigma ,\sigma '} {\sigma \sigma '} ( - 1)\left\langle {\psi _{ - \sigma } ( - {\bf{p}}, - \omega _p ,o')\bar \psi _{\sigma '} ({\bf{k}} + {\bf{Q}}_i^{\rm{P}} ,\omega _k ,\tilde o)} \right\rangle _0 \left\langle {\psi _\sigma  ({\bf{p}} + {\bf{Q}}_i^{\rm{P}} ,\omega _p ,o)\bar \psi _{ - \sigma '} ( - {\bf{k}}, - \omega _k ,\tilde o')} \right\rangle _0 \\
&= 2\delta _{{\bf{p}},{\bf{k}}} \delta _{\omega _p ,\omega _k } G_{o'\tilde o'}^0 ( - {\bf{p}}, - \omega _p )G_{o\tilde o}^0 ({\bf{p}} + {\bf{Q}}_i^{\rm{P}} ,\omega _p )
 + 2\delta _{{\bf{p}}, - {\bf{k}} - {\bf{Q}}_i^{\rm{P}} } \delta _{\omega _p , - \omega _k } G_{o'\tilde o}^0 ( - {\bf{p}}, - \omega _p )G_{o\tilde o'}^0 ({\bf{p}} + {\bf{Q}}_i^{\rm{P}} ,\omega _p ).
\end{split}
\end{equation}
Inserting this result into Eq. (\ref{eqA24}), we obtain
\begin{equation}\nonumber
\begin{split}
&\left\langle {O_\alpha ^{{\rm{sSC}}} ({\bf{Q}}_i^{\rm{P}} )S_{{\rm{HS}}}^{{\rm{sSC}}} [\psi ,\bar \psi ;\Delta _{{\rm{sSC}}}^* ,\Delta _{{\rm{sSC}}} ]} \right\rangle _0  =  - \frac{1}{\hbar }\sum\limits_{\beta  = 1}^{M_{{\rm{sSC}},i} } {\Delta _\beta ^{{\rm{sSC}}} ({\bf{Q}}_i^{\rm{P}} )} \sum\limits_{o,o',m} {[\varphi _{oo'm}^{{\rm{sSC}},\alpha } ({\bf{Q}}_i^{\rm{P}} )]^* } \sum\limits_{\tilde o,\tilde o',n} {\varphi _{\tilde o\tilde o'n}^{{\rm{sSC}},\beta } ({\bf{Q}}_i^{\rm{P}} )} \\
& \times 2\sum\limits_{{\bf{p}},\omega _p } {e^{i{\bf{R}}_m  \cdot {\bf{p}}} } \left[ {e^{ - i{\bf{R}}_n  \cdot {\bf{p}}} } \right.G_{o'\tilde o'}^0 ( - {\bf{p}}, - \omega _p )G_{o\tilde o}^0 ({\bf{p}} + {\bf{Q}}_i^{\rm{P}} ,\omega _p )
 + \left. {e^{ - i{\bf{R}}_n  \cdot ( - {\bf{p}} - {\bf{Q}}_i^{\rm{P}} )} G_{o'\tilde o}^0 ( - {\bf{p}}, - \omega _p )G_{o\tilde o'}^0 ({\bf{p}} + {\bf{Q}}_i^{\rm{P}} ,\omega _p )} \right] \\
& =  - \frac{2}{\hbar }\sum\limits_{\beta  = 1}^{M_{{\rm{sSC}},i} } {\Delta _\beta ^{{\rm{sSC}}} ({\bf{Q}}_i^{\rm{P}} )} \sum\limits_{o,o',m} {[\varphi _{oo'm}^{{\rm{sSC}},\alpha } ({\bf{Q}}_i^{\rm{P}} )]^* } \sum\limits_{{\bf{p}},\omega _p } {e^{i{\bf{R}}_m  \cdot {\bf{p}}} } \\
& \times \sum\limits_{\tilde o,\tilde o',n} {\left( {\varphi _{\tilde o\tilde o'n}^{{\rm{sSC}},\beta } ({\bf{Q}}_i^{\rm{P}} ) + \varphi _{\tilde o'\tilde o\bar n}^{{\rm{sSC}},\beta } ({\bf{Q}}_i^{\rm{P}} )e^{ - i{\bf{R}}_n  \cdot {\bf{Q}}_i^{\rm{P}} } } \right)} e^{ - i{\bf{R}}_n  \cdot {\bf{p}}} G_{o'\tilde o'}^0 ( - {\bf{p}}, - \omega _p )G_{o\tilde o}^0 ({\bf{p}} + {\bf{Q}}_i^{\rm{P}} ,\omega _p ).
\end{split}
\end{equation}
Due to the relation (\ref{eqA06}), the above equation becomes
\begin{equation}\nonumber
\begin{split}
&\left\langle {O_\alpha ^{{\rm{sSC}}} ({\bf{Q}}_i^{\rm{P}} )S_{{\rm{HS}}}^{{\rm{sSC}}} [\psi ,\bar \psi ;\Delta _{{\rm{sSC}}}^* ,\Delta _{{\rm{sSC}}} ]} \right\rangle _0  =  - \frac{4}{\hbar }\sum\limits_{\beta  = 1}^{M_{{\rm{sSC}},i} } {\Delta _\beta ^{{\rm{sSC}}} ({\bf{Q}}_i^{\rm{P}} )} \sum\limits_{o,o',m} {[\varphi _{oo'm}^{{\rm{sSC}},\alpha } ({\bf{Q}}_i^{\rm{P}} )]^* } \sum\limits_{\tilde o,\tilde o',n} {\varphi _{\tilde o\tilde o'n}^{{\rm{sSC}},\beta } ({\bf{Q}}_i^{\rm{P}} )} \\
& \times \sum\limits_{\bf{p}} {e^{i{\bf{R}}_m  \cdot {\bf{p}}} } e^{ - i{\bf{R}}_n  \cdot {\bf{p}}} \sum\limits_{\omega _p } {G_{o\tilde o}^0 ({\bf{p}} + {\bf{Q}}_i^{\rm{P}} ,\omega _p )G_{o'\tilde o'}^0 ( - {\bf{p}}, - \omega _p )} \\
& = 4N\beta \hbar \sum\limits_{\gamma  = 1}^{M_{{\rm{sSC}},i} } {\Delta _\gamma ^{{\rm{sSC}}} ({\bf{Q}}_i^{\rm{P}} )} \sum\limits_{o,o',m} {[\varphi _{oo'm}^{{\rm{sSC}},\alpha } ({\bf{Q}}_i^{\rm{P}} )]^* } \sum\limits_{\tilde o,\tilde o',n} {\varphi _{\tilde o\tilde o'n}^{{\rm{sSC}},\gamma } ({\bf{Q}}_i^{\rm{P}} )} \chi _{oo'm,\tilde o\tilde o'n}^{{\rm{pp(}}\Omega  = 0{\rm{)}}} ({\bf{Q}}_i^{\rm{P}}).
\end{split}
\end{equation}
Taking into account the definition (\ref{eqA11}), we have the final result
\begin{equation}\label{eqA25}
\begin{split}
\left\langle {O_\alpha ^{{\rm{sSC}}} ({\bf{Q}}_i^{\rm{P}} )S_{{\rm{HS}}}^{{\rm{sSC}}} [\psi ,\bar \psi ;\Delta _{{\rm{sSC}}}^* ,\Delta _{{\rm{sSC}}} ]} \right\rangle _0  = 4N\beta \hbar \sum\limits_{\gamma  = 1}^{M_{{\rm{sSC}},i} } {X_{\alpha \gamma }^{{\rm{pp,sSC}}} ({\bf{Q}}_i^{\rm{P}} )\Delta _\gamma ^{{\rm{sSC}}} ({\bf{Q}}_i^{\rm{P}} )} .
\end{split}
\end{equation}
In a similar way, one can verify the following relations:
\begin{equation}\label{eqA26}
\begin{split}
&\left\langle {O_\alpha ^{{\rm{sSC}}} ({\bf{Q}}_i^{\rm{P}} )S_{{\rm{HS}}}^{{\rm{tSC}}} [\psi ,\bar \psi ;\vec \Delta _{{\rm{tSC}}}^* ,\vec \Delta _{{\rm{tSC}}} ]} \right\rangle _0  = \left\langle {O_\alpha ^{{\rm{sSC}}} ({\bf{Q}}_i^{\rm{P}} )S_{{\rm{HS}}}^{{\rm{SPN}}} [\psi ,\bar \psi ;\vec \Delta _{{\rm{SPN}}}^* ,\vec \Delta _{{\rm{SPN}}} ]} \right\rangle _0 \\
& = \left\langle {O_\alpha ^{{\rm{sSC}}} ({\bf{Q}}_i^{\rm{P}} )S_{{\rm{HS}}}^{{\rm{CHG}}} [\psi ,\bar \psi ;\Delta _{{\rm{CHG}}}^* ,\Delta _{{\rm{CHG}}} ]} \right\rangle _0  = 0.
\end{split}
\end{equation}
Inserting Eqs. (\ref{eqA25}) and (\ref{eqA26}) into Eq. (\ref{eqA22}), we get
\begin{equation}\label{eqA27}
\begin{split}
\left\langle {O_\alpha ^{{\rm{sSC}}} ({\bf{Q}}_i^{\rm{P}} )} \right\rangle _\Delta   =  - 4N\beta \hbar \sum\limits_{\gamma  = 1}^{M_{{\rm{sSC}},i} } {X_{\alpha \gamma }^{{\rm{pp,sSC}}} ({\bf{Q}}_i^{\rm{P}} )\Delta _\gamma ^{{\rm{sSC}}} ({\bf{Q}}_i^{\rm{P}} )} .
\end{split}
\end{equation}
Finally, the first equation in Eq. (\ref{eqA19}) becomes
\begin{equation}\label{eqA28}
\begin{split}
\Delta _\alpha ^{{\rm{sSC}}} ({\bf{Q}}_i^{\rm{P}} ) =  - \Lambda ^{{\rm{sSC}},\alpha } ({\bf{Q}}_i^{\rm{P}} )\sum\limits_{\beta  = 1}^{M_{{\rm{sSC}},i} } {X_{\alpha \beta }^{{\rm{pp,sSC}}} ({\bf{Q}}_i^{\rm{P}} )\Delta _\beta ^{{\rm{sSC}}} ({\bf{Q}}_i^{\rm{P}} )} ,
\end{split}
\end{equation}
or equivalently, with $Y_\alpha   \equiv \frac{{\Delta _\alpha ^{{\rm{sSC}}} ({\bf{Q}}_i^{\rm{P}} )}}{{\sqrt {\Lambda ^{{\rm{sSC}},\alpha } ({\bf{Q}}_i^{\rm{P}} )} }}$,
\begin{equation}\label{eqA29}
\begin{split}
&\sum\limits_{\beta  = 1}^{M_{{\rm{sSC}},i} } {\left\{ {\sqrt {\Lambda ^{{\rm{sSC}},\alpha } ({\bf{Q}}_i^{\rm{P}} )} \left[ { - X_{\alpha \beta }^{{\rm{pp,sSC}}} ({\bf{Q}}_i^{\rm{P}} )} \right]\sqrt {\Lambda ^{{\rm{sSC}},\beta } ({\bf{Q}}_i^{\rm{P}} )} } \right\}} Y_\beta   = Y_\alpha ,\\
&\textrm{(Critical condition of the MF theory in the spin-singlet pairing channel)}.
\end{split}
\end{equation}
This is exactly the same as the critical condition (\ref{eqA17}).

We consider the mean value $\left\langle {\vec O_\alpha ^{{\rm{tSC}}} ({\bf{Q}}_i^{\rm{P}} )} \right\rangle _\Delta$. It is clear that $\left\langle {\vec O_\alpha ^{{\rm{tSC}}} ({\bf{Q}}_i^{\rm{P}} )} \right\rangle _0  = 0$. So we have
\begin{equation}\label{eqA30}
\begin{split}
\left\langle {\vec O_\alpha ^{{\rm{tSC}}} ({\bf{Q}}_i^{\rm{P}} )} \right\rangle _\Delta   =  - \left\langle {\vec O_\alpha ^{{\rm{tSC}}} ({\bf{Q}}_i^{\rm{P}} )} \right. &\left( {S_{{\rm{HS}}}^{{\rm{sSC}}} [\psi ,\bar \psi ;\Delta _{{\rm{sSC}}}^* ,\Delta _{{\rm{sSC}}} ]} \right. + S_{{\rm{HS}}}^{{\rm{tSC}}} [\psi ,\bar \psi ;\vec \Delta _{{\rm{tSC}}}^* ,\vec \Delta _{{\rm{tSC}}} ]\\
& + S_{{\rm{HS}}}^{{\rm{SPN}}} [\psi ,\bar \psi ;\vec \Delta _{{\rm{SPN}}}^* ,\vec \Delta _{{\rm{SPN}}} ] + \left. {\left. {S_{{\rm{HS}}}^{{\rm{CHG}}} [\psi ,\bar \psi ;\Delta _{{\rm{CHG}}}^* ,\Delta _{{\rm{CHG}}} ]} \right)} \right\rangle _0.
\end{split}
\end{equation}
Through the procedure similar to the one described above, we obtain the following results:
\begin{equation}\label{eqA31}
\begin{split}
\left\langle {\vec O_\alpha ^{{\rm{tSC}}} ({\bf{Q}}_i^{\rm{P}} )S_{{\rm{HS}}}^{{\rm{tSC}}} [\psi ,\bar \psi ;\vec \Delta _{{\rm{tSC}}}^* ,\vec \Delta _{{\rm{tSC}}} ]} \right\rangle _0  = 4N\beta \hbar \sum\limits_{\gamma  = 1}^{M_{{\rm{tSC}},i} } {X_{\alpha \gamma }^{{\rm{pp,tSC}}} ({\bf{Q}}_i^{\rm{P}} )\vec \Delta _\gamma ^{{\rm{tSC}}} ({\bf{Q}}_i^{\rm{P}} )} ,
\end{split}
\end{equation}
\begin{equation}\label{eqA32}
\begin{split}
&\left\langle {\vec O_\alpha ^{{\rm{tSC}}} ({\bf{Q}}_i^{\rm{P}} )S_{{\rm{HS}}}^{{\rm{sSC}}} [\psi ,\bar \psi ;\Delta _{{\rm{sSC}}}^* ,\Delta _{{\rm{sSC}}} ]} \right\rangle _0  = \left\langle {\vec O_\alpha ^{{\rm{tSC}}} ({\bf{Q}}_i^{\rm{P}} )S_{{\rm{HS}}}^{{\rm{SPN}}} [\psi ,\bar \psi ;\vec \Delta _{{\rm{SPN}}}^* ,\vec \Delta _{{\rm{SPN}}} ]} \right\rangle _0 \\
& = \left\langle {\vec O_\alpha ^{{\rm{tSC}}} ({\bf{Q}}_i^{\rm{P}} )S_{{\rm{HS}}}^{{\rm{CHG}}} [\psi ,\bar \psi ;\Delta _{{\rm{CHG}}}^* ,\Delta _{{\rm{CHG}}} ]} \right\rangle _0  = 0.
\end{split}
\end{equation}
Inserting Eqs. (\ref{eqA31}) and (\ref{eqA32}) into Eq. (\ref{eqA30}), we get
\begin{equation}\label{eqA33}
\begin{split}
\left\langle {\vec O_\alpha ^{{\rm{tSC}}} ({\bf{Q}}_i^{\rm{P}} )} \right\rangle _\Delta   =  - 4N\beta \hbar \sum\limits_{\gamma  = 1}^{M_{{\rm{tSC}},i} } {X_{\alpha \gamma }^{{\rm{pp,tSC}}} ({\bf{Q}}_i^{\rm{P}} )\vec \Delta _\gamma ^{{\rm{tSC}}} ({\bf{Q}}_i^{\rm{P}} )} .
\end{split}
\end{equation}
Then, the second equation in Eq. (\ref{eqA19}) becomes
\begin{equation}\label{eqA34}
\begin{split}
\vec \Delta _\alpha ^{{\rm{tSC}}} ({\bf{Q}}_i^{\rm{P}} ) =  - \Lambda ^{{\rm{tSC}},\alpha } ({\bf{Q}}_i^{\rm{P}} )\sum\limits_{\beta  = 1}^{M_{{\rm{tSC}},i} } {X_{\alpha \beta }^{{\rm{pp,tSC}}} ({\bf{Q}}_i^{\rm{P}} )\vec \Delta _\beta ^{{\rm{tSC}}} ({\bf{Q}}_i^{\rm{P}} )} ,
\end{split}
\end{equation}
or equivalently, with $\vec Y_\alpha   \equiv \frac{{\vec \Delta _\alpha ^{{\rm{tSC}}} ({\bf{Q}}_i^{\rm{P}} )}}{{\sqrt {\Lambda ^{{\rm{tSC}},\alpha } ({\bf{Q}}_i^{\rm{P}} )} }}$,
\begin{equation}\label{eqA35}
\begin{split}
&\sum\limits_{\beta  = 1}^{M_{{\rm{tSC}},i} } {\left\{ {\sqrt {\Lambda ^{{\rm{tSC}},\alpha } ({\bf{Q}}_i^{\rm{P}} )} \left[ { - X_{\alpha \beta }^{{\rm{pp,tSC}}} ({\bf{Q}}_i^{\rm{P}} )} \right]\sqrt {\Lambda ^{{\rm{tSC}},\beta } ({\bf{Q}}_i^{\rm{P}} )} } \right\}} \vec Y_\beta   = \vec Y_\alpha ,\\
&\textrm{(Critical condition of the MF theory in the spin-triplet pairing channel)}.
\end{split}
\end{equation}
This is exactly the same as the critical condition (\ref{eqA18}).

Now let us consider the mean value $\left\langle {\vec O_\alpha ^{{\rm{SPN}}} ({\bf{Q}}_i^{\rm{C}} )} \right\rangle _\Delta $. Evidently, $\left\langle {\vec O_\alpha ^{{\rm{SPN}}} ({\bf{Q}}_i^{\rm{C}} )} \right\rangle _0  = 0$, and we get
\begin{equation}\label{eqA36}
\begin{split}
\left\langle {\vec O_\alpha ^{{\rm{SPN}}} ({\bf{Q}}_i^{\rm{C}} )} \right\rangle _\Delta   =  - \left\langle {\vec O_\alpha ^{{\rm{SPN}}} ({\bf{Q}}_i^{\rm{C}} )} \right. &\left( {S_{{\rm{HS}}}^{{\rm{sSC}}} [\psi ,\bar \psi ;\Delta _{{\rm{sSC}}}^* ,\Delta _{{\rm{sSC}}} ]} \right. + S_{{\rm{HS}}}^{{\rm{tSC}}} [\psi ,\bar \psi ;\vec \Delta _{{\rm{tSC}}}^* ,\vec \Delta _{{\rm{tSC}}} ]\\
& + S_{{\rm{HS}}}^{{\rm{SPN}}} [\psi ,\bar \psi ;\vec \Delta _{{\rm{SPN}}}^* ,\vec \Delta _{{\rm{SPN}}} ] + \left. {\left. {S_{{\rm{HS}}}^{{\rm{CHG}}} [\psi ,\bar \psi ;\Delta _{{\rm{CHG}}}^* ,\Delta _{{\rm{CHG}}} ]} \right)} \right\rangle _0 .
\end{split}
\end{equation}
Inserting Eq. (\ref{eq076}) into the above equation, one can rewrite the third term in it as follows:
\begin{equation}\nonumber
\begin{split}
&\left\langle {\vec O_\alpha ^{{\rm{SPN}}} ({\bf{Q}}_i^{\rm{C}} )S_{{\rm{HS}}}^{{\rm{SPN}}} [\psi ,\bar \psi ;\vec \Delta _{{\rm{SPN}}}^* ,\vec \Delta _{{\rm{SPN}}} ]} \right\rangle _0  \\
&\hspace{2pc} =  - \frac{1}{\hbar }\sum\limits_{j = 1}^{N_{\rm{C}} } {\sum\limits_{\beta  = 1}^{M_{{\rm{C}},j} } {\left\langle {\vec O_\alpha ^{{\rm{SPN}}} ({\bf{Q}}_i^{\rm{C}} )\left\{ {\left[ {\vec \Delta _\beta ^{{\rm{SPN}}} ({\bf{Q}}_j^{\rm{C}} )} \right]^*  \cdot \vec O_\beta ^{{\rm{SPN}}} ({\bf{Q}}_j^{\rm{C}} )} \right.} \right.} } 
 + \left. {\left. {\vec \Delta _\beta ^{{\rm{SPN}}} ({\bf{Q}}_j^{\rm{C}} ) \cdot \left[ {\vec O_\beta ^{{\rm{SPN}}} ({\bf{Q}}_j^{\rm{C}} )} \right]^* } \right\}} \right\rangle _0 .
\end{split}
\end{equation}
Concretely, its $x$-component reads
\begin{equation}\nonumber
\begin{split}
&\left\langle {O_{\alpha ,x}^{{\rm{SPN}}} ({\bf{Q}}_i^{\rm{C}} )S_{{\rm{HS}}}^{{\rm{SPN}}} [\psi ,\bar \psi ;\vec \Delta _{{\rm{SPN}}}^* ,\vec \Delta _{{\rm{SPN}}} ]} \right\rangle _0  \\
& = - \frac{1}{\hbar }\sum\limits_{j = 1}^{N_{\rm{C}} } {\sum\limits_{\beta  = 1}^{M_{{\rm{C}},j} } {\left\langle {O_{\alpha ,x}^{{\rm{SPN}}} ({\bf{Q}}_i^{\rm{C}} )\left\{ {\left[ {\vec \Delta _\beta ^{{\rm{SPN}}} ({\bf{Q}}_j^{\rm{C}} )} \right]^*  \cdot \vec O_\beta ^{{\rm{SPN}}} ({\bf{Q}}_j^{\rm{C}} )} \right.} \right.} }
+ \left. {\left. {\vec \Delta _\beta ^{{\rm{SPN}}} ({\bf{Q}}_j^{\rm{C}} ) \cdot \left[ {\vec O_\beta ^{{\rm{SPN}}} ({\bf{Q}}_j^{\rm{C}} )} \right]^* } \right\}} \right\rangle _0 \\
&=  - \frac{1}{\hbar }\sum\limits_{j = 1}^{N_{\rm{C}} } {\sum\limits_{\beta  = 1}^{M_{{\rm{C}},j} } {\left\{ {\left[ {\vec \Delta _\beta ^{{\rm{SPN}}} ({\bf{Q}}_j^{\rm{C}} )} \right]^*  \cdot } \right.} } \left\langle {O_{\alpha ,x}^{{\rm{SPN}}} ({\bf{Q}}_i^{\rm{C}} )\vec O_\beta ^{{\rm{SPN}}} ({\bf{Q}}_j^{\rm{C}} )} \right\rangle _0  + \left. {\vec \Delta _\beta ^{{\rm{SPN}}} ({\bf{Q}}_j^{\rm{C}} ) \cdot \left\langle {O_{\alpha ,x}^{{\rm{SPN}}} ({\bf{Q}}_i^{\rm{C}} )\left[ {\vec O_\beta ^{{\rm{SPN}}} ({\bf{Q}}_j^{\rm{C}} )} \right]^* } \right\rangle _0 } \right\}\\
&=  - \frac{1}{\hbar }\sum\limits_{\beta  = 1}^{M_{{\rm{C}},i} } {\left\{ {\left[ {\vec \Delta _\beta ^{{\rm{SPN}}} ( - {\bf{Q}}_i^{\rm{C}} )} \right]^* } \right.}  \cdot \left\langle {O_{\alpha ,x}^{{\rm{SPN}}} ({\bf{Q}}_i^{\rm{C}} )\vec O_\beta ^{{\rm{SPN}}} ( - {\bf{Q}}_i^{\rm{C}} )} \right\rangle _0  + \vec \Delta _\beta ^{{\rm{SPN}}} ({\bf{Q}}_i^{\rm{C}} ) \cdot \left. {\left\langle {O_{\alpha ,x}^{{\rm{SPN}}} ({\bf{Q}}_i^{\rm{C}} )\left[ {\vec O_\beta ^{{\rm{SPN}}} ({\bf{Q}}_i^{\rm{C}} )} \right]^* } \right\rangle _0 } \right\}.
\end{split}
\end{equation}
By using the relations
\begin{equation}\nonumber
\begin{split}
&\left\langle {O_{\alpha ,x}^{{\rm{SPN}}} ({\bf{Q}}_i^{\rm{C}} )\vec O_\beta ^{{\rm{SPN}}} ({\bf{Q}}_j^{\rm{C}} )} \right\rangle _0  = \delta _{{\bf{Q}}_i^{\rm{C}} , - {\bf{Q}}_j^{\rm{C}} } \left\langle {O_{\alpha ,x}^{{\rm{SPN}}} ({\bf{Q}}_i^{\rm{C}} )\vec O_\beta ^{{\rm{SPN}}} ( - {\bf{Q}}_i^{\rm{C}} )} \right\rangle _0 ,\\
&\left\langle {O_{\alpha ,x}^{{\rm{SPN}}} ({\bf{Q}}_i^{\rm{C}} )\left[ {\vec O_\beta ^{{\rm{SPN}}} ({\bf{Q}}_j^{\rm{C}} )} \right]^* } \right\rangle _0  = \delta _{ij} \left\langle {O_{\alpha ,x}^{{\rm{SPN}}} ({\bf{Q}}_i^{\rm{C}} )\left[ {\vec O_\beta ^{{\rm{SPN}}} ({\bf{Q}}_i^{\rm{C}} )} \right]^* } \right\rangle _0 ,\\
&\left[ {\vec \Delta _\beta ^{{\rm{SPN}}} ( - {\bf{Q}}_i^{\rm{C}} )} \right]^*  = \vec \Delta _\beta ^{{\rm{SPN}}} ({\bf{Q}}_i^{\rm{C}} ),\vec O_\beta ^{{\rm{SPN}}} ( - {\bf{Q}}_i^{\rm{C}} ) = \left[ {\vec O_\beta ^{{\rm{SPN}}} ({\bf{Q}}_i^{\rm{C}} )} \right]^* ,
\end{split}
\end{equation}
we obtain
\begin{equation}\label{eqA37}
\begin{split}
\left\langle {O_{\alpha ,x}^{{\rm{SPN}}} ({\bf{Q}}_i^{\rm{C}} )S_{{\rm{HS}}}^{{\rm{SPN}}} [\psi ,\bar \psi ;\vec \Delta _{{\rm{SPN}}}^* ,\vec \Delta _{{\rm{SPN}}} ]} \right\rangle _0  =  - \frac{2}{\hbar }\sum\limits_{\beta  = 1}^{M_{{\rm{C}},i} } {\vec \Delta _\beta ^{{\rm{SPN}}} ({\bf{Q}}_i^{\rm{C}} )}  \cdot \left\langle {O_{\alpha ,x}^{{\rm{SPN}}} ({\bf{Q}}_i^{\rm{C}} )\left[ {\vec O_\beta ^{{\rm{SPN}}} ({\bf{Q}}_i^{\rm{C}} )} \right]^* } \right\rangle _0 .
\end{split}
\end{equation}
Inserting the third equation in Eq. (\ref{eq069}) into Eq. (\ref{eqA37}), we have
\begin{equation}\label{eqA38}
\begin{split}
&\left\langle {O_{\alpha ,x}^{{\rm{SPN}}} ({\bf{Q}}_i^{\rm{C}} )S_{{\rm{HS}}}^{{\rm{SPN}}} [\psi ,\bar \psi ;\vec \Delta _{{\rm{SPN}}}^* ,\vec \Delta _{{\rm{SPN}}} ]} \right\rangle _0  =  - \frac{2}{\hbar }\sum\limits_{\beta  = 1}^{M_{{\rm{C}},i} } {\vec \Delta _\beta ^{{\rm{SPN}}} ({\bf{Q}}_i^{\rm{C}} )}  \cdot \left\langle {O_{\alpha ,x}^{{\rm{SPN}}} ({\bf{Q}}_i^{\rm{C}} )\left[ {\vec O_\beta ^{{\rm{SPN}}} ({\bf{Q}}_i^{\rm{C}} )} \right]^* } \right\rangle _0 \\
& =  - \frac{2}{\hbar }\sum\limits_{\beta  = 1}^{M_{{\rm{C}},i} } {\sum\limits_{\gamma  = x,y,z} {\Delta _{\beta ,\gamma }^{{\rm{SPN}}} ({\bf{Q}}_i^{\rm{C}} )} } \sum\limits_{{\bf{p}},\omega _p } {\sum\limits_{o,o',m} {[\varphi _{oo'm}^{{\rm{C}},\alpha } ({\bf{Q}}_i^{\rm{C}} )]^* e^{i{\bf{R}}_m  \cdot {\bf{p}}} } } \sum\limits_{{\bf{k}},\omega _k } {\sum\limits_{\tilde o,\tilde o',n} {\varphi _{\tilde o\tilde o'n}^{{\rm{C}},\beta } ({\bf{Q}}_i^{\rm{C}} )e^{ - i{\bf{R}}_n  \cdot {\bf{k}}} } } \\
& \times \sum\limits_{\sigma ,\sigma '} {\sigma _{\sigma \sigma '}^x \sum\limits_{s,s'} {\left[ {\sigma _{ss'}^\gamma  } \right]^* } } \left\langle {\bar \psi _\sigma  ({\bf{p}},\omega _p ,o')\psi _{\sigma '} ({\bf{p}} + {\bf{Q}}_i^{\rm{C}} ,\omega _p ,o)\bar \psi _{s'} ({\bf{k}} + {\bf{Q}}_i^{\rm{C}} ,\omega _k ,\tilde o)\psi _s ({\bf{k}},\omega _k ,\tilde o')} \right\rangle _0 .
\end{split}
\end{equation}
The mean value in the above equation is evaluated using Wick's theorem
\begin{equation}\nonumber
\begin{split}
&\sum\limits_{\sigma ,\sigma '} {\sigma _{\sigma \sigma '}^x \sum\limits_{s,s'} {\left[ {\sigma _{ss'}^\gamma  } \right]^* } } \left\langle {\bar \psi _\sigma  ({\bf{p}},\omega _p ,o')\psi _{\sigma '} ({\bf{p}} + {\bf{Q}}_i^{\rm{C}} ,\omega _p ,o)\bar \psi _{s'} ({\bf{k}} + {\bf{Q}}_i^{\rm{C}} ,\omega _k ,\tilde o)\psi _s ({\bf{k}},\omega _k ,\tilde o')} \right\rangle _0 \\
&= \sum\limits_{\sigma ,\sigma '} {\sum\limits_{s,s'} {\sigma _{\sigma \sigma '}^x \sigma _{s's}^\gamma  } } \left\langle {\bar \psi _\sigma  ({\bf{p}},\omega _p ,o')\psi _{\sigma '} ({\bf{p}} + {\bf{Q}}_i^{\rm{C}} ,\omega _p ,o)} \right\rangle _0 \left\langle {\bar \psi _{s'} ({\bf{k}} + {\bf{Q}}_i^{\rm{C}} ,\omega _k ,\tilde o)\psi _s ({\bf{k}},\omega _k ,\tilde o')} \right\rangle _0 \\
& \hspace{1pc} + \sum\limits_{\sigma ,\sigma '} {\sum\limits_{s,s'} {\sigma _{\sigma \sigma '}^x \sigma _{s's}^\gamma  } } \left\langle {\bar \psi _\sigma  ({\bf{p}},\omega _p ,o')\psi _s ({\bf{k}},\omega _k ,\tilde o')} \right\rangle _0 \left\langle {\psi _{\sigma '} ({\bf{p}} + {\bf{Q}}_i^{\rm{C}} ,\omega _p ,o)\bar \psi _{s'} ({\bf{k}} + {\bf{Q}}_i^{\rm{C}} ,\omega _k ,\tilde o)} \right\rangle _0 \\
& = {\rm{Tr}}\{ \sigma ^x \} {\rm{Tr}}\{ \sigma ^\gamma  \} \delta _{{\bf{Q}}_i^{\rm{C}} ,0} G_{oo'}^0 ({\bf{p}},\omega _p )G_{\tilde o'\tilde o}^0 ({\bf{k}},\omega _k )
 - {\rm{Tr}}\{ \sigma ^x \sigma ^\gamma  \} \delta _{{\bf{p}},{\bf{k}}} \delta _{\omega _p ,\omega _k } G_{\tilde o'o'}^0 ({\bf{p}},\omega _p )G_{o\tilde o}^0 ({\bf{p}} + {\bf{Q}}_i^{\rm{C}} ,\omega _p ) \\
& =  - 2\delta _{x\gamma } \delta _{{\bf{p}},{\bf{k}}} \delta _{\omega _p ,\omega _k } G_{\tilde o'o'}^0 ({\bf{p}},\omega _p )G_{o\tilde o}^0 ({\bf{p}} + {\bf{Q}}_i^{\rm{C}} ,\omega _p ).
\end{split}
\end{equation}
Inserting this result into Eq. (\ref{eqA38}), we obtain
\begin{equation}\nonumber
\begin{split}
&\left\langle {O_{\alpha ,x}^{{\rm{SPN}}} ({\bf{Q}}_i^{\rm{C}} )S_{{\rm{HS}}}^{{\rm{SPN}}} [\psi ,\bar \psi ;\vec \Delta _{{\rm{SPN}}}^* ,\vec \Delta _{{\rm{SPN}}} ]} \right\rangle _0  =  - \frac{2}{\hbar }\sum\limits_{\beta  = 1}^{M_{{\rm{C}},i} } {\Delta _{\beta ,x}^{{\rm{SPN}}} ({\bf{Q}}_i^{\rm{C}} )} \sum\limits_{o,o',m} {[\varphi _{oo'm}^{{\rm{C}},\alpha } ({\bf{Q}}_i^{\rm{C}} )]^* } \sum\limits_{\tilde o,\tilde o',n} {\varphi _{\tilde o\tilde o'n}^{{\rm{C}},\beta } ({\bf{Q}}_i^{\rm{C}} )} \\
& \hspace{4pc} \times ( - 2)\sum\limits_{\bf{p}} {e^{i{\bf{R}}_m  \cdot {\bf{p}}} e^{ - i{\bf{R}}_n  \cdot {\bf{p}}} } \sum\limits_{\omega _p } {G_{o\tilde o}^0 ({\bf{p}} + {\bf{Q}}_i^{\rm{C}} ,\omega _p )G_{\tilde o'o'}^0 ({\bf{p}},\omega _p )} \\
&\hspace{2pc} =  - 4N\beta \hbar \sum\limits_{\gamma  = 1}^{M_{{\rm{C}},i} } {\Delta _{\gamma ,x}^{{\rm{SPN}}} ({\bf{Q}}_i^{\rm{C}} )} \sum\limits_{o,o',m} {[\varphi _{oo'm}^{{\rm{C}},\alpha } ({\bf{Q}}_i^{\rm{C}} )]^* } \sum\limits_{\tilde o,\tilde o',n} {\varphi _{\tilde o\tilde o'n}^{{\rm{C}},\gamma } ({\bf{Q}}_i^{\rm{C}} )} \chi _{oo'm,\tilde o\tilde o'n}^{{\rm{ph(}}\Omega  = 0{\rm{)}}} ({\bf{Q}}_i^{\rm{C}} ).
\end{split}
\end{equation}
Taking into account the definition (\ref{eqA11}), we obtain the final result
\begin{equation}\label{eqA39}
\begin{split}
\left\langle {O_{\alpha ,x}^{{\rm{SPN}}} ({\bf{Q}}_i^{\rm{C}} )S_{{\rm{HS}}}^{{\rm{SPN}}} [\psi ,\bar \psi ;\vec \Delta _{{\rm{SPN}}}^* ,\vec \Delta _{{\rm{SPN}}} ]} \right\rangle _0  =  - 4N\beta \hbar \sum\limits_{\gamma  = 1}^{M_{{\rm{C}},i} } {X_{\alpha \gamma }^{{\rm{ph,C}}} ({\bf{Q}}_i^{\rm{C}} )\Delta _{\gamma ,x}^{{\rm{SPN}}} ({\bf{Q}}_i^{\rm{C}} )} .
\end{split}
\end{equation}
In a similar way, one can verify the following relations:
\begin{equation}\label{eqA40}
\begin{split}
&\left\langle {O_{\alpha ,x}^{{\rm{SPN}}} ({\bf{Q}}_i^{\rm{C}} )S_{{\rm{HS}}}^{{\rm{sSC}}} [\psi ,\bar \psi ;\Delta _{{\rm{sSC}}}^* ,\Delta _{{\rm{sSC}}} ]} \right\rangle _0  = \left\langle {O_{\alpha ,x}^{{\rm{SPN}}} ({\bf{Q}}_i^{\rm{C}} )S_{{\rm{HS}}}^{{\rm{tSC}}} [\psi ,\bar \psi ;\vec \Delta _{{\rm{tSC}}}^* ,\vec \Delta _{{\rm{tSC}}} ]} \right\rangle _0 \\
&= \left\langle {O_{\alpha ,x}^{{\rm{SPN}}} ({\bf{Q}}_i^{\rm{C}} )S_{{\rm{HS}}}^{{\rm{CHG}}} [\psi ,\bar \psi ;\Delta _{{\rm{CHG}}}^* ,\Delta _{{\rm{CHG}}} ]} \right\rangle _0  = 0.
\end{split}
\end{equation}
Inserting Eqs. (\ref{eqA39}) and (\ref{eqA40}) into Eq. (\ref{eqA36}), we get
\begin{equation}\label{eqA41}
\begin{split}
\left\langle {O_{\alpha ,x}^{{\rm{SPN}}} ({\bf{Q}}_i^{\rm{C}} )} \right\rangle _\Delta   = 4N\beta \hbar \sum\limits_{\gamma  = 1}^{M_{{\rm{C}},i} } {X_{\alpha \gamma }^{{\rm{ph,C}}} ({\bf{Q}}_i^{\rm{C}} )\Delta _{\gamma ,x}^{{\rm{SPN}}} ({\bf{Q}}_i^{\rm{C}} )}.
\end{split}
\end{equation}
Finally, the third equation in Eq. (\ref{eqA19}) becomes
\begin{equation}\label{eqA42}
\begin{split}
\Delta _{\alpha ,x}^{{\rm{SPN}}} ({\bf{Q}}_i^{\rm{C}} ) = \Lambda ^{{\rm{C}},\alpha } ({\bf{Q}}_i^{\rm{C}} )\sum\limits_{\beta  = 1}^{M_{{\rm{C}},i} } {X_{\alpha \beta }^{{\rm{ph,C}}} ({\bf{Q}}_i^{\rm{C}} )\Delta _{\beta ,x}^{{\rm{SPN}}} ({\bf{Q}}_i^{\rm{C}} )}.
\end{split}
\end{equation}
This equation is also valid for $y, z$-components, giving
\begin{equation}\label{eqA43}
\begin{split}
\vec \Delta _\alpha ^{{\rm{SPN}}} ({\bf{Q}}_i^{\rm{C}} ) = \Lambda ^{{\rm{C}},\alpha } ({\bf{Q}}_i^{\rm{C}} )\sum\limits_{\beta  = 1}^{M_{{\rm{C}},i} } {X_{\alpha \beta }^{{\rm{ph,C}}} ({\bf{Q}}_i^{\rm{C}} )\vec \Delta _\beta ^{{\rm{SPN}}} ({\bf{Q}}_i^{\rm{C}} )},
\end{split}
\end{equation}
or equivalently, with $\vec Y_\alpha   \equiv \frac{{\vec \Delta _\alpha ^{{\rm{SPN}}} ({\bf{Q}}_i^{\rm{C}} )}}{{\sqrt {\Lambda ^{{\rm{C}},\alpha } ({\bf{Q}}_i^{\rm{C}} )} }}$,
\begin{equation}\label{eqA44}
\begin{split}
&\sum\limits_{\beta  = 1}^{M_{{\rm{C}},i} } {\left\{ {\sqrt {\Lambda ^{{\rm{C}},\alpha } ({\bf{Q}}_i^{\rm{C}} )} X_{\alpha \beta }^{{\rm{ph,C}}} ({\bf{Q}}_i^{\rm{C}} )\sqrt {\Lambda ^{{\rm{C}},\beta } ({\bf{Q}}_i^{\rm{C}} )} } \right\}} \vec Y_\beta   = \vec Y_\alpha ,\\
&\textrm{(Critical condition of the MF theory in the spin channel)}.
\end{split}
\end{equation}
This is exactly the same as the critical condition (\ref{eqA13}).

We consider the mean value $\left\langle {O_\alpha ^{{\rm{CHG}}} ({\bf{Q}}_i^{\rm{W}} )} \right\rangle _\Delta$, assuming that $\left\langle {O_\alpha ^{{\rm{CHG}}} ({\bf{Q}}_i^{\rm{W}} )} \right\rangle _0  = 0$.
\begin{equation}\label{eqA45}
\begin{split}
\left\langle {O_\alpha ^{{\rm{CHG}}} ({\bf{Q}}_i^{\rm{W}} )} \right\rangle _\Delta   =  - \left\langle {O_\alpha ^{{\rm{CHG}}} ({\bf{Q}}_i^{\rm{W}} )} \right. & \left( {S_{{\rm{HS}}}^{{\rm{sSC}}} [\psi ,\bar \psi ;\Delta _{{\rm{sSC}}}^* ,\Delta _{{\rm{sSC}}} ]} \right. + S_{{\rm{HS}}}^{{\rm{tSC}}} [\psi ,\bar \psi ;\vec \Delta _{{\rm{tSC}}}^* ,\vec \Delta _{{\rm{tSC}}} ]\\
& + S_{{\rm{HS}}}^{{\rm{SPN}}} [\psi ,\bar \psi ;\vec \Delta _{{\rm{SPN}}}^* ,\vec \Delta _{{\rm{SPN}}} ] + \left. {\left. {S_{{\rm{HS}}}^{{\rm{CHG}}} [\psi ,\bar \psi ;\Delta _{{\rm{CHG}}}^* ,\Delta _{{\rm{CHG}}} ]} \right)} \right\rangle _0 .
\end{split}
\end{equation}
Through the procedure similar to the one for $\left\langle {\vec O_\alpha ^{{\rm{SPN}}} ({\bf{Q}}_i^{\rm{C}} )} \right\rangle _\Delta$, we obtain the following results:
\begin{equation}\label{eqA46}
\begin{split}
\left\langle {O_\alpha ^{{\rm{CHG}}} ({\bf{Q}}_i^{\rm{W}} )S_{{\rm{HS}}}^{{\rm{CHG}}} [\psi ,\bar \psi ;\Delta _{{\rm{CHG}}}^* ,\Delta _{{\rm{CHG}}} ]} \right\rangle _0  =  - 4N\beta \hbar \sum\limits_{\gamma  = 1}^{M_{{\rm{W}},i} } {X_{\alpha \gamma }^{{\rm{ph,W}}} ({\bf{Q}}_i^{\rm{W}} )\Delta _\gamma ^{{\rm{CHG}}} ({\bf{Q}}_i^{\rm{W}} )} ,
\end{split}
\end{equation}
\begin{equation}\label{eqA47}
\begin{split}
&\left\langle {O_\alpha ^{{\rm{CHG}}} ({\bf{Q}}_i^{\rm{W}} )S_{{\rm{HS}}}^{{\rm{sSC}}} [\psi ,\bar \psi ;\Delta _{{\rm{sSC}}}^* ,\Delta _{{\rm{sSC}}} ]} \right\rangle _0  = \left\langle {O_\alpha ^{{\rm{CHG}}} ({\bf{Q}}_i^{\rm{W}} )S_{{\rm{HS}}}^{{\rm{tSC}}} [\psi ,\bar \psi ;\vec \Delta _{{\rm{tSC}}}^* ,\vec \Delta _{{\rm{tSC}}} ]} \right\rangle _0 \\
& = \left\langle {O_\alpha ^{{\rm{CHG}}} ({\bf{Q}}_i^{\rm{W}} )S_{{\rm{HS}}}^{{\rm{SPN}}} [\psi ,\bar \psi ;\vec \Delta _{{\rm{SPN}}}^* ,\vec \Delta _{{\rm{SPN}}} ]} \right\rangle _0  = 0.
\end{split}
\end{equation}
Inserting Eqs. (\ref{eqA46}) and (\ref{eqA47}) into Eq. (\ref{eqA45}), we get
\begin{equation}\label{eqA48}
\begin{split}
\left\langle {O_\alpha ^{{\rm{CHG}}} ({\bf{Q}}_i^{\rm{W}} )} \right\rangle _\Delta   = 4N\beta \hbar \sum\limits_{\gamma  = 1}^{M_{{\rm{W}},i} } {X_{\alpha \gamma }^{{\rm{ph,W}}} ({\bf{Q}}_i^{\rm{W}} )\Delta _\gamma ^{{\rm{CHG}}} ({\bf{Q}}_i^{\rm{W}} )}.
\end{split}
\end{equation}
Then, the fourth equation in Eq. (\ref{eqA19}) becomes
\begin{equation}\label{eqA49}
\begin{split}
\Delta _\alpha ^{{\rm{CHG}}} ({\bf{Q}}_i^{\rm{W}} ) = \Lambda ^{{\rm{W}},\alpha } ({\bf{Q}}_i^{\rm{W}} )\sum\limits_{\beta  = 1}^{M_{{\rm{W}},i} } {X_{\alpha \beta }^{{\rm{ph,W}}} ({\bf{Q}}_i^{\rm{W}} )\Delta _\beta ^{{\rm{CHG}}} ({\bf{Q}}_i^{\rm{W}} )},
\end{split}
\end{equation}
or equivalently, with $Y_\alpha   \equiv \frac{{\Delta _\alpha ^{{\rm{CHG}}} ({\bf{Q}}_i^{\rm{W}} )}}{{\sqrt {\Lambda ^{{\rm{W}},\alpha } ({\bf{Q}}_i^{\rm{W}} )} }}$,
\begin{equation}\label{eqA50}
\begin{split}
&\sum\limits_{\beta  = 1}^{M_{{\rm{W}},i} } {\left\{ {\sqrt {\Lambda ^{{\rm{W}},\alpha } ({\bf{Q}}_i^{\rm{W}} )} X_{\alpha \beta }^{{\rm{ph,W}}} ({\bf{Q}}_i^{\rm{W}} )\sqrt {\Lambda ^{{\rm{W}},\beta } ({\bf{Q}}_i^{\rm{W}} )} } \right\}} Y_\beta   = Y_\alpha ,\\
&\textrm{(Critical condition of the MF theory in the charge channel)}.
\end{split}
\end{equation}
This is exactly the same as the critical condition (\ref{eqA14}).

\section{Comparison of the renormalized MF and our TUFRG + MF approaches} \label{appendB}

Here we compare the critical conditions in the renormalized MF \cite{ref26} and our TUFRG + MF approaches. First, let us consider the renormalized MF in the framework of the TUFRG. The bosonic propagators $P^\Omega, C^\Omega$, and $W^\Omega$ are obtained by solving the TUFRG flow with the sharp momentum cutoff regulator. In this regularization the fermionic propagator is given by
\begin{equation}\label{eqB01}
G_{oo'}^{0,\Omega } (\omega ,{\bf{k}}) = \sum\limits_b {T_{ob} ({\bf{k}})[T_{o'b} ({\bf{k}})]^* } \frac{1}{{i\omega  - (\varepsilon _b ({\bf{k}}) - \mu )/\hbar }}\Theta (|\varepsilon_b ({\bf{k}})-\mu|-\Omega),
\end{equation}
where $\varepsilon_b({\bf{k}})$ and $\Theta (x)$ denote the energy level of the band $b$ and the Heaviside step function, respectively, while the unitary matrix $T_{ob} ({\bf{k}})$ connects two operators in the orbital ($\hat c_{{\bf{k}}o\sigma }$) and the band ($\hat d_{{\bf{k}}b\sigma }$) pictures, according to
\begin{equation}\nonumber
\hat c_{{\bf{k}}o\sigma }  = \sum\limits_b {T_{ob} ({\bf{k}})\hat d_{{\bf{k}}b\sigma } } ,\hat d_{{\bf{k}}b\sigma }  = \sum\limits_o {[T_{ob} ({\bf{k}})]^* \hat c_{{\bf{k}}o\sigma}} .
\end{equation}
At the divergence scale $\Omega _D$, the bosonic propagators are expressed in terms of the singular modes $\left| {\phi ^{{\rm{X}},\alpha } ({\bf{Q}}_i^{\rm{X}} )} \right\rangle $ as given in Eq. (\ref{eq051}), and the effective action $\Gamma ^{\Omega _D }$ as in Eq. (\ref{eq052}). The modes $\left| {\phi ^{{\rm{X}},\alpha } ({\bf{Q}}_i^{\rm{X}} )} \right\rangle $ should satisfy the constraints (\ref{eq064}), (\ref{eq066}), and (\ref{eq067}).

In the renormalized MF approach, only the low-energy modes with $|\varepsilon _b ({\bf{k}})-\mu| \le \Omega _D$ are involved in the MF calculation. Thus, we have to address the following action:
\begin{equation}\label{eqB02}
S[\psi ,\bar \psi ] = S_0 [\psi ,\bar \psi ] + \Gamma ^{{\rm{sSC}}} [\psi ,\bar \psi ] + \Gamma ^{{\rm{tSC}}} [\psi ,\bar \psi ] + \Gamma ^{{\rm{SPN}}} [\psi ,\bar \psi ] + \Gamma ^{{\rm{CHG}}} [\psi ,\bar \psi ].
\end{equation}
The noninteracting part ($S_0$) of the action is represented as Eq. (\ref{eq003}), with $H_{oo'}^0 ({\bf{k}})$ given by
\begin{equation}\label{eqB03}
H_{oo'}^0 ({\bf{k}}) = \sum\limits_b {T_{ob} ({\bf{k}})\varepsilon _b ({\bf{k}})[T_{o'b} ({\bf{k}})]^* } .
\end{equation}
The expressions of the interaction parts, $\Gamma ^{{\rm{sSC}}} ,\Gamma ^{{\rm{tSC}}} ,\Gamma ^{{\rm{SPN}}}$, and $\Gamma ^{{\rm{CHG}}}$, are given by a replacement of $\Lambda ^{{\rm{X}},\alpha } ({\bf{Q}}_i^{\rm{X}} ) \to \lambda ^{{\rm{X}},\alpha } ({\bf{Q}}_i^{\rm{X}} ),O_\alpha ^{\rm{X}} ({\bf{Q}}_i^{\rm{X}} ,\omega ) \to B_\alpha ^{\rm{X}} ({\bf{Q}}_i^{\rm{X}} ,\omega )$ in Eq. (\ref{eq068})
\begin{equation}\label{eqB04}
\begin{split}
\Gamma ^{{\rm{sSC}}} [\psi ,\bar \psi ] =&  - \frac{{\rm{1}}}{2}\frac{1}{{2N\beta \hbar ^2 }}\sum\limits_{i = 1}^{N_{\rm{P}} } {\sum\limits_{\alpha  = 1}^{M_{{\rm{P}},i} } {\lambda ^{{\rm{P}},\alpha } ({\bf{Q}}_i^{\rm{P}} )} } \sum\limits_\omega  {[B_\alpha ^{{\rm{sSC}}} ({\bf{Q}}_i^{\rm{P}} ,\omega )]^* B_\alpha ^{{\rm{sSC}}} ({\bf{Q}}_i^{\rm{P}} ,\omega )},\\
\Gamma ^{{\rm{tSC}}} [\psi ,\bar \psi ] =&  - \frac{{\rm{1}}}{2}\frac{1}{{2N\beta \hbar ^2 }}\sum\limits_{i = 1}^{N_{\rm{P}} } {\sum\limits_{\alpha  = 1}^{M_{{\rm{P}},i} } {\lambda ^{{\rm{P}},\alpha } ({\bf{Q}}_i^{\rm{P}} )} } \sum\limits_\omega  {[\vec B_\alpha ^{{\rm{tSC}}} ({\bf{Q}}_i^{\rm{P}} ,\omega )]^*  \cdot \vec B_\alpha ^{{\rm{tSC}}} ({\bf{Q}}_i^{\rm{P}} ,\omega )} ,\\
\Gamma ^{{\rm{SPN}}} [\psi ,\bar \psi ] =&  - \frac{{\rm{1}}}{2}\frac{1}{{2N\beta \hbar ^2 }}\sum\limits_{i = 1}^{N_{\rm{C}} } {\sum\limits_{\alpha  = 1}^{M_{{\rm{C}},i} } {\lambda ^{{\rm{C}},\alpha } ({\bf{Q}}_i^{\rm{C}} )} } \sum\limits_\omega  {[\vec B_\alpha ^{{\rm{SPN}}} ({\bf{Q}}_i^{\rm{C}} ,\omega )]^*  \cdot \vec B_\alpha ^{{\rm{SPN}}} ({\bf{Q}}_i^{\rm{C}} ,\omega )},\\
\Gamma ^{{\rm{CHG}}} [\psi ,\bar \psi ] =&  - \frac{{\rm{1}}}{2}\frac{1}{{2N\beta \hbar ^2 }}\sum\limits_{i = 1}^{N_{\rm{W}} } {\sum\limits_{\alpha  = 1}^{M_{{\rm{W}},i} } {\lambda ^{{\rm{W}},\alpha } ({\bf{Q}}_i^{\rm{W}} )} } \sum\limits_\omega  {[B_\alpha ^{{\rm{CHG}}} ({\bf{Q}}_i^{\rm{W}} ,\omega )]^* B_\alpha ^{{\rm{CHG}}} ({\bf{Q}}_i^{\rm{W}} ,\omega )},
\end{split}
\end{equation}
with the fermion bilinears in the X-channel, $B_\alpha ^{\rm{X}}$ $({\rm{X}} \in \{ {\rm{sSC,  tSC,  SPN,  CHG}}\} )$, defined as
\begin{equation}\label{eqB05}
\begin{split}
B_\alpha ^{{\rm{sSC}}} ({\bf{Q}}_i^{\rm{P}} ,\omega ) \equiv& \sum\limits_{{\bf{p}},\omega _p } {\sum\limits_{o,o',m} {[\phi _{oo'm}^{{\rm{P}},\alpha } ({\bf{Q}}_i^{\rm{P}} )]^* e^{i{\bf{R}}_m\cdot {\bf{p}}}}}
\sum\limits_{\tilde o,\tilde o'} {\rho _{o'\tilde o'} ( - {\bf{p}})\rho _{o\tilde o} ({\bf{p}} + {\bf{Q}}_i^{\rm{P}} )} \sum\limits_\sigma  {\sigma \psi _{ - \sigma } ( - {\bf{p}}, - \omega _p ,\tilde o')[\psi _\sigma  ({\bf{p}} + {\bf{Q}}_i^{\rm{P}} ,\omega _p  + \omega ,\tilde o)}],\\
\vec B_\alpha ^{{\rm{tSC}}} ({\bf{Q}}_i^{\rm{P}} ,\omega ) \equiv& (B_{\alpha ,x}^{{\rm{tSC}}} ({\bf{Q}}_i^{\rm{P}} ,\omega ),B_{\alpha ,y}^{{\rm{tSC}}} ({\bf{Q}}_i^{\rm{P}} ,\omega ),B_{\alpha ,z}^{{\rm{tSC}}} ({\bf{Q}}_i^{\rm{P}} ,\omega ))\\
\equiv& \sum\limits_{{\bf{p}},\omega _p } {\sum\limits_{o,o',m} {[\phi _{oo'm}^{{\rm{P}},\alpha } ({\bf{Q}}_i^{\rm{P}} )]^* e^{i{\bf{R}}_m  \cdot {\bf{p}}} } } 
\sum\limits_{\tilde o,\tilde o'} {\rho _{o'\tilde o'} ( - {\bf{p}})\rho _{o\tilde o} ({\bf{p}} + {\bf{Q}}_i^{\rm{P}} )} 
\left( { - \sum\limits_\sigma  {\sigma \psi _\sigma  ( - {\bf{p}}, - \omega _p ,\tilde o')\psi _\sigma  ({\bf{p}} + {\bf{Q}}_i^{\rm{P}} ,\omega _p  + \omega ,\tilde o)} ,} \right.\\
& \left. { - i\sum\limits_\sigma  {\psi _\sigma  ( - {\bf{p}}, - \omega _p ,\tilde o')\psi _\sigma  ({\bf{p}} + {\bf{Q}}_i^{\rm{P}} ,\omega _p  + \omega ,\tilde o)} ,\sum\limits_\sigma  {\psi _{ - \sigma } ( - {\bf{p}}, - \omega _p ,\tilde o')\psi _\sigma  ({\bf{p}} + {\bf{Q}}_i^{\rm{P}} ,\omega _p  + \omega ,\tilde o)} } \right),\\
\vec B_\alpha ^{{\rm{SPN}}} ({\bf{Q}}_i^{\rm{C}} ,\omega ) \equiv& \sum\limits_{{\bf{p}},\omega _p } {\sum\limits_{o,o',m} {[\phi _{oo'm}^{{\rm{C}},\alpha } ({\bf{Q}}_i^{\rm{C}} )]^* e^{i{\bf{R}}_m \cdot {\bf{p}}}}}
\sum\limits_{\tilde o,\tilde o'} {\rho _{\tilde o'o'} ({\bf{p}})\rho _{o\tilde o} ({\bf{p}} + {\bf{Q}}_i^{\rm{C}} )} \sum\limits_{\sigma ,\sigma '} {\bar \psi _\sigma  ({\bf{p}},\omega _p ,\tilde o')\vec \sigma _{\sigma \sigma '} \psi _{\sigma '} ({\bf{p}} + {\bf{Q}}_i^{\rm{C}} ,\omega _p  + \omega ,\tilde o)},\\ 
B_\alpha ^{{\rm{CHG}}} ({\bf{Q}}_i^{\rm{W}} ,\omega ) \equiv& \sum\limits_{{\bf{p}},\omega _p } {\sum\limits_{o,o',m} {[\phi _{oo'm}^{{\rm{W}},\alpha } ({\bf{Q}}_i^{\rm{W}} )]^* e^{i{\bf{R}}_m \cdot {\bf{p}}}}} 
\sum\limits_{\tilde o,\tilde o'} {\rho _{\tilde o'o'} ({\bf{p}})\rho _{o\tilde o} ({\bf{p}} + {\bf{Q}}_i^{\rm{W}} )} \sum\limits_\sigma  {\bar \psi _\sigma  ({\bf{p}},\omega _p ,\tilde o')\psi _\sigma  ({\bf{p}} + {\bf{Q}}_i^{\rm{W}} ,\omega _p  + \omega ,\tilde o)}.
\end{split}
\end{equation}
We introduced here the low-energy density matrix $\rho ({\bf{k}})$:
\begin{equation}\label{eqB06}
\rho _{oo'} ({\bf{k}}) \equiv \sum\limits_b {T_{ob} ({\bf{k}})\Theta (\Omega _D  - |\varepsilon _b ({\bf{k}}) - \mu |)[T_{o'b} ({\bf{k}})]^* }.
\end{equation}

The self-consistency condition is obtained from Eq. (\ref{eq083}) by using the same replacement ($\Lambda ^{{\rm{X}},\alpha } ({\bf{Q}}_i^{\rm{X}} ) \to \lambda ^{{\rm{X}},\alpha } ({\bf{Q}}_i^{\rm{X}} ),O_\alpha ^{\rm{X}} ({\bf{Q}}_i^{\rm{X}} ,\omega ) \to B_\alpha ^{\rm{X}} ({\bf{Q}}_i^{\rm{X}} ,\omega )$),
\begin{equation}\label{eqB07}
\begin{split}
\Delta _\alpha ^{{\rm{sSC}}} ({\bf{Q}}_i^{\rm{P}} ) =& \frac{{\lambda ^{{\rm{sSC}},\alpha } ({\bf{Q}}_i^{\rm{P}} )}}{{4N\beta \hbar }}\left\langle {B_\alpha ^{{\rm{sSC}}} ({\bf{Q}}_i^{\rm{P}} )} \right\rangle _\Delta,
\vec \Delta _\alpha ^{{\rm{tSC}}} ({\bf{Q}}_i^{\rm{P}} ) = \frac{{\lambda ^{{\rm{tSC}},\alpha } ({\bf{Q}}_i^{\rm{P}} )}}{{4N\beta \hbar }}\left\langle {\vec B_\alpha ^{{\rm{tSC}}} ({\bf{Q}}_i^{\rm{P}} )} \right\rangle _\Delta,\\
\vec \Delta _\alpha ^{{\rm{SPN}}} ({\bf{Q}}_i^{\rm{C}} ) =& \frac{{\lambda ^{{\rm{C}},\alpha } ({\bf{Q}}_i^{\rm{C}} )}}{{4N\beta \hbar }}\left\langle {\vec B_\alpha ^{{\rm{SPN}}} ({\bf{Q}}_i^{\rm{C}} )} \right\rangle _\Delta,
\Delta _\alpha ^{{\rm{CHG}}} ({\bf{Q}}_i^{\rm{W}} ) = \frac{{\lambda ^{{\rm{W}},\alpha } ({\bf{Q}}_i^{\rm{W}} )}}{{4N\beta \hbar }}\left\langle {B_\alpha ^{{\rm{CHG}}} ({\bf{Q}}_i^{\rm{W}} )} \right\rangle _\Delta .
\end{split}
\end{equation}
Following the discussions in Eqs. (\ref{eqA19}) to (\ref{eqA50}), one can easily derive the critical conditions of the renormalized MF scheme. The result is given by a replacement of $\Lambda ^{{\rm{X}},\alpha } ({\bf{Q}}_i^{\rm{X}} ) \to \lambda ^{{\rm{X}},\alpha } ({\bf{Q}}_i^{\rm{X}} ),X_{\alpha \beta }^{\rm{X}} ({\bf{Q}}_i^{\rm{X}} ) \to D_{\alpha \beta }^{\rm{X}} ({\bf{Q}}_i^{\rm{X}} )$ in the critical conditions of the MF theory (Eqs. (\ref{eqA29}), (\ref{eqA35}), (\ref{eqA44}), and (\ref{eqA50})). Here $D_{\alpha \beta }^{\rm{X}} ({\bf{Q}}_i^{\rm{X}})$ is defined as
\begin{equation}\label{eqB08}
\begin{split}
D_{\alpha \beta }^{{\rm{pp,sSC}}} ({\bf{Q}}_i^{\rm{P}} ) \equiv& \left\langle {\phi ^{{\rm{sSC}},\alpha } ({\bf{Q}}_i^{\rm{P}} )} \right|\bar \chi ^{{\rm{pp}}(\Omega _D )} ({\bf{Q}}_i^{\rm{P}} )\left| {\phi ^{{\rm{sSC}},\beta } ({\bf{Q}}_i^{\rm{P}} )} \right\rangle,
D_{\alpha \beta }^{{\rm{pp,tSC}}} ({\bf{Q}}_i^{\rm{P}} ) \equiv \left\langle {\phi ^{{\rm{tSC}},\alpha } ({\bf{Q}}_i^{\rm{P}} )} \right|\bar \chi ^{{\rm{pp}}(\Omega _D )} ({\bf{Q}}_i^{\rm{P}} )\left| {\phi ^{{\rm{tSC}},\beta } ({\bf{Q}}_i^{\rm{P}} )} \right\rangle ,\\
D_{\alpha \beta }^{{\rm{ph,C}}} ({\bf{Q}}_i^{\rm{C}} ) \equiv& \left\langle {\phi ^{{\rm{C}},\alpha } ({\bf{Q}}_i^{\rm{C}} )} \right|\bar \chi ^{{\rm{ph}}(\Omega _D )} ({\bf{Q}}_i^{\rm{C}} )\left| {\phi ^{{\rm{C}},\beta } ({\bf{Q}}_i^{\rm{C}} )} \right\rangle,
D_{\alpha \beta }^{{\rm{ph,W}}} ({\bf{Q}}_i^{\rm{W}} ) \equiv \left\langle {\phi ^{{\rm{W}},\alpha } ({\bf{Q}}_i^{\rm{W}} )} \right|\bar \chi ^{{\rm{ph}}(\Omega _D )} ({\bf{Q}}_i^{\rm{W}} )\left| {\phi ^{{\rm{W}},\beta } ({\bf{Q}}_i^{\rm{W}} )} \right\rangle,
\end{split}
\end{equation}
with $\bar \chi ^{{\rm{pp}}(\Omega _D )} ({\bf{q}})$ and $\bar \chi ^{{\rm{ph}}(\Omega _D )} ({\bf{q}})$, given by
\begin{equation}\label{eqB09}
\begin{split}
\bar \chi _{o'_1 o'_2 m,o_1 o_2 n}^{{\rm{pp(}}\Omega _D {\rm{)}}} ({\bf{q}}) \equiv&  - \int {\frac{{d{\bf{k}}}}{{S_{{\rm{BZ}}} }}} f_m ({\bf{k}})f_n^* ({\bf{k}})\left[ {\frac{1}{{\beta \hbar ^2 }}\sum\limits_\omega  {\bar G_{o'_1 o_1 }^{\Omega _D } ({\bf{k}} + {\bf{q}},\omega )\bar G_{o'_2 o_2 }^{\Omega _D } ( - {\bf{k}}, - \omega )} } \right],\\
\bar \chi _{o'_1 o'_2 m,o_1 o_2 n}^{{\rm{ph(}}\Omega _D {\rm{)}}} ({\bf{q}}) \equiv&  - \int {\frac{{d{\bf{k}}}}{{S_{{\rm{BZ}}} }}} f_m ({\bf{k}})f_n^* ({\bf{k}})\left[ {\frac{1}{{\beta \hbar ^2 }}\sum\limits_\omega  {\bar G_{o'_1 o_1 }^{\Omega _D } ({\bf{k}} + {\bf{q}},\omega )\bar G_{o_2 o'_2 }^{\Omega _D } ({\bf{k}},\omega )} } \right],
\end{split}
\end{equation}
and the low-energy propagator $\bar G_{oo'}^{\Omega _D } ({\bf{k}},\omega ) \equiv \sum\limits_{\tilde o,\tilde o'} {\rho _{o\tilde o} ({\bf{k}})} G_{\tilde o\tilde o'}^0 ({\bf{k}},\omega )\rho _{\tilde o'o'} ({\bf{k}})$. From Eqs. (\ref{eqB01}) and (\ref{eqB06}) one can easily verify the relation
\begin{equation}\label{eqB10}
\bar G_{oo'}^{\Omega _D } ({\bf{k}},\omega ) = G_{oo'}^0 ({\bf{k}},\omega ) - G_{oo'}^{0,\Omega _D } ({\bf{k}},\omega ).
\end{equation}

The critical conditions of the renormalized MF theory are represented as follows:
\begin{equation}\label{eqB11}
\begin{split}
&\sum\limits_{\beta  = 1}^{M_{{\rm{sSC}},i} } {\left\{ {\sqrt {\lambda ^{{\rm{sSC}},\alpha } ({\bf{Q}}_i^{\rm{P}} )} \left[ { - D_{\alpha \beta }^{{\rm{pp,sSC}}} ({\bf{Q}}_i^{\rm{P}} )} \right]\sqrt {\lambda ^{{\rm{sSC}},\beta } ({\bf{Q}}_i^{\rm{P}} )} } \right\}} Y_\beta   = Y_\alpha \\
&\hspace{2pc} \textrm{(Critical condition in the spin-singlet pairing channel),}
\end{split}
\end{equation}
\begin{equation}\label{eqB12}
\begin{split}
&\sum\limits_{\beta  = 1}^{M_{{\rm{tSC}},i} } {\left\{ {\sqrt {\lambda ^{{\rm{tSC}},\alpha } ({\bf{Q}}_i^{\rm{P}} )} \left[ { - D_{\alpha \beta }^{{\rm{pp,tSC}}} ({\bf{Q}}_i^{\rm{P}} )} \right]\sqrt {\lambda ^{{\rm{tSC}},\beta } ({\bf{Q}}_i^{\rm{P}} )} } \right\}} \vec Y_\beta   = \vec Y_\alpha \\
&\hspace{2pc} \textrm{(Critical condition in the spin-tripglet pairing channel),}
\end{split}
\end{equation}
\begin{equation}\label{eqB13}
\sum\limits_{\beta  = 1}^{M_{{\rm{C}},i} } {\left\{ {\sqrt {\lambda ^{{\rm{C}},\alpha } ({\bf{Q}}_i^{\rm{C}} )} D_{\alpha \beta }^{{\rm{ph,C}}} ({\bf{Q}}_i^{\rm{C}} )\sqrt {\lambda ^{{\rm{C}},\beta } ({\bf{Q}}_i^{\rm{C}} )} } \right\}} \vec Y_\beta   = \vec Y_\alpha 
\hspace{1pc} \textrm{(Critical condition in the spin channel),}
\end{equation}
\begin{equation}\label{eqB14}
\sum\limits_{\beta  = 1}^{M_{{\rm{W}},i} } {\left\{ {\sqrt {\lambda ^{{\rm{W}},\alpha } ({\bf{Q}}_i^{\rm{W}} )} D_{\alpha \beta }^{{\rm{ph,W}}} ({\bf{Q}}_i^{\rm{W}} )\sqrt {\lambda ^{{\rm{W}},\beta } ({\bf{Q}}_i^{\rm{W}} )} } \right\}} Y_\beta   = Y_\alpha 
\hspace{1pc} \textrm{(Critical condition in the charge channel).}
\end{equation}
We consider in more detail Eq. (\ref{eqB11}), which can be expressed as
\begin{equation}\nonumber
\sum\limits_{\beta  = 1}^{M_{{\rm{sSC}},i} } {\sqrt {\lambda ^{{\rm{sSC}},\alpha } ({\bf{Q}}_i^{\rm{P}} )} \left[ {\frac{1}{{\lambda ^{{\rm{sSC}},\alpha } ({\bf{Q}}_i^{\rm{P}} )}}\delta _{\alpha \beta }  + D_{\alpha \beta }^{{\rm{pp,sSC}}} ({\bf{Q}}_i^{\rm{P}} )} \right]} \sqrt {\lambda ^{{\rm{sSC}},\beta } ({\bf{Q}}_i^{\rm{P}} )} Y_\beta   = 0.
\end{equation}
With $X_\beta   \equiv \sqrt {\lambda ^{{\rm{sSC}},\beta } ({\bf{Q}}_i^{\rm{P}} )} Y_\beta $, we can take into account Eqs. (\ref{eq051}) and (\ref{eqB08}) to rewrite the above equation as
\begin{equation}\nonumber
\left\langle {\phi ^{{\rm{sSC}},\alpha } ({\bf{Q}}_i^{\rm{P}} )} \right|\left[ {\left( { - P^{\Omega _D } ({\bf{Q}}_i^{\rm{P}} )} \right)^{ - 1}  + \bar \chi ^{{\rm{pp}}(\Omega _D )} ({\bf{Q}}_i^{\rm{P}} )} \right]\left( {\sum\limits_{\beta  = 1}^{M_{{\rm{sSC}},i} } {X_\beta  \left| {\phi ^{{\rm{sSC}},\beta } ({\bf{Q}}_i^{\rm{P}} )} \right\rangle } } \right) = 0
\hspace{1pc} \left( {\alpha  = 1, ... , M_{{\rm{sSC}},i} } \right).
\end{equation}
Therefore, the critical condition (\ref{eqB11}) requires the matrix
\begin{equation}\label{eqB15}
\left[ { - \bar P({\bf{Q}}_i^{\rm{P}} )} \right]^{ - 1}  \equiv \left[ { - P^{\Omega _D } ({\bf{Q}}_i^{\rm{P}} )} \right]^{ - 1}  + \bar \chi ^{{\rm{pp}}(\Omega _D )} ({\bf{Q}}_i^{\rm{P}} ),
\end{equation}
to have an eigenvector associated with zero eigenvalue in the $M_{{\rm{sSC}},i} $-dimensional space consisting of the bases $\left\{ {\left| {\phi ^{{\rm{sSC}},\alpha } ({\bf{Q}}_i^{\rm{P}} )} \right\rangle } \right\}$. Likewise, the critical condition (\ref{eqB12}) requires the zero eigenvalue of the matrix $\left( { - \bar P({\bf{Q}}_i^{\rm{P}} )} \right)^{ - 1} $ in the $M_{{\rm{tSC}},i} $-dimensional space consisting of $\left\{ {\left| {\phi ^{{\rm{tSC}},\alpha } ({\bf{Q}}_i^{\rm{P}} )} \right\rangle } \right\}$. In a similar way, one can extract the meaning of critical conditions (\ref{eqB13}) and (\ref{eqB14}). Namely, the condition (\ref{eqB13}) requires the zero eigenvalue of the matrix
\begin{equation}\label{eqB16}
\left[ {\bar C({\bf{Q}}_i^{\rm{C}} )} \right]^{ - 1}  \equiv \left[ {C^{\Omega _D } ({\bf{Q}}_i^{\rm{C}} )} \right]^{ - 1}  - \bar \chi ^{{\rm{ph}}(\Omega _D )} ({\bf{Q}}_i^{\rm{C}} ),
\end{equation}
in the $M_{{\rm{C}},i} $-dimensional space consisting of $\left\{ {\left| {\phi ^{{\rm{C}},\alpha } ({\bf{Q}}_i^{\rm{C}} )} \right\rangle } \right\}$, while Eq. (\ref{eqB14}) requires the zero eigenvalue of
\begin{equation}\label{eqB17}
\left[ {\bar W({\bf{Q}}_i^{\rm{W}} )} \right]^{ - 1}  \equiv \left[ {W^{\Omega _D } ({\bf{Q}}_i^{\rm{W}} )} \right]^{ - 1}  - \bar \chi ^{{\rm{ph}}(\Omega _D )} ({\bf{Q}}_i^{\rm{W}} )
\end{equation}
in the $M_{{\rm{W}},i} $-dimensional space consisting of $\left\{ {\left| {\phi ^{{\rm{W}},\alpha } ({\bf{Q}}_i^{\rm{W}} )} \right\rangle } \right\}$.

Second, let us reconsider our novel scheme. The critical conditions of our TUFRG + MF approach are given by Eqs. (\ref{eqA17}), (\ref{eqA18}), (\ref{eqA13}), and (\ref{eqA14}). Following the arguments above, one can draw the meaning of these conditions. The critical condition (\ref{eqA17}) (or Eq. (\ref{eqA18})) requires the zero eigenvalue of the matrix
\begin{equation}\label{eqB18}
\left[ { - P^{\Omega  = 0} ({\bf{Q}}_i^{\rm{P}} )} \right]^{ - 1}  = \left[ { - \tilde P({\bf{Q}}_i^{\rm{P}} )} \right]^{ - 1}  + \chi ^{{\rm{pp}}(\Omega  = 0)} ({\bf{Q}}_i^{\rm{P}} ),
\end{equation}
in the $M_{{\rm{sSC}},i} $ (or $M_{{\rm{tSC}},i} $)-dimensional space consisting of the bases $\left\{ {\left| {\varphi ^{{\rm{sSC}},\alpha } ({\bf{Q}}_i^{\rm{P}} )} \right\rangle } \right\}$ (or $\left\{ {\left| {\varphi ^{{\rm{tSC}},\alpha } ({\bf{Q}}_i^{\rm{P}} )} \right\rangle } \right\}$). And the critical conditions (\ref{eqA13}) and (\ref{eqA14}) require the zero eigenvalues of the matrices $\left[ {C^{\Omega  = 0} ({\bf{Q}}_i^{\rm{C}} )} \right]^{ - 1} $ and $\left[ {W^{\Omega  = 0} ({\bf{Q}}_i^{\rm{W}} )} \right]^{ - 1} $, expressed as
\begin{equation}\label{eqB19}
\left[ {C^{\Omega  = 0} ({\bf{Q}}_i^{\rm{C}} )} \right]^{ - 1}  = \left[ {\tilde C({\bf{Q}}_i^{\rm{C}} )} \right]^{ - 1}  - \chi ^{{\rm{ph}}(\Omega  = 0)} ({\bf{Q}}_i^{\rm{C}} ),
\end{equation}
\begin{equation}\label{eqB20}
\left[ {W^{\Omega  = 0} ({\bf{Q}}_i^{\rm{W}} )} \right]^{ - 1}  = \left[ {\tilde W({\bf{Q}}_i^{\rm{W}} )} \right]^{ - 1}  - \chi ^{{\rm{ph}}(\Omega  = 0)} ({\bf{Q}}_i^{\rm{W}} ),
\end{equation}
in the $M_{{\rm{C}},i} $- and $M_{{\rm{W}},i} $-dimensional spaces, consisting of $\left\{ {\left| {\varphi ^{{\rm{C}},\alpha } ({\bf{Q}}_i^{\rm{C}} )} \right\rangle } \right\}$ and $\left\{ {\left| {\varphi ^{{\rm{W}},\alpha } ({\bf{Q}}_i^{\rm{W}} )} \right\rangle } \right\}$, respectively.

Now we compare the critical conditions in the renormalized MF and our TUFRG + MF approaches. Note that two $M_{{\rm{X}},i} $-dimensional spaces, each consisting of the bases $\left\{ {\left| {\phi ^{{\rm{X}},\alpha } ({\bf{Q}}_i^{\rm{X}} )} \right\rangle } \right\}$ and $\left\{ {\left| {\varphi ^{{\rm{X}},\alpha } ({\bf{Q}}_i^{\rm{X}} )} \right\rangle } \right\}$, are identical because these two bases are related to each other by Eq. (\ref{eq060}). Let us compare two matrices in Eqs. (\ref{eqB15}) and (\ref{eqB18}). Combining these equations with the definition (\ref{eq049}), we obtain
\begin{equation}\label{eqB21}
\left[ { - \bar P({\bf{Q}}_i^{\rm{P}} )} \right]^{ - 1}  = \left[ { - P^{\Omega  = 0} ({\bf{Q}}_i^{\rm{P}} )} \right]^{ - 1}  - \chi ^{{\rm{pp - mix}}(\Omega _D )} ({\bf{Q}}_i^{\rm{P}} ),
\end{equation}
where $\chi ^{{\rm{pp - mix}}(\Omega _D )} ({\bf{Q}}_i^{\rm{P}} )$ is the contribution from the interplay between the high- and low-energy modes, defined by
\begin{equation}\label{eqB22}
\begin{split}
\chi ^{{\rm{pp - mix}}(\Omega _D )} ({\bf{q}}) \equiv& \chi ^{{\rm{pp}}(\Omega  = 0)} ({\bf{q}}) - \bar \chi ^{{\rm{pp}}(\Omega _D )}  - \chi ^{{\rm{pp}}(\Omega _D )} ({\bf{q}}),\\
\chi _{o'_1 o'_2 m,o_1 o_2 n}^{{\rm{pp - mix}}(\Omega _D )} ({\bf{q}}) =&  - \int {\frac{{d{\bf{k}}}}{{S_{{\rm{BZ}}} }}} f_m ({\bf{k}})f_n^* ({\bf{k}})\left[ {\frac{1}{{\beta \hbar ^2 }}} \right.\sum\limits_\omega  {\left( {G_{o'_1 o_1 }^{\Omega _D } ({\bf{k}} + {\bf{q}},\omega )\bar G_{o'_2 o_2 }^{\Omega _D } ( - {\bf{k}}, - \omega )} \right.} 
 + \left. {\left. {\bar G_{o'_1 o_1 }^{\Omega _D } ({\bf{k}} + {\bf{q}},\omega )G_{o'_2 o_2 }^{\Omega _D } ( - {\bf{k}}, - \omega )} \right)} \right].
\end{split}
\end{equation}
Similar equations for the matrices, $\bar C({\bf{Q}}_i^{\rm{C}} )$ and $\bar W({\bf{Q}}_i^{\rm{W}} )$, are easily derived
\begin{equation}\label{eqB23}
\left[ {\bar C({\bf{Q}}_i^{\rm{C}} )} \right]^{ - 1}  = \left[ {C^{\Omega  = 0} ({\bf{Q}}_i^{\rm{C}} )} \right]^{ - 1}  + \chi ^{{\rm{ph - mix}}(\Omega _D )} ({\bf{Q}}_i^{\rm{C}} ),
\end{equation}
\begin{equation}\label{eqB24}
\left[ {\bar W({\bf{Q}}_i^{\rm{W}} )} \right]^{ - 1}  = \left[ {W^{\Omega  = 0} ({\bf{Q}}_i^{\rm{W}} )} \right]^{ - 1}  + \chi ^{{\rm{ph - mix}}(\Omega _D )} ({\bf{Q}}_i^{\rm{W}} ).
\end{equation}
Here $\chi ^{{\rm{ph - mix}}(\Omega _D )} ({\bf{Q}}_i^{\rm{C}} )$ is defined by
\begin{equation}\label{eqB25}
\begin{split}
\chi ^{{\rm{ph - mix}}(\Omega _D )} ({\bf{q}}) \equiv& \chi ^{{\rm{ph}}(\Omega  = 0)} ({\bf{q}}) - \bar \chi ^{{\rm{ph}}(\Omega _D )}  - \chi ^{{\rm{ph}}(\Omega _D )} ({\bf{q}}),\\
\chi _{o'_1 o'_2 m,o_1 o_2 n}^{{\rm{ph - mix}}(\Omega _D )} ({\bf{q}}) =&  - \int {\frac{{d{\bf{k}}}}{{S_{{\rm{BZ}}} }}} f_m ({\bf{k}})f_n^* ({\bf{k}})\left[ {\frac{1}{{\beta \hbar ^2 }}} \right.\sum\limits_\omega  {\left( {G_{o'_1 o_1 }^{\Omega _D } ({\bf{k}} + {\bf{q}},\omega )\bar G_{o_2 o'_2 }^{\Omega _D } ({\bf{k}},\omega )} \right.} 
 + \left. {\left. {\bar G_{o'_1 o_1 }^{\Omega _D } ({\bf{k}} + {\bf{q}},\omega )G_{o_2 o'_2 }^{\Omega _D } ({\bf{k}},\omega )} \right)} \right].
\end{split}
\end{equation}
Due to the positivity of $ - \chi ^{{\rm{pp - mix}}(\Omega _D )} ({\bf{q}})$ and $\chi ^{{\rm{ph - mix}}(\Omega _D )} ({\bf{q}})$, the ordering tendencies are underestimated in the renormalized MF than in the novel TUFRG + MF approach. However, in some special cases, e.g., in the case of the SC order in the single-band systems, the two critical conditions become identical due to the vanishing $\chi ^{{\rm{pp - mix}}(\Omega _D )} ({\bf{Q}} = 0)$.

\end{widetext}

\bibliographystyle{apsrev}
\bibliography{SJO_PRB_2023V2}

\end{document}